# Highlights and Discoveries from the *Chandra* X-ray Observatory[1]

H Tananbaum[1], M C Weisskopf[2], W Tucker[1], B Wilkes[1] and P Edmonds[1]

[1]**Smithsonian Astrophysical Observatory, 60 Garden Street, Cambridge, MA 02138.**
[2] **NASA/Marshall Space Flight Center, ZP12, 320 Sparkman Drive, Huntsville, AL 35805.**

**Abstract.** Within 40 years of the detection of the first extrasolar X-ray source in 1962, NASA's *Chandra* X-ray Observatory has achieved an increase in sensitivity of 10 orders of magnitude, comparable to the gain in going from naked-eye observations to the most powerful optical telescopes over the past 400 years. *Chandra* is unique in its capabilities for producing sub-arcsecond X-ray images with 100-200 eV energy resolution for energies in the range 0.08<E<10 keV, locating X-ray sources to high precision, detecting extremely faint sources, and obtaining high resolution spectra of selected cosmic phenomena. The extended *Chandra* mission provides a long observing baseline with stable and well-calibrated instruments, enabling temporal studies over time-scales from milliseconds to years. In this report we present a selection of highlights that illustrate how observations using *Chandra*, sometimes alone, but often in conjunction with other telescopes, have deepened, and in some instances revolutionized, our understanding of topics as diverse as protoplanetary nebulae; massive stars; supernova explosions; pulsar wind nebulae; the superfluid interior of neutron stars; accretion flows around black holes; the growth of supermassive black holes and their role in the regulation of star formation and growth of galaxies; impacts of collisions, mergers, and feedback on growth and evolution of groups and clusters of galaxies; and properties of dark matter and dark energy.

# Table of Contents

# 1 Introduction

Earth's atmosphere is an efficient absorber of X-rays, so the observation of cosmic X-rays had to await the dawning of the space age. The first hint that cosmic X-rays exist came in 1949 [1], when radiation detectors aboard rockets were briefly carried above the atmosphere where they detected X-rays coming from the Sun. It took more than a decade before a greatly improved detector discovered X-rays coming from sources beyond the solar system in 1962 [2].

The development of telescopes that can focus X-rays has led to an extraordinary leap in sensitivity. Within 4 decades of the detection of the first extrasolar X-ray source, NASA's *Chandra* X-ray Observatory, launched in 1999 aboard Space Shuttle Columbia, has achieved an increase in sensitivity comparable to going from naked-eye observations to the most powerful optical telescopes over the past 400 yr. As of mid-2013, ~500,000 cosmic X-ray sources have been detected, the most distant of which is about 12.5 billion light years from Earth.

The *Chandra* X-ray Observatory operates from ~80 eV to 10 keV with unique capabilities for producing sub-arcsecond X-ray images with 100-200 eV energy resolution, locating X-ray sources to high precision, detecting extremely faint sources, and obtaining high resolution spectra of selected cosmic phenomena. These qualities have established *Chandra* as one of the most versatile and powerful tools for astrophysical research in the 21st century. *Chandra* explores the hot, high-energy regions of the universe, observing X-ray sources with fluxes spanning more than 10 orders of magnitude. Sources range from the first discovered and X-ray brightest, Sco X-1, to the faintest in the *Chandra* Deep Field South survey, from which *Chandra* detects ~1 photon every 4 days. The extended *Chandra* mission provides a long observing baseline with stable and well-calibrated instruments, enabling temporal studies over time-scales from milliseconds to years.

Both thermal and non-thermal X-ray emission processes play important roles in the production of cosmic X-rays in a wide variety of settings. Neutron star surfaces and optically thick accretion disks around neutron stars and black holes produce blackbody radiation with a temperature $T \sim$ 1MK – 10 MK. Optically thin, hot gases with $T$ ranging from a few MK to a few 10 MK in stellar coronae and supernovae remnants (SNR) produce a rich spectrum of X-ray lines from cosmically abundant elements such as oxygen, silicon and iron. Clusters of galaxies, the largest gravitationally bound structures in the universe, are filled with 10 MK to 200 MK gas that is a strong source of thermal bremsstrahlung X-rays, and there is good evidence that a significant fraction of the baryons in the universe are in a diffuse cosmic web of hot gas with $T > $ 1 MK.

Cosmic non-thermal X-radiation is produced primarily by the synchrotron process, in which relativistic effects boost the frequency of the observed radiation from electrons by a factor $\gamma^2$, where $\gamma$ is the Lorentz factor. The production of electron synchrotron X-rays with energies ~ 10 keV requires $B\gamma^2 \sim 6 \times 10^{11}$, where B is the magnetic field strength in gauss. For example, the observed X-radiation from supernova shock waves with $B \sim 10^{-4}$ gauss, requires electrons with $\gamma \sim 8 \times 10^7$, corresponding to 40 TeV electrons.

Another important source of cosmic X-radiation occurs when low-energy photons undergo Compton scattering with high energy electrons. This occurs in hot coronae above accretion disks around black holes, where lower energy photons from the accretion disk are Compton-scattered into the X-ray band by hot electrons in the corona, and on a much larger scale, when cosmic microwave background photons scatter off the relativistic electrons that produce synchrotron radio emission in giant radio sources.

*Chandra* is part of a larger context where, for the first time, sub-arcsecond imaging of many cosmic sources is available across a wide band of wavelengths. The synergy with NASA's Hubble Space

Telescope and Spitzer Space Telescope, as well as large ground-based optical, millimeter, and radio telescopes is providing a more complete view of the cosmic processes at work in shaping the universe

In this report we present an overview that illustrates how observations using *Chandra*, sometimes alone, but often in conjunction with other telescopes, have deepened our understanding of topics as diverse as protoplanetary nebulae and exoplanets, massive stars, supernova shock waves, superfluid interior of neutron stars (NS), accretion flows around black holes (BH), co-evolution of supermassive black holes (SMBH) and galaxies, collisions and mergers as well as feedback processes in groups and clusters of galaxies, and the properties of dark matter and dark energy. In the remainder of this section and in Table 1 we give a brief summary of these highlights.

Section 2 gives an overview of the observatory and its current operational status, which after 14 yr, remains excellent, with no known limitations that preclude a mission of 25 yr, perhaps longer. Sections 3 through 7 present selected *Chandra* highlights on the following topics: stars and solar system objects - Section 3; SNR, NS and BH - Section 4; SMBH & active galactic nuclei (AGN) - Section 5; galaxies, groups and clusters - Section 6; and large-scale structure, dark matter, dark energy and cosmology - Section 7. Section 8 concludes with a brief set of metrics that illustrate the scientific impact of *Chandra* along with a few thoughts about prospects for the future.

*Chandra* observations of hot stellar coronae and shocked stellar winds are used to investigate the evolution of stars and the properties of clusters of recently formed stars. High-resolution grating spectra allow detailed modeling of mass loss from massive stars, suggesting that the outflowing gas is clumpy and implying that the mass loss is 3-4 times less than inferred from smooth wind models. Observations of the X-ray emission from four of the most massive known stars in a nearby galaxy support estimated masses >150 $M_\odot$, where $M_\odot$ (= $2 \times 10^{33}$ g) is the solar mass, for 3 of the stars, while significantly lowering the mass of a fourth. Combined infrared and X-ray observations of young star clusters provide insight into the formation and survivability of protoplanetary disks around stars. The data indicate that disks around massive stars have a shorter dissipation timescale than those around lower mass stars. *Chandra* observations of CoRoT-2, an unusual planetary system containing a transiting "hot Jupiter" with an inflated radius and a young (~200 Myr) main sequence host star indicate that X-ray flux from the star is eroding the planet's atmosphere. In our solar system, high resolution *Chandra* spectra have established that charge exchange between neutral (mostly hydrogen) atoms with solar wind ions is the primary mechanism for X-ray emission from comets and is also an important source of X-rays from the exospheres of Earth, Venus, and Mars.

At the other end of the stellar evolutionary time line, *Chandra* images and spectra of SNR provide insight into the nature of the progenitor stars and the supernova mechanism. The spatially resolved spectra make it possible to track the production and dispersal of heavy elements by supernovae. The forward and reverse shock waves are resolved, thus confirming the basic gas dynamical model for supernova shocks. *Chandra* observations of non-thermal synchrotron radiation from the forward shock, plus the detection of a pattern of X-ray stripes in the Tycho SNR provide key insights into the acceleration of protons to energies ~1000 TeV by a supernova shock wave. Images of jets and rings produced by the flow of relativistic particles powered by highly magnetized, rapidly rotating NS demonstrate that these objects, and by analogy rotating BH, can convert rotational energy into relativistic particles and transmit this energy over very large distances from the central object. Observations with *Chandra* probe the behavior of matter in the gravitational fields around NS and BH. These data provide evidence for a neutron superfluid in the interior of a NS, and enable the best measurement to date of the spin of a stellar mass BH. *Chandra* also plays a key role in an exciting, and controversial area of research that involves the so-called ultraluminous X-ray sources (ULX), which may be stellar mass BH accreting matter at an extraordinarily rapid rate or may indicate a new class of intermediate-mass BH, with masses ~ $10^2$-$10^5$ $M_\odot$. It is also possible that both types of source exist.

*Chandra*'s ability to pinpoint actively growing SMBH through the X-radiation they generate makes it a unique tool for studying the environment of SMBH and for tracking their growth and evolution across cosmic time. With *Chandra*'s angular resolution, it has been possible to detect many thousands of SMBH, and to resolve the cosmic X-ray background radiation into discrete sources, most of which are SMBH. Monitoring of the *Chandra* light curves for gravitationally lensed quasars establishes the extent of the regions producing X-ray and optical emission from a SMBH at ~10 to 70 $R_g$ respectively for one such system, where $R_g = GM/c^2$ is the gravitational radius for a BH of mass M. These results may require modifications to the standard thin disk model. Absorption studies confirm the general picture of an AGN as a SMBH surrounded by an accretion disk inside a thick torus of absorbing material. Many, but by no means all, of the differences between AGN can be explained by the orientation of the torus to the line of sight. The accretion process is complex, with some matter falling into the SMBH, while some flows out in highly-ionized, high-speed winds with velocities ranging from $10^2$-$10^5$ km s$^{-1}$ and some shoots away as jets of relativistic particles. In a few cases, *Chandra* measures the density and temperature profiles of the accretion flow near SMBH, and at the other extreme, traces jets of X-ray emitting particles over distances of hundreds of kpc (1 kpc = 3,262 light yr), well outside the confines of the host galaxies.

Beyond the scale of galaxies, X-ray and radio data demonstrate convincingly that SMBH and intense bursts of star formation can affect their environment over hundreds of kpc. Combined images of galaxy clusters and groups made with *Chandra* and radio telescopes provide strong evidence for a large-scale feedback mechanism connecting explosive activity associated with SMBH, cooling of intergalactic gas, the rate of star formation, the growth of the central BH, and the development of the host galaxy's structure. Jets from SMBH and winds driven by bursts of star formation also enrich the intergalactic medium with heavy elements. *Chandra* has measured the hydrodynamic and thermodynamic properties of the gas in groups and clusters of galaxies, and determined the metal abundance and baryonic fraction in the cluster gas. These observations provide an accurate estimate of the relative amounts of dark and luminous matter in the universe. The dark matter component has a density ~5 times that of the baryonic matter. In turn, the dominant baryonic component in the clusters is hot gas with temperature typically 10-100 MK and mass roughly 6 times the mass in stars.

*Chandra* images and spectra of the hot gas in galaxy clusters play a key role in the investigating the nature of dark matter and dark energy. X-ray and optical data have revealed the separation of dark and baryonic matter in colliding clusters of galaxies. This observation is the best evidence yet for the existence of dark matter in that alternative theories of gravity are very unlikely to explain the separation of dark and baryonic matter. Measurement of the gas mass for large samples of galaxy clusters generates the cluster mass function which is used to determine the rate for growth of structure. This method establishes new constraints on cosmological parameters and the evolution of dark energy equation of state that are among the tightest and most robust available from current data.

Table 1 gives a subjective and by-no-means complete list of significant discoveries made with *Chandra*. These and other highlights are discussed in greater detail in the following sections.

Table 1. Selected Highlights from the *Chandra* X-ray Observatory.

| Highlight | Section in text |
|---|---|
| Combined *Chandra* and infrared observations of young star clusters show that the fraction of stars with protoplanetary disks drops to 20% at 5 Myr and near zero at ~ 10 Myr, implying that the formation of large gas planets has to be completed within this time frame. | 3 |
| High resolution spectra establish that charge exchange with solar wind ions is the primary mechanism for X-ray emission from comets and the exospheres of Earth and other planets. | 3 |
| High-resolution images uncover a central compact object in the CasA SNR. Monitoring over a decade shows cooling and provides first direct evidence for neutron superfluidity in the NS core. | 4 |
| Images and spectra of forward and reverse shock waves in SNR confirm the basic gas dynamic model. Spatial, spectral and time-variable features in several SNR provide insight into the acceleration of cosmic rays. | 4 |
| Images resolve jets and rings in pulsar wind nebulae produced by flow of relativistic particles from highly magnetized, rapidly rotating NS. | 4 |
| Two methods measure near-maximal spin for stellar mass BH and SMBH. | 4, 5 |
| Space density of SMBH measured over large range of redshifts resolves the X-ray background radiation into discrete sources, mostly SMBH. Data suggest cosmic down-sizing whereby the most massive SMBH grow at the earliest times while less massive SMBH are still growing today. | 5 |
| Surveys of SMBH indicate that both accretion and mergers play a role in their growth over cosmic time scales. | 5 |
| Detection of relativistic X-ray jets from SMBH on scales ranging from a few tens of pc to hundreds of kpc requires reacceleration of particles, possibly from shocks at discrete knots seen in images. | 5 |
| Evidence for heating of hot gas in galaxies and clusters by SMBH outbursts supports concept that SMBH can regulate the growth of galaxies. | 6 |
| X-rays images and spectra of galaxies and clusters show enrichment of the interstellar and the intergalactic medium via winds driven by bursts of star formation and jets from SMBH. | 6 |
| *Chandra* measures hydrodynamic and thermodynamic properties of gas in clusters of galaxies, detecting cold fronts and shock waves produced by cluster mergers. *Chandra* also measures metal abundance and baryonic fraction in outer regions of clusters. | 6 |
| Images and spectra show that dark matter component of clusters of galaxies has density ~5 times that of the baryonic matter and that dominant baryonic component of clusters is hot gas with temperature ~10-100 MK and mass ~6 times that of the mass in stars. | 6, 7 |
| Determination of the cluster mass function by *Chandra* and comparison with models for growth of structure provide one of the most precise measures to date for the dark energy equation of state. | 7 |

## 2    Overview and Status of the *Chandra* X-ray Observatory

The *Chandra* X-ray Observatory, the third of NASA's four Great Observatories and its flagship mission for X-ray astronomy, was launched by NASA's Space Shuttle Columbia, Eileen Collins commanding. The launch took place on July 23, 1999 and *Chandra* was boosted to a high earth elliptical orbit by a separable Inertial Upper Stage followed by several burns of engines integral to the spacecraft. The other Great Observatories are the Hubble Space Telescope (HST), the Compton Gamma-Ray Observatory (no longer operating but succeeded in 2008 by the Fermi Gamma-ray Telescope) and the Spitzer Infrared Space Telescope.

The key to *Chandra*'s success is the great advance in angular resolution. The mirrors produce images with half-power-diameter ~0.5 arcsec for X-ray energies in the range $0.08 < E < 10$ keV. This angular resolution represents a 10-fold improvement over the two previous best X-ray telescopes – the US-led Einstein X-ray Observatory (1978-1981) and the German-led Röntgensatellit (*ROSAT* – 1990 -1999). The 10-meter focal length High Resolution Mirror Assembly (HRMA) consists of four nested pairs (paraboloid-hyperboloid), grazing-incidence, glass-ceramic mirrors coated with iridium to enhance their reflectivity at X-ray wavelengths. The Observatory's highly eccentric orbit makes possible continuous observations of up to ~185 ks. The observing efficiency, ranging from 65 to 75%, is limited primarily by the need to protect the instruments from particles, especially protons, during *Chandra*'s passages through Earth's radiation belts.

NASA's Marshall Space Flight Center provides overall project management and project science oversight. The Project Scientist is Martin C. Weisskopf of the Marshall Space Flight Center (MSFC). Day-to-day responsibility for *Chandra* lies with the *Chandra* X-ray Center (CXC), Harvey Tananbaum, Director until April 20, 2014 and Belinda Wilkes now Director. The CXC is located at the Cambridge Massachusetts facilities of the Smithsonian Astrophysical Observatory (SAO).

The observatory (Figure 2-1) consists of three principal elements: (1) the telescope comprised of the HRMA, two X-ray transmission gratings that can be inserted into the X-ray path, and a 10-meter-long optical bench; (2) a spacecraft module that provides electrical power, communications, and attitude control; and (3) a Science Instrument Module that holds two focal-plane cameras – the Advanced Charged-Coupled-Device (CCD) Imaging Spectrometer (ACIS) and the High Resolution Camera (HRC) – and mechanisms to adjust the camera's position and focus. The Observatory is 13.8 m in length, has a mass of 4,800 kg, and the furthest ends of its solar panels are 19 m apart.

A system of gyroscopes, reaction wheels, reference lights, and a CCD-based star camera enables *Chandra* to maneuver between targets and point stably while also providing data for accurately determining the sky positions of observed objects. The blurring of images due to pointing uncertainty is <0.10 arcsecond, negligibly affecting the resolution of the telescope. Absolute positions can be determined to ≤0.6 arcsec for 90% of sources, providing an unrivaled capability for X-ray source localization.

The ACIS Instrument Principal Investigator (PI) is Gordon Garmire. ACIS was developed by collaboration amongst Pennsylvania State University (PSU), the Massachusetts Institute of Technology (MIT) Kavli Institute for Astrophysics and Space Research, and the Jet Propulsion Laboratory (JPL). The ACIS was built by Lockheed Martin and MIT, with the CCDs developed by MIT's Lincoln Laboratory. ACIS contains two arrays of CCDs that provide position and energy information for each detected X-ray photon. The imaging array is optimized for spectrally resolved ($E/\Delta E \sim 10\text{-}40$ for the energy range 0.1 - 10 keV) high-resolution imaging over a wide field of view (17arcmin×17arcmin). In conjunction with the High Energy Transmission Grating (HETG), the spectroscopy array provides high-resolution spectroscopy with a resolving power ($E/\Delta E$) up to 1000 over the 0.4-8 keV band.

The HRC Instrument PI is Stephen Murray. The HRC was built at SAO and uses two micro-channel plate detectors, one for wide-field imaging and the other serving as readout for the Low Energy Transmission Grating (LETG). When used with the HRC's spectral array, the LETG provides spectral resolution >1000 at low (0.08 – 0.2 keV) energies, while covering the full *Chandra* energy band. The HRC detectors have the highest spatial resolution (0.13 arcsec/pixel) on *Chandra* and, in certain operating modes, the fastest time resolution (16 μs).

A.C. Brinkman of the Laboratory for Space Research in Utrecht, the Netherlands was the original PI of the LETG. The instrument was built in collaboration with the Max Planck Institut für Extraterrestrische Physik in Garching, Germany, and the grating was manufactured in collaboration with Heidenhaim GmbH. The Instrument PI for the HETG is Claude Canizares of MIT. The HETG is comprised of two grating assemblies – the High Energy Grating (HEG) and the Medium Energy Grating (MEG). The HEG intercepts X-rays from only the two inner mirror shells of the HRMA and the MEG intercepts X-rays from only the two outer mirror shells. The dispersion directions of the HEG and MEG are offset by 10 degrees so the two patterns can be easily distinguished.

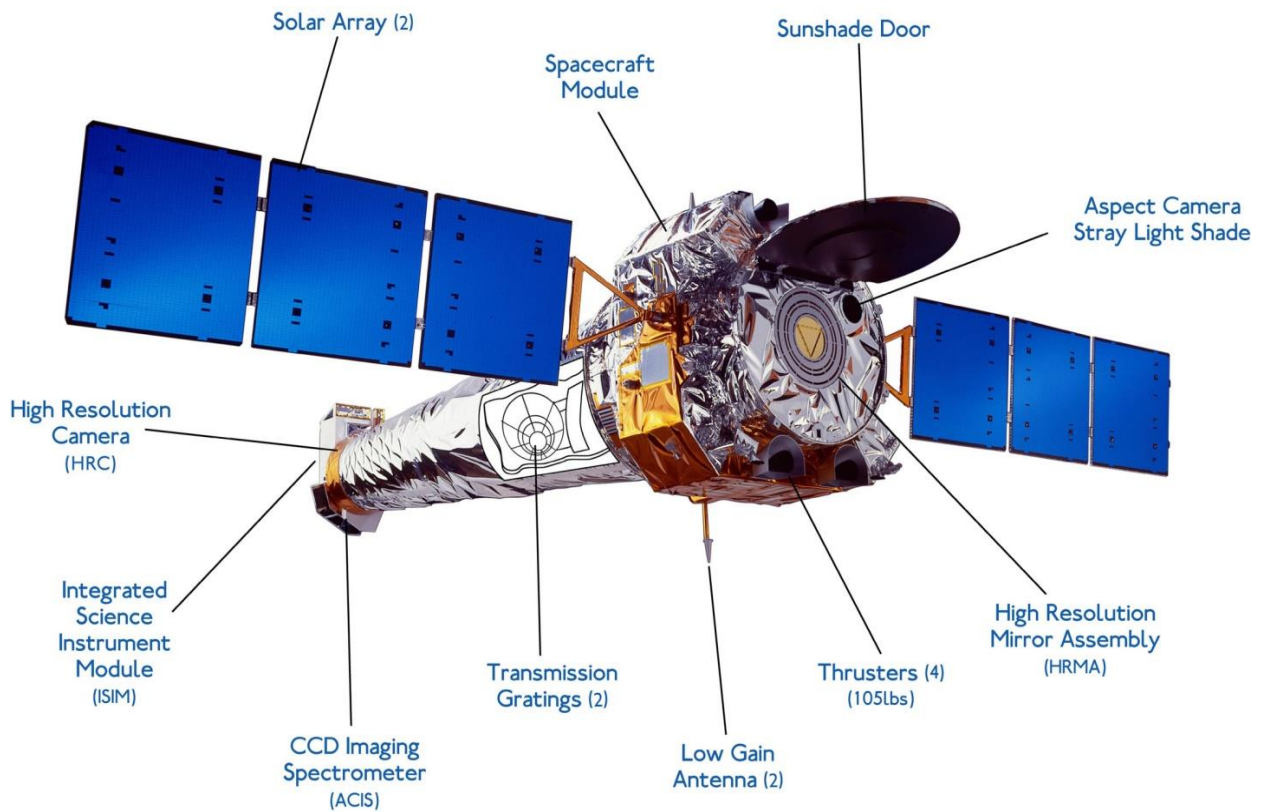

Figure 2-1. Cutaway representation of the *Chandra* X-ray observatory showing major elements. Credit: NASA/CXC/NGST.

TRW was the prime contractor and spacecraft developer, Hughes-Danbury Optical Systems polished the mirrors and Optical Coating Laboratories, Inc. coated them. Eastman Kodak integrated the optics, and Ball Aerospace and Technology Corporation built the Science Instrument Module and Aspect System.

*Chandra* was designed with an official lifetime of 3 yr, based on life-testing of various electronics and other components, and its design goal was five years. However, *Chandra*'s highly elliptical, high-earth orbit provides certain benefits for mission longevity. Unlike satellites in low-earth orbit, which cycle from sun-light to darkness as many as 15 times per day, a satellite in high orbit like *Chandra* incurs much less power and thermal cycling which introduces much less stress on sub-systems. As a result, *Chandra* has operated for nearly a decade and a half with almost all of its redundant sub-systems still operating on the prime or initial side. There were two exceptions. The first was a package of gyroscopes which was switched to the redundant set after a few years when 1 of the 2 primary gyro rotors showed an increase in its bias current. The second was the fine sun sensor which was switched in 2013 to the redundant unit due to noise near the edge of the field of view. The Observatory's thermal insulation system has degraded somewhat during *Chandra*'s operation due to impacts of solar particles and ultra-violet radiation. The main effect is that some parts of the Observatory run hotter than at launch while a few others run colder.

The scheduling of targets has become more challenging as the team folds in thermal considerations when observing plans for each week are assembled or when an unexpected astronomical event triggers a peer-review-approved Target of Opportunity or a request for Director's Discretionary Time. The output of the solar arrays has decreased as expected, and after 14 yr they still supply more than twice the required power. An on-board propellant is used for the thrusters that unload momentum from the reaction wheels which adjust the pointing of the Observatory, and the supply is sufficient for several more decades of operation. An assessment by the Aerospace Systems sector of the Northrop Grumman Corporation has determined that there are no known limitations which would preclude a mission of at least 25 yr.

# 3 Stars and Solar System Objects

## 3.1 Stars

Although the X-ray emission from stellar coronae is a small fraction of a normal star's bolometric (i.e. total) luminosity, X-ray observations provide indicators of important physical parameters such as magnetic activity and age. High-resolution X-ray spectroscopy with *Chandra*'s grating spectrometers has enabled measurement of many spectral lines — including resolved multiplets — and, in some cases, measurements of line profiles and line shifts. These measures provide critical diagnostics of the density, temperature, composition and dynamics of hot coronal plasmas and have enabled tests of models for the X-ray emission.

Furthermore, *Chandra*'s high spatial resolution, sensitivity, broad band pass, and large field of view give astronomers an important window for viewing the drama of stellar evolution, from the formation of stars in dense clouds of dust and gas to their demise, either quietly as white dwarfs or violently as supernovae. Coupled with infrared and optical observations, *Chandra* data may identify young stars, with and without protoplanetary disks, as well as investigate the process of the triggering of star formation by winds and radiation from massive stars.

### 3.1.1 *Stellar Winds*

Stellar winds carry away a significant portion of a massive star's material as the star ages. The winds deposit energy, momentum and matter into the interstellar medium. X-ray emission-line profile analysis provides a way, independent from more traditional techniques, to measure these mass-loss rates. Unlike ultraviolet absorption line diagnostics, X-ray line profile analysis is not very sensitive to the ionization balance, and relies on the continuum-opacity rather than line-opacity. Because of this reliance, the X-ray analysis is not subject to the uncertainty associated with saturated absorption. For example, *Chandra* HETG observations of two O supergiants, (ζ Pup (Figure 3-1) and HD 93129A [3, 4]), have led to a reevaluation of the mass-loss rates for these stars. These data indicate mass-loss rates 3-4 times lower than those determined from smooth-wind models and are consistent with estimates which allow for small-scale clumpiness.

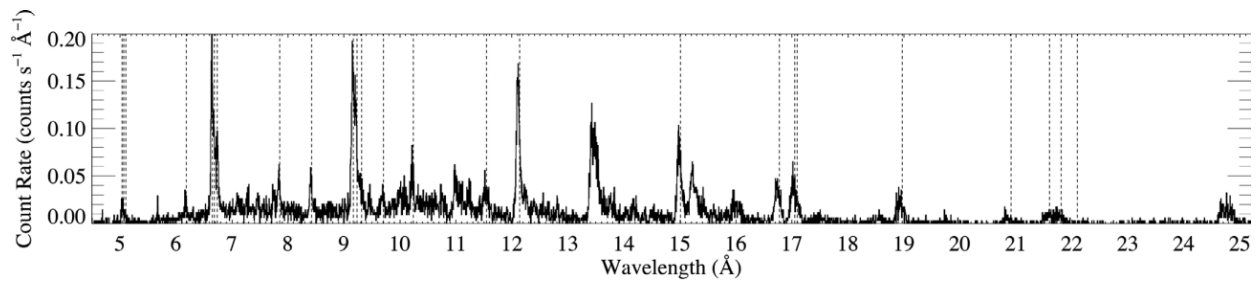

Figure 3-1. A *Chandra* HETG spectrum of ζ Pup [3]. The most prominent lines are He-like complexes from Si XIII (6.65, 6.69 & 6.74 Å) and Mg XI (9.17, 9.23 & 9.31 Å), Lyα lines from Ne X (12.13 Å), and OVIII (18.97 Å) and Fe XVII lines (15.01, 16.78, 17.05 & 17.10 Å). Vertical dashed lines show the laboratory rest wavelengths of lines fitted with a wind profile model. The rather modest asymmetry in the line profiles is fit by assuming a moderate mass-loss rate of $3.5 \pm 0.3 \times 10^{-6}$ $M_\odot$/ yr . Credit: Adapted from [3].

The X-ray emission from most massive O stars is likely due to interacting shock waves embedded in powerful winds. *Chandra* data revealed relatively hard (flat) spectra, which have been interpreted as due to absorption of low energy X-rays by cooler material that envelope the hot, X-ray-emitting gas. In some cases collisions of confined gas streams flowing from opposite hemispheres along magnetic loops appear to explain the data. In still other sources, the X-ray emission is produced by colliding winds in a binary system (see, e.g., the discussion [5] of *Chandra* observations of $\theta^1$ Ori C). *Chandra* grating spectra provide rich diagnostics of plasma conditions and allow detailed modeling. For example, magnetically channeled wind shock mechanisms successfully model the spectroscopy of $\theta^1$ Ori C [5,6] (Figure 3-2). Once the many unresolved point sources are removed from the *Chandra* images, the long-predicted diffuse X-rays from shocked O star winds come into view. *Chandra* has detected diffuse X-ray emission from ~10 MK gas at levels in the range from ~1-200×10$^{33}$ erg s$^{-1}$ [7, 8]. In most cases the diffuse gas appears to be generated by stellar winds from massive stars colliding with other winds or with the surrounding low-density clouds of partially ionized gas in which star formation has recently taken place.

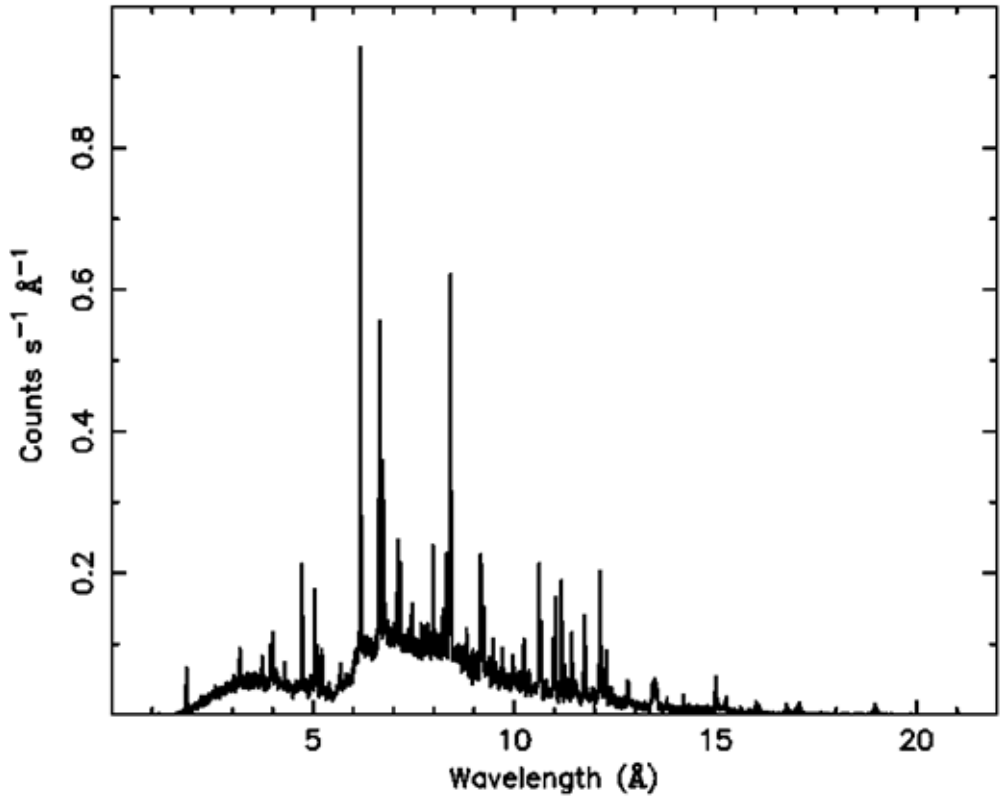

Figure 3-2. *Chandra* grating spectrum of $\theta^1$ Ori C, the brightest source in the Orion image of Figure 3-4. The data reveal spectral lines from various ionization states of Fe, Ca, Ar, Si, S, Mg, and O (see Table 2 of [5]. Credit: Courtesy M. Gagne, N. Schulz and the HETG GTO Program. Adapted from [6].

3.1.2 *Stellar Mass*

The determination of the masses of very massive stars using traditional approaches can be unreliable, due in part to severe spatial crowding within the cores of the star clusters where such objects are found. An upper limit ~150 times the mass of the sun ($M_\odot$) is generally accepted, but it is uncertain whether this limit is statistical or physical [9]. X-ray observations offer a means of not only avoiding spatial crowding but also helping to determine whether a star is single or a part of a binary system. This is an important distinction as, if a candidate is part of a binary system, interactions between the two stars' stellar winds produces stronger X-ray emission than would be expected from a single star. For example, *Chandra* observations of four massive star candidates in the star cluster R136 in 30 Doradus [10], a star-forming cloud in the nearby Large Magellanic Cloud galaxy, discovered that the X-ray luminosity of one of the stars, R136a4, is consistent with that expected from a colliding wind binary system [11] and so significantly lowered the mass estimate initially based on optical observations and evolutionary models. Conversely, the X-ray emission from the other three stars was indicative of a single star system, lending further credence to the initial mass estimates of 320, 240, and 165 $M_\odot$, respectively, for these stars.

### 3.1.3 *Young Star Clusters*

*Chandra*'s spatial resolution has also opened up the X-ray window to allow the study of massive young star clusters. For a typical *Chandra* observation of a single cluster, hundreds to thousands of lower-mass pre-main sequence stars are detected with sub-arcsecond position accuracy. *Chandra* images often penetrate deeper into molecular clouds than the largest ground-based telescopes, and trace previously unseen embedded populations. Tens of thousands of such stars have been detected using *Chandra* in several surveys. Moreover, a large fraction of these stars was missed by infrared-only observations because their disks have dissipated or been destroyed so they show no infrared excess (the property traditionally used to distinguish cluster members from field stars). Figure 3-3 and Figure 3-4 illustrate the powerful impact of *Chandra* on the study of young star clusters. The *Chandra* X-ray data of Eta Carina yielded a catalog of >14,000 X-ray point sources [12], more than 12,000 of which are young stars with ages between 1 and 10 Myr. Removing the point sources further allows a study of the diffuse X-ray emission that pervades the region. The spectra of this diffuse emission suggest that it is generated by charge exchange at the interfaces between Carina's hot rarefied gas and its many cold neutral pillars, ridges, and clumps [13]. Other contributions to the diffuse emission are from stellar winds and SNR.

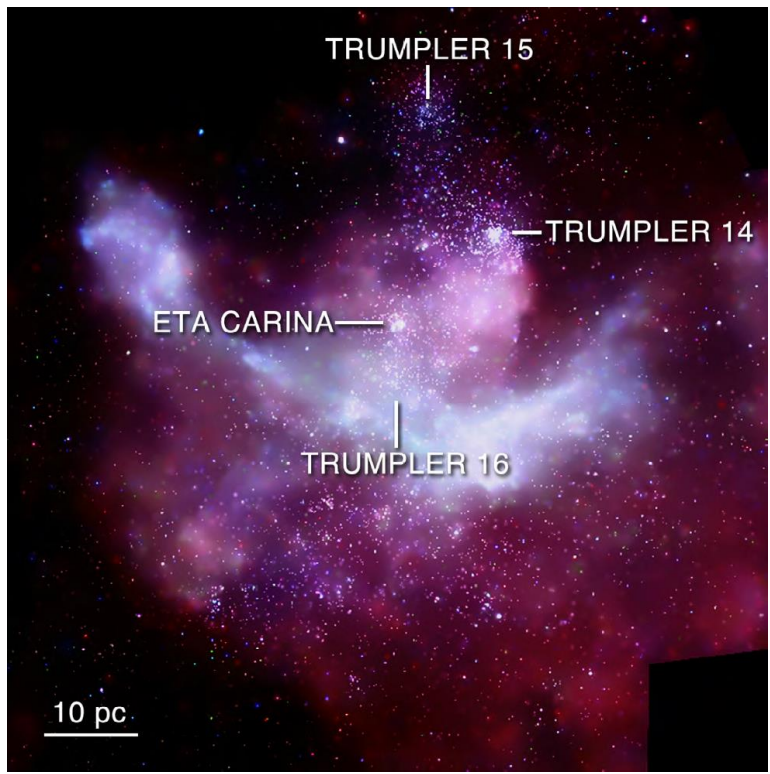

Figure 3-3. Diffuse X-ray emission and point sources from young stars are shown in this *Chandra* mosaic image of the Carina Nebula [12]. The image is 68 arcmin across (45 pc). The positions of Eta Carina and three star clusters are labeled. Credit: NASA/CXC/PSU/L. Townsley et al.

Combining X-ray and infrared data for a star cluster also allows one to infer the numbers of young stars with and without disks in the star cluster. Dusty protoplanetary disks emit infrared radiation and the young stars can be identified through their X-ray emission. *Chandra* and Spitzer surveys of the OB association IC 1795 were combined for such a study [14]. The surveys found that the disk fraction for sources with masses >2 $M_\odot$ is ~20%, while the fraction for lower mass objects (0.8–2 $M_\odot$) is higher, ~50%. One concludes that disks around massive stars have a shorter dissipation timescale than those around lower mass stars.

Comparative studies of many star clusters mapping the general trend of disk dissipation in low-mass and high-mass environments suggest that the fraction of stars with disks is about 80%-90% in young clusters (1 Myr old). By 5 Myr the disk fraction drops to 20% and after 10 Myr almost all disks have dissipated. Consequently one may infer that the formation of gas planets has to be completed within this time-scale as the material to form the planets is no longer available. Further, using the percentage of disks around stars as a proxy for age, the chronology of star formation can be mapped across star forming regions. *Chandra*, optical, and infrared observations of the massive young cluster NGC 6611 and its parental cloud (the Eagle Nebula) yield evidence [15] for sequential star formation going from the southeast (age ~2.6 Myr) to the northwest (age ~0.3 Myr). A giant molecular shell ~200 pc wide and likely created by a wave of supernova activity 6 Myr ago is also seen in the southern portion of the Eagle Nebula [16]. Its position and motion are consistent with an interaction of the interstellar medium with multiple supernova explosions and/or stellar winds of massive stars near the Galactic plane 2–3 Myr ago, which, in turn, triggered the observed wave of star formation.

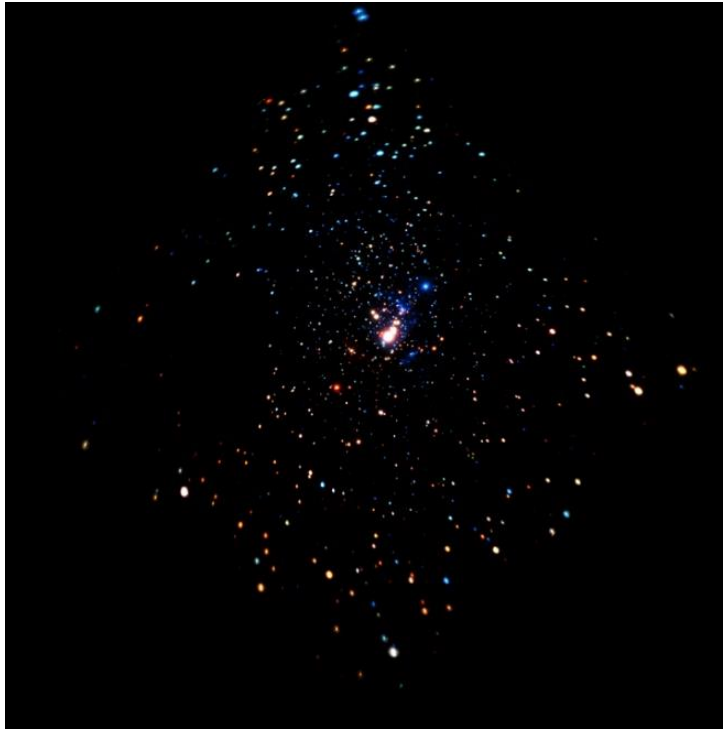

Figure 3-4. A *Chandra* deep image of the Orion Nebula Cluster containing over 1,600 X-ray sources of which ~1,400 are young stars in the nebula [17.] The rest are either background galaxies or foreground stars. The red X-ray sources are mostly young stars with little absorption by intervening gas and the blue are mostly young stars with larger amounts of gas absorption. The luminous source in the middle of the cluster, Theta 1 Orionis C, is the brightest and most massive star in the cluster. The image is 16 arcmin across (2.1 pc). Credit: NASA/CXC/PSU/E. Feigelson & K. Getman et al.

Protoplanetary disk evolution likely depends on the stellar environment. Evidence from neighboring star forming regions in the Gould Belt, a large (~600 pc in length) ring of gas and young stars in our galaxy, indicates that massive stars can erode and evaporate disks surrounding nearby low-mass stars (see e.g. [18]). The implications are that planet formation might be impeded or even prevented in massive stellar clusters. A *Chandra* survey of the Cygnus OB2 association was undertaken to investigate the erosion of protoplanetary disks by hot young stars. At 1450 pc away, Cygnus OB2 is the closest massive star forming region, and hosts over a thousand OB stars. *Chandra* [19] surveyed 1 square degree, covering the entire central cluster and its immediate environs (Figure 3-5) pinpointing more than 5000 young stars. Combined with multiwavelength optical and infrared observations, the *Chandra* study of the association indicates that the fraction of stars with disks is several times smaller than in nearby lower mass clusters of similar age. Low disk fractions are especially found close to the more massive OB stars, indicating the corrosive power of their intense radiation for planet-forming disks.

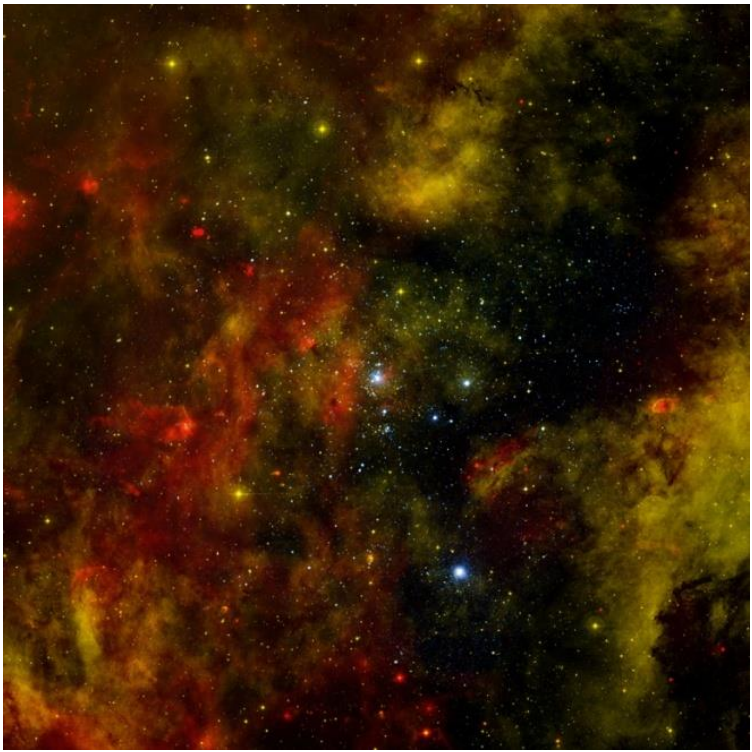

Figure 3-5. A composite image of the massive Cygnus OB2 association, showing X-rays from *Chandra* (blue), infrared data from Spitzer (red), and optical data from the Isaac Newton Telescope (orange). The *Chandra* study [19] has detected more than 5000 stellar X-ray sources in this system and finds evidence for severe erosion of planet-forming disks due to the radiation from numerous massive stars. The image is 11.8 arcmin across (4.9 pc). Credit: X-ray: NASA/CXC/SAO/J. Drake et al, Optical: Univ. of Hertfordshire/INT/IPHAS, Infrared: NASA/JPL-Caltech.

3.1.4  *Planetary Nebula*

*Chandra* has played a pivotal role in the study of planetary nebulae. The name derives from the planetary disk-like appearance of these objects when viewed with a small telescope. In fact they are expanding glowing shells of ionized gas ejected from giant stars. Planetary nebulae represent a brief phase when an intermediate mass (~1–8 $M_\odot$) star is running out of nuclear fuel and is making the transition from a giant star to a white dwarf. By revealing both the point-like X-ray sources and diffuse X-ray emission, *Chandra* is able to provide new insight into the nature of planetary nebulae. In particular the *Chandra* Planetary Nebula Survey [20] has established that ~50% of planetary nebulae harbor point X-ray sources, while ~30% display emission from hot bubbles generated via shocks and wind interactions (Figure 3-6). The high frequency of X-ray sources with a hard-x-ray excess (i.e. a flat spectrum component) that are associated with the central stars points to the frequent presence of binary companions that are likely responsible for nonspherical morphologies. The extended hot bubbles of X-rays appear within optically dark central cavities, indicating the presence of the superheated ($>10^6$ K) gas whose supersonic expansion in the planetary nebula interior generates a "tsunami" that reshapes the system. Thus, X-ray observations serve as complementary probes of the mechanisms underlying planetary nebula structures and their structural evolution.

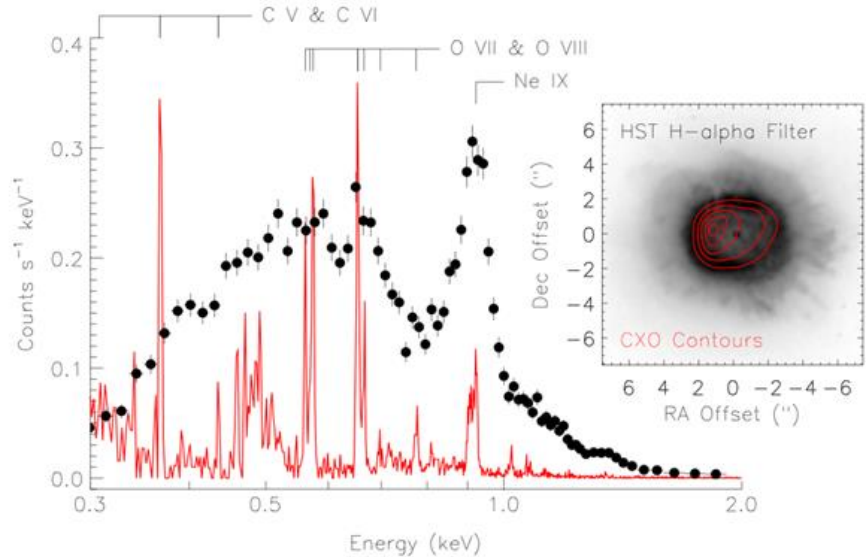

Figure 3-6. The inset shows a Hubble Space Telescope Hα image of the planetary nebula BD+30°3639 with *Chandra* contours of the X-ray emitting bubble [21]. Strong Ne emission from the bubble is apparent in the *Chandra* ACIS-S3 CCD spectrum (black circles). Higher spectral resolution LETG data further reveal that the low-energy end of the hot bubble spectrum is dominated by emission lines of C and O, and constrain its temperature to lie in the range 1.7-2.9 MK [22]. The forest of lines near 0.5 keV is likely due to C6+ ions from the hot bubble penetrating exterior (cooler) nebular gas before recombining [23]. Credit: Adapted from [21] and [23].

## 3.2 Stellar X-rays and Exoplanets

*Chandra* observations of CoRoT-2, an unusual planetary system containing a transiting hot Jupiter with an inflated radius, and a young (~200 Myr) main sequence host star (CoRoT-2a) [24] indicate that the star is an exceptionally active X-ray emitter. The X-ray flux at the distance of the planet is estimated to be roughly five orders of magnitude larger than the solar X-ray flux received by Earth. This high X-ray flux is likely eroding the planet's atmosphere at a rate $\sim 5\times 10^{12}$ g s$^{-1}$. The relatively strong X-ray luminosity may be due to the star's interaction with the planet, which could have spun the star up, or enhanced its magnetic activity, or both [25]. Ongoing *Chandra* observations of solar analogs with nearly identical stellar properties and those that host planets spanning a mass range $M \sin i = 0.1–9.7\ M_{Jup}$, (where *i* is the angle of inclination to the plane of the sky) and separations $a = 0.04–3.8$ AU are testing whether X-ray emission from stars with close-in planets is systematically enhanced. These measures will evaluate or constrain the efficiency of any star-planet interaction. If star-planet interaction is indeed a ubiquitous phenomenon in hot Jupiter systems, X-ray enhancements can provide a unique, direct probe of exoplanet magnetic fields.

*3.3    Solar System Objects*

The interaction of energetic particles and radiation from the Sun with the atmospheres of planets, their satellites, and comets produces X-radiation through a number of physical processes: scattering, fluorescence, charge exchange, or the stimulation of auroral activity. *Chandra* has observed — in many cases discovered — X-rays from Venus [26]; the Earth [27] and its moon [28]; Mars [29]; Jupiter [30, 31], its aurorae [32], some of its moons [31, 32] and the Io plasma torus [31]; Saturn [33, 34] and its rings [35]; and numerous comets [36, 37]. For overviews of these observations, see e.g. references [30, 31, 32].

An important mechanism for producing X-rays from solar system objects is charge exchange, which occurs when a highly ionized solar wind atom collides with a neutral atom (gas or solid) and captures an electron, usually in an excited state. As the ion relaxes, it radiates an X-ray characteristic of the solar wind ion. Lines produced by charge exchange with solar wind ions such as CV, CVI, OVII, OVIII, and NeIX have all been detected.

The X-ray spectrum of a comet measures its out-gassing rate (e.g. [36] ) and probes the solar wind in situ — in essence, a laboratory for highly ionized particles in low-density plasmas. Prior to *Chandra*, the origin of cometary X-rays was debated. Now, observational and theoretical work has demonstrated that charge-exchange collisions of highly charged solar wind ions with cometary neutral species are the best explanation for the observed X-ray emission [32]. *Chandra* has also been used to discover X-ray signatures of charge exchange in the exospheres of Venus and Mars, enabling measurement of the planetary outgassing rates. In addition, *Chandra* (and *XMM-Newton*) also probe planetary atmospheres through the detection of fluorescence caused by solar X-rays.

Planets, which have magnetic fields e.g., Earth, Jupiter, and Saturn, generate X-rays by auroral activity. *Chandra* observations discovered [38] time-variable X-ray flux from Jupiter that originates, not from the ultra-violet (UV) auroral zone as had been expected, but from higher latitudes near Jupiter's pole – regions that map to the outer boundary of Jupiter's magnetosphere. Charged particles precipitating into the Jovian atmosphere must therefore originate from much farther than the Io plasma torus. Altogether these discoveries suggest a complicated current system in Jupiter's polar magnetosphere, requiring a revision of models.

Charge exchange X-rays may be produced from the interaction of the particles in the solar wind with the Earth's exosphere (geocorona), with material in and outside the magnetosheath, and possibly even with the heliosphere. The magnetosheath is the region of space between the magnetopause (the abrupt boundary between the earth's magnetosphere and the surrounding plasma) and the bow shock of a planet's magnetosphere. The heliosphere is a region of space about our Sun containing charged particles. It extends well beyond the orbit of Pluto. *Chandra* data suggest that at energies near 0.75 keV, the total local soft-x-ray intensity is from the heliosphere, while near 0.25 keV, the heliospheric component is significant, but does not explain all the observations. A mix of solar wind charge-exchange emission and a warm (~0.6 MK), rather than hot, local bubble ~ 100 pc in diameter in the interstellar medium appears to be required to explain these observations [39]. The presence of such a bubble can affect the search for X-ray emission from the warm, hot, intergalactic medium (WHIM), a 0.1 MK - 10 MK plasma that has been proposed to contain over half the baryons at the current epoch. Ignoring the contribution of solar wind charge exchange emission to the background could lead to the wrong interpretation of observed emission as due to the WHIM. The WHIM is discussed further in 7.1.

## 4 Supernova Remnants, Neutron Stars, & Black Holes

### 4.1 *Supernovae and Supernova Remnants*

Every 50 yr or so, in a galaxy the size of our Milky Way galaxy, a supernova occurs. Supernovae are among the most violent events in the universe, and signal the catastrophic collapse of the central regions of a star. The supernova is extremely bright, with an optical luminosity $L \sim 10^8\text{-}10^{10}\ L_\odot$ for a few months ($L_\odot = 4 \times 10^{33}$ erg s$^{-1}$ = solar luminosity). For thousands of years the expanding stellar ejecta disperse heavy elements that enrich the interstellar gas, drive shock waves that heat the interstellar gas, accelerate particles to relativistic energies, and trigger the formation of new stars.

The final stage in the evolution of a star is determined by its initial mass. Stars with initial masses less than the Chandrasekhar limit of 1.4 $M_\odot$ eventually collapse to form a white dwarf, which is stabilized by electron -degenerate pressure at a radius $R_{wd} \sim 5000$ km. Stars with initial masses in the range 1.4 – 8 $M_\odot$ lose sufficient mass through stellar winds in the course of their evolution, so they eventually form white dwarfs, too.

The core of a star with mass > 8-10 $M_\odot$ cannot be stabilized by electron-degenerate pressure, and collapses catastrophically to form a NS, and then a BH if the star is sufficiently massive [40, 41]. Just prior to collapse the star consists of different layers with the products of the different consecutive thermonuclear fusion stages. From the core to the outside one expects: iron-group elements, then silicon-group elements, oxygen, neon and magnesium, carbon, helium, and, finally, unprocessed hydrogen-rich material.

The creation of the iron-group core, which lasts about a day, is the beginning of the end of the star, as no energy can be gained from nuclear fusion of iron. The details of what happens next are the subject of considerable debate, but there is general agreement that the collapse of the core to form a NS releases an enormous amount of energy. Most of the gravitational energy liberated ($E \sim GM^2/R_{ns} \sim 10^{53}$ erg, with $R_{ns}$ = the neutron star radius ~ 10 km) is in the form of neutrinos, but a substantial amount of energy is carried away in the ejecta at speeds of thousands of kilometers per second. The energy of the ejected matter is ~ $10^{51}$ ergs and is set, not by the neutron star's binding energy, but by the binding energy of the degenerate iron core, which is similar to that of a white dwarf. Supernovae produced by this process are called core-collapse supernovae.

Supernovae were originally classified on the basis of their observed optical properties. Type II supernovae show conspicuous evidence for hydrogen in the expanding debris ejected in the explosion, whereas Type I supernovae do not. By the 1980's evidence had accumulated that, except for the absence of hydrogen in their spectra, some Type I supernovae showed many of the characteristics of Type II supernovae. These supernovae, called Type Ib and Type Ic, apparently differ from Type II because they lost their outer hydrogen envelope prior to the explosion. The hydrogen envelope could have been lost by a vigorous outflow of matter prior to the explosion, or because it was pulled away by a companion star. Both scenarios seem possible.

Type II supernovae occur in regions with an abundance of bright, young stars, such as the spiral arms of galaxies. They rarely occur in elliptical galaxies which are dominated by old, low-mass stars. Since bright young stars are typically stars with masses greater than about 10 $M_\odot$, this and other evidence led to the conclusion that Type II and similar supernovae are core-collapse supernovae.

Type Ia supernovae, in contrast, are observed in all kinds of galaxies, and are thought to be produced when a white dwarf star is driven over the Chandrasekhar limit by accretion of matter from a companion star or by the merger with another white dwarf. The ensuing gravitational collapse triggers a thermonuclear explosion that releases $10^{51}$ ergs, and completely disrupts the white dwarf [42, 43]. The expanding cloud of ejecta glows brightly for many weeks as radioactive nickel produced in the explosion decays into cobalt and then iron.

Because Type Ia supernovae all occur in a star that has a mass of about 1.4 $M_\odot$, they all produce about the same amount of light. This property, and their extreme luminosity, ~ $10^{10}$ $L_\odot$, make Type Ia supernovae extremely useful as a distance indicator for probing the distant reaches of the universe. In recent years, Type Ia supernova have been used to show that the expansion of the universe is accelerating, presumably because the universe is filled with a mysterious substance called dark energy [44, 45, 46].

The intense radiation emitted by both core-collapse and Type Ia supernovae lasts from several months to a few years before fading away. In the meantime, the rapidly expanding (thousands of km/s) ejecta drive a forward shock wave into the circumstellar gas, and a reverse shock into the supernova ejecta. These shock waves create a SNR consisting of hot gas and high-energy particles that glows in radio through X-ray and γ-ray wavelengths for thousands of years. The forward shock wave also accelerates electrons and other charged particles to extremely high energies. Electrons spiraling around the magnetic field behind the forward shock wave produce synchrotron radiation over a wide range of wavelengths.

With *Chandra* it has become possible to observe the forward shock wave and the ejecta heated by the reverse shock wave. The study of SNR with radio, infrared, optical and X-ray telescopes enables astronomers to trace the progress of the shock waves and distribution of elements ejected in the explosion, and in some cases, to use high-resolution spectra to make three dimensional maps of the ejecta. With this information it is possible to do forensic research and investigate the cause and circumstances of death of the stars that produced the remnants.

### 4.1.1 *Type Ia SNR.*

X-ray spectra of SNR provide information on which elements are enhanced in the ejecta, and can be used to distinguish between the remnants of thermonuclear and core-collapse supernovae.

Thermonuclear white dwarf explosions are predicted to produce about a factor ten more mass in Fe-group elements than core-collapse supernovae, enabling *Chandra* spectra to be used to classify 17 SNR in the Galaxy and the nearby Large and Small Magellanic Cloud galaxies as Type Ia remnants [47]. Moreover, for the remnant of the supernova observed by Tycho Brahe in 1572 AD, and for SNR B0506-67.5 in the Large Magellanic Cloud, the Type Ia identification has been confirmed by optical spectra of light echoes [48, 49].

In a complementary approach, *Chandra*'s combined spatial and spectral resolution has been used to make detailed images in Si XIII line emission in 5 Type Ia and 12 core-collapse SNR. An analysis of the morphological differences using a multipole statistical analysis of the surface brightness distribution shows that the types of supernovae can be distinguished by their shape [50, 51]. SN Ia remnants are found to be more spherical than core-collapse remnants and the distribution of the elements in the ejecta more stratified, with heavier elements concentrated toward the center, as observed in the Type Ia SNR 0519-69 [52].

The observed structural difference is consistent with theoretical expectations. In Type Ia supernovae, the thermonuclear origin requires a regular burning front to propagate through the white dwarf (see however, the discussion of SNR G1.9+0.3 below), whereas core collapse supernovae are driven by the gravitational energy released by the collapse of the stellar core, so much of the energy is deposited in the inner regions, leading to a more chaotic, irregular explosion.

Going beyond the classification of a supernova as a Type Ia to determine whether the progenitor system was two white dwarfs (double degenerate model), or a white dwarf plus a normal solar-type star, or red giant (single degenerate model) involves more subtle and indirect methods. The specific characteristics of the progenitor binary system are expected to have an important imprint on the circumstellar medium into which the supernova shock ejecta expand, and could have a detectable effect on the evolution of the shock wave, which would show up in the morphology and spectra of the SNR. The density of the circumstellar medium affects both the size of the SNR and the ionization state of the post-shock gas. Ionization in the post-shock gas is primarily due to collisions with electrons, so the ionization time $\sim [n_e C_i (T, Z)]^{-1}$, where $C_i (T, Z)$ is the collisional ionization rate at electron temperature $T$ for a given ion $i$ of element $Z$. If the remnant expanded into a relatively low density medium, it would have a larger radius and longer ionization times for the various ions. The latter would have significant effects on the strength of observed X-ray lines, reflecting a departure from ionization equilibrium. Radii and ionization times derived from X-ray spectra for a sample of 7 Type Ia SNR studied by *Chandra* shows that this is not the case These results imply that the progenitors could not have had a fast wind, contrary to expectations for some single degenerate models [53].

In contrast, the Kepler SNR (Figure 4-1) shows an anomalous (for a Type Ia supernova) amount of nitrogen in the spectrum of an arc-like feature of the northern region of the remnant. On the basis of simulations of the expansion, it has been argued that the progenitor system was a binary consisting of a white dwarf with a giant secondary star of 4-5 $M_\odot$ [54, 55]. The arc-like structure is interpreted as a bow shock produced by the motion of the remnant through the material ejected by the progenitor star during its giant phase. Kepler's interaction with this material indicates that its progenitor was a moderate mass star that exploded only about 100 million yr after it was formed. If Type Ia supernovae can occur so quickly, they can be found at much higher redshifts, enabling their use to measure cosmic expansion at these epochs The existence of prompt Type Ia SNe adds another layer of complexity to their use as standard candles for studying dark energy. Recent studies of the effects of a changing mix of Type Ia SNe types with redshift based on star-formation rate indicates that the neglect of prompt Type Ia SNe can lead to a systematic error $\sim$ 4% in the value of the dark matter equation of state parameter [56].

Spatially resolved X-ray spectroscopy of SNR is especially valuable for investigating both the nature of the progenitor and the explosion process. *Chandra* observations of the youngest known (~110 yr-old) Galactic SNR, G1.9+0.3, have yielded the first definite detection of a 4.1 keV line attributed to $^{44}$Sc. The emission is produced by the decay of $^{44}$Ti by electron capture, which leaves a K-shell vacancy in $^{44}$Sc. The inferred mass of $^{44}$Ti is $\sim 4\times 10^{-5}$ $M_\odot$, consistent with the predicted yields of both core collapse and Type Ia delayed detonation models. In these latter models nuclear reactions occur in a slowly expanding wave front, producing iron-group elements. The energy from these reactions causes the star to expand, changing its density and allowing a much faster-moving detonation front to propagate through the star. The spatial distribution of both iron and the intermediate mass elements silicon and sulfur suggests that the explosion itself was asymmetric, perhaps consistent with the greater production of $^{44}$Ti found in asymmetric Type Ia explosion models. The prominent Fe K line and the extremely high shock velocities > 18,000 km/s also support a Type Ia origin [57].

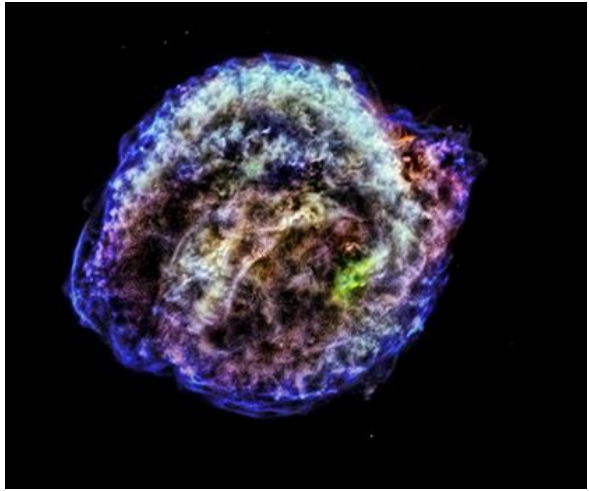

Figure 4-1. Kepler's SNR, the remnant of the supernova discovered by Johannes Kepler in 1604. The red, green and blue colors show low (0.3-0.72 keV), intermediate (0.72-1.7 keV) and high (1.7-8.0 keV) energy X-rays observed with *Chandra*. Note the bright arc at the top of the image which has been interpreted as a bow shock produced by the motion of the remnant through the progenitor star's red giant wind [54]. Scale: Image is 2.5 arcmin across (about 8 pc). Credit: NASA/CXC/NCSU/M.Burkey et al.

4.1.2   *Remnants of Core Collapse Supernovae*

The expected yields of various elements in a core-collapse supernova are uncertain, and vary substantially from one set of models to the other. In general they are dominated by carbon, oxygen, neon and magnesium. Prominent lines from oxygen ions fall in both the optical and X-ray bands, so one of the primary ways to identify the remnant of a core-collapse supernova is through the detection of large oxygen abundance.

X-ray emission from supernovae that are a few years to a few decades old can provide a useful probe of the nature and mass loss history of the progenitor stars. Mass loss from a pre-supernova star forms the medium into which the supernova ejecta initially expand, so X-ray emission from the interaction of the supernova shock wave with the circumstellar medium provides a probe of the mass-loss parameters and hence the nature of the pre-supernova star.

For SN1987A, which occurred in the nearby (55 kpc) Large Magellanic Cloud galaxy, the progenitor is known with certainty -- it was blue supergiant star—but the rest of the story is complicated, and illustrates the difficulties of stellar forensics. An evolved blue supergiant is expected to have a fast wind and to create a low-density cavity within a red supergiant wind of a previous mass loss phase. What in fact was observed in the optical band was a striking triple-ring system (see Figure 4-2) that might be explained by the merger of a 15-20 $M_\odot$ star with a 5 $M_\odot$ star about 20,000 yr before the explosion [58]. In this picture, the circumstellar environment was sculpted by the collision of a fast blue supergiant wind colliding with a slow, asymmetric red giant wind. Another possibility is that the triple-ring structure was produced by mass-loss from a fast-rotating star [59]. In both cases an equatorial ring was produced and the SN 1987A blast wave is now moving into that ring.

SN 1987A has been monitored by *Chandra* with about two observations per year since 1999. A multi-component component model [60, 61] has been proposed to explain the high resolution X-ray spectra of SN1987A: (1) a dense, clumpy ring heated to T ~ 3-5 MK by a slow shock (500 km/s); (2) hot (~30 MK) gas with broad emission lines characteristic of velocities ~ 10,000 km/sec, coming from the lower density, shock-heated wind located above and below the equatorial ring; and (3) a T ~ 30 MK component with lower density, which could be from reflected shocks in the vicinity of clumps in the ring. Based on a recent recalculation of fluxes using an updated calibration for *Chandra* data, there seems to be no evidence that the X-ray luminosity is leveling off over time, so the shock wave is presumably still moving through the equatorial ring [62].

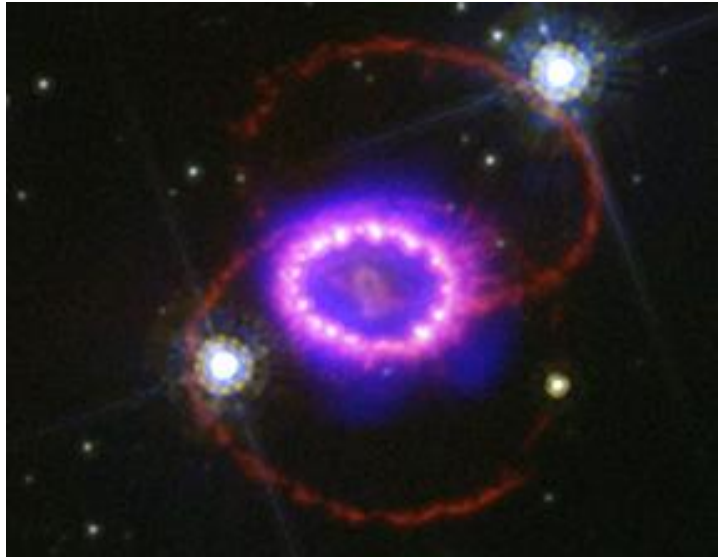

Figure 4-2. Composite Optical (pink-white) and X-ray (blue-purple) image of SN 1987A, showing the triple ring structure, with the bright equatorial ring [58, 59, 60, 61, 62]. Scale: Image is 6 arcmin across (about 16 pc). Credit: X-ray: NASA/CXC/PSU/S. Park & D. Burrows; Optical: NASA/STScI/CfA/P. Challis.

Radiation from the NS that is expected to exist in the central regions of SN1987A has yet to be detected. Given the sensitivity of *Chandra* for detecting faint sources and its ability to resolve a central source, the data indicate that the central source is still obscured by surrounding dust and gas. It may take ~ 100 yr until this obscuring veil lifts.

Studies of nearby core-collapse SNR with ages ranging from a few hundred to a few thousand years illustrate some general trends as well as the complexities involved. A 3-D model of the Cassiopeia A (Cas A) SNR, which has an estimated age ~ 340 yr, has been constructed using X-ray data from *Chandra* and infrared data from the Spitzer Space Telescope [63]. This reconstruction revealed that the structure of Cas A can be characterized by a spherical component, a tilted thick disk, and multiple ejecta jets/pistons and fast-moving knots all populating the thick disk plane. Although interaction with the circumstellar medium affects the detailed appearance of the remnant, the bulk of the symmetries and asymmetries in Cas A are intrinsic to the explosion.

An X-ray image of the silicon-rich ejecta in Cas A (Figure 4-3) reveals a distinct bipolar structure which appears to have resulted from jets associated with the supernova [64]. Hydrodynamic simulations [65] show that the Cas A jets could not have survived interaction with a circumstellar shell of the type produced by the expansion of a fast blue supergiant wind into a slow red supergiant wind. This strongly suggests that the Cas A progenitor exploded while in the red supergiant phase, consistent with an initial mass below ~ 25 $M_\odot$.

Other evidence is consistent with this interpretation. A study of the diffuse thermal X-ray emission in the outer regions of Cas A showed that the SNR is expanding into a slow (~ 15 km/s) wind rather than into a uniform medium [66]. This suggests that the progenitor of Cas A had an initial mass ~ 16 $M_\odot$ and that the mass prior to the explosion was ~ 5 $M_\odot$.

The approximately 3,000 year-old galactic SNR G292.0+1.8 illustrates many aspects of the evolution of the remnant of a core-collapse supernova: a blast wave moving through previously ejected circumstellar material, ejecta enriched with heavy elements, a central NS or pulsar, and a pulsar wind nebula powered by relativistic particles produced by the pulsar (Figure 4-4). The detection of the NS and its nebula conclusively associates this young, oxygen-rich SNR with a supernova produced by the core-collapse of a massive star. The offset of the pulsar from the center of the nebula may be due to the recoil from an asymmetric explosion. Estimates of the mass lost in the wind, the mass of the ejecta, and the existence of a NS combine to indicate a progenitor mass in the range 20-35 $M_\odot$ [67, 68].

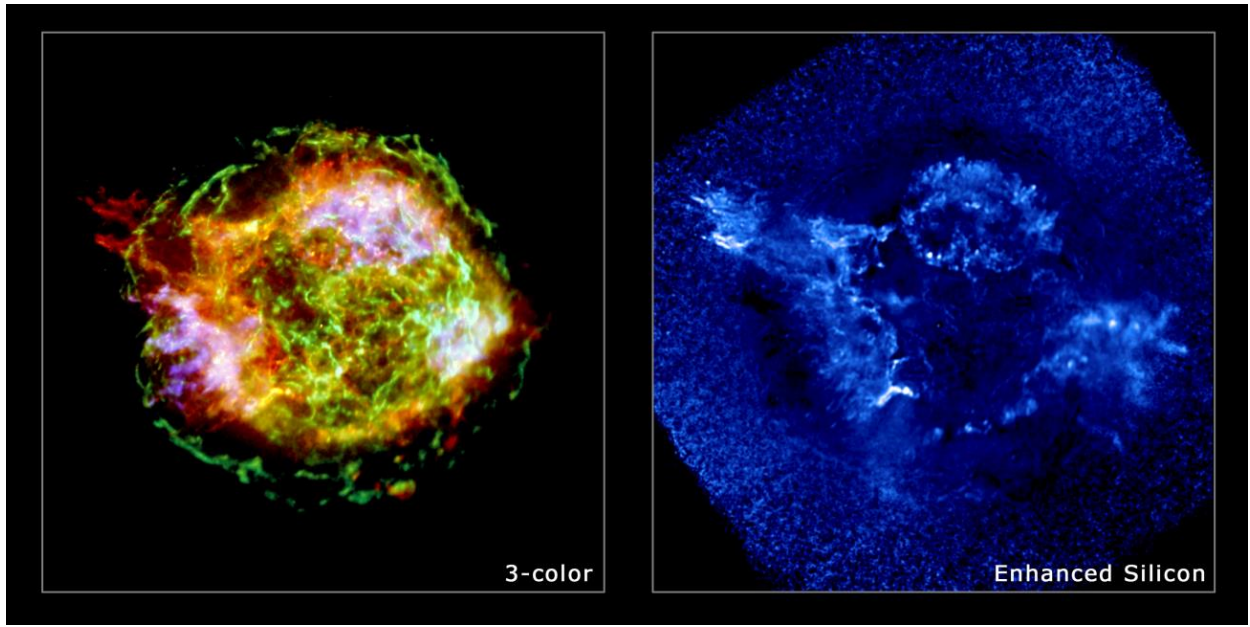

Figure 4-3. Left Panel: *Chandra* image of the Cas A SNR with red = Si XIII He α (1.78-2.0 keV), blue = Fe K (6.52-6.95 keV), and green = 4.2-6.4 keV continuum emission. The outer green ring is from the forward shock. Inside this ring are the ejecta heated to $T$ ~10 MK by the reverse shock [64]. Right panel: the ratio of the Si He α and the 1.3-1.5 keV band images, which highlights the jet and counter jet traced by the Si emission [64]. Images are 8.4 arcmin across (about 8 pc). Credit NASA/CXC/GSFC/U. Hwang et al.

Out of the complexity and diversity of the remnants of core-collapse supernovae, some general similarities seem to be emerging. The masses of their progenitors are in the range 15-35 $M_\odot$. In contrast to Type Ia SNR, the core-collapse remnants are very irregular and show large deviations from spherical symmetry, with evidence of bipolar structure and jets. The origin of these structures is unknown. They could be due to a companion star and accretion disk, though none has been found, or to accretion instabilities in the explosion process; to rapid rotation; to strong magnetic fields, or to a combination of all of the above [47].

The fate of the cores of stars with masses above ~ 30 $M_\odot$ is uncertain [69, 70]. They could stabilize as NS with the ejection of large, massive envelopes, or they could collapse to BH with or without the ejection of envelopes. The SNR W49B has been proposed as a candidate BH-producing supernova [71]. The shape of W49B indicates that it is the product of an explosion in which matter is ejected at high speeds along the poles of a rotating star, in line with some models of supernovae in rapidly rotating stars. Further evidence comes from the abundance of iron, which is enhanced along the jet, consistent with the predicted yields in models of bipolar/jet driven supernovae wherein the enhanced kinetic energy at the polar axis causes nickel to be produced more efficiently there. Strict upper limits on the X-ray luminosity of any undetected point source exclude the presence of a NS in W49B, and suggest that the supernova left behind a BH. The estimated age of W49B is 1,000 yr, which would make the remnant BH the youngest known BH in the Galaxy.

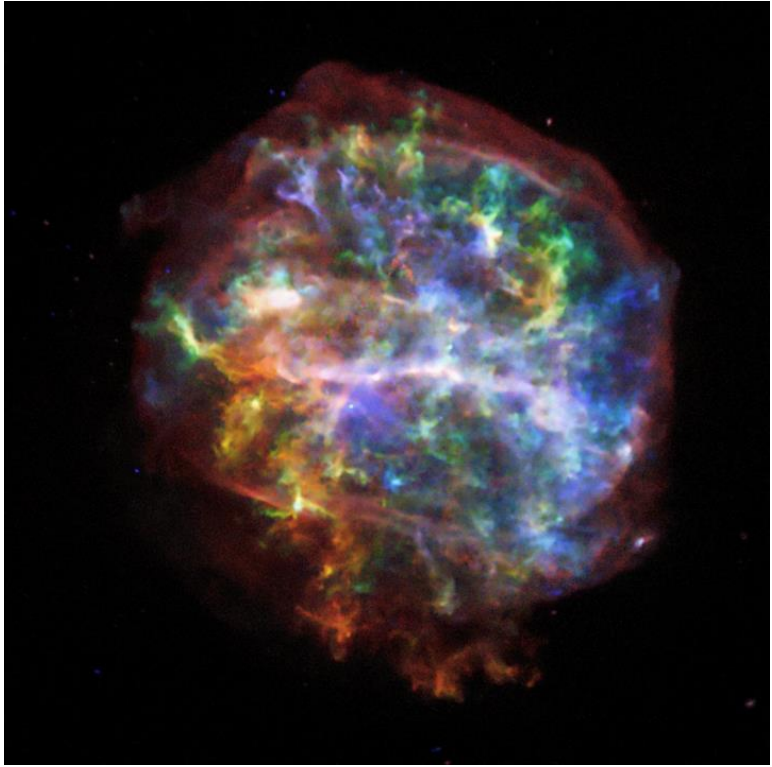

Figure 4-4. *Chandra* image of G292.0+1.8, the remnant of a core-collapse supernova. In this image, red represents the 0.580-0.710 and 0.880-0.950 keV energy bands, orange 0.980-1.100 keV, green 1.280-1.430 keV and blue 1.81-2.050-2.400 and 2.620 keV. The bright horizontal belt across the middle of the remnant is due to shocked circumstellar material, and the pulsar can be seen as the white spot in the purple cloud below the belt at roughly the 8 o'clock position [67, 68]. Scale: Image is 11.5 arcmin across (about 20 pc). Credit: NASA/CXC/PSU/S. Park et al.

*4.2   Cosmic Ray Acceleration in SN Shock Waves*

Energetic arguments suggest that acceleration of particles in supernova shock waves is the source of cosmic rays (mostly protons and helium nuclei) up to the knee in the cosmic ray spectrum at 1000 TeV [72]. The recent detection of 100 GeV gamma rays from the Tycho SNR by the Fermi Gamma-Ray Space Telescope supports this model [73], but an explanation of the γ-ray data in terms of energetic electrons rather than protons is also possible [74].

*Chandra* observations of Tycho reveal a strikingly ordered pattern of nonthermal stripes which may be evidence for the acceleration of particles up to $10^{15}$ eV (Figure 4-5). The spacing between the stripes is

~0.2 pc, corresponding to the gyroradii of $10^{14}$ to $10^{15}$ eV protons for magnetic field strengths of a few to a few tens of microgauss [75]. Simulations show that a cosmic-ray current-driven instability can amplify the magnetic field and produce narrow peaks in the magnetic turbulence with a separation ~ a gyroradius [76].

More generally, *Chandra's* ability to distinguish the X-ray spectra of forward and reverse shocks has shown that X-ray emission from the forward shock is probably non-thermal synchrotron radiation from relativistic electrons accelerated in the forward shock. Synchrotron X-ray emission requires acceleration of electrons up to tens of TeV and suggests acceleration of ions to even higher energies. Further evidence for the acceleration of ions in SNR comes from the observed temperature behind the forward shock, which in some remnants is lower than expected from standard shock-wave theory. This implies that a significant portion of the post-shock energy is going into the acceleration of ions and electrons [77].

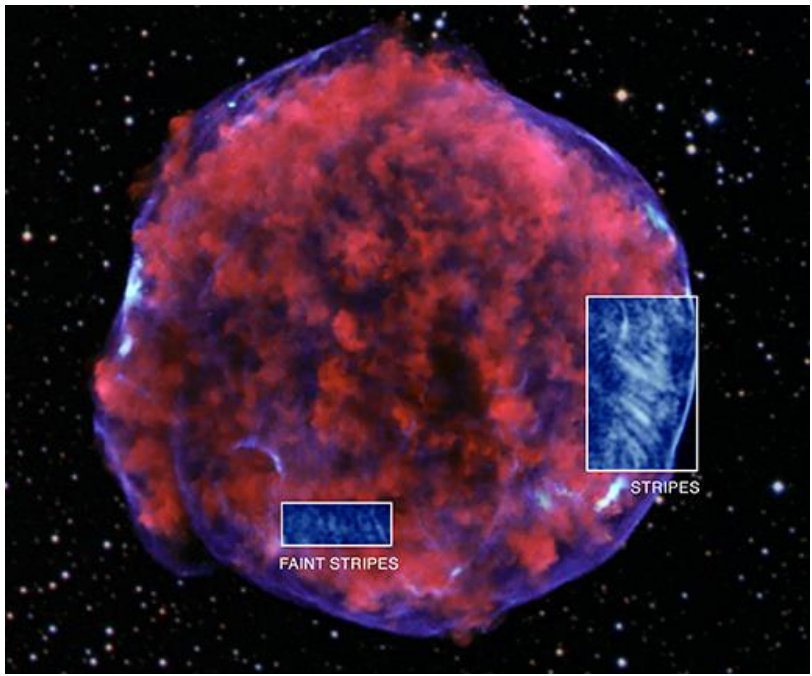

Figure 4-5. Tycho's SNR, the remnant of a supernova discovered by Tycho Brahe in 1572. Low-energy X-rays (red, 1.6-2.15 keV) in the image show expanding debris from the supernova; high energy X-rays (blue, 4-6 keV) show the blast wave, and a pattern of X-ray "stripes" (see inset boxes [75]). The image is 9 arcmin across (about 17 pc). Credit: X-ray NASA/CXC/Rutgers/K. Eriksen et al; Optical: DSSPulsar Wind Nebulae

*4.3 Pulsar Wind Nebulae*

Core collapse SNR should contain the object produced by the collapsed core, i.e. a NS or a BH. Although theoretical predictions vary on which, if any, core collapses produce BH, there is general agreement that most collapses will produce a NS. The existence of a NS has been confirmed for many SNR, and their manifestations have been as varied as the remnants themselves. On the basis on conservation of angular momentum and magnetic flux, the NS is expected to be rapidly rotating and highly magnetized. The rapid rotation of a NS magnetosphere gives rise to strong electric fields, particle acceleration and pulsed radiation across the whole electromagnetic spectrum. The rotational energy loss of the NS, or pulsar, is primarily via a Poynting flux at a rate $dE/dt \sim \mu^2 \Omega^4/c^3$, where $\mu$ is the magnetic moment of the NS, $\Omega$ is its angular frequency and $c$ is the speed of light [78, 79]. The corresponding spin down rate of the NS is $d\Omega/dt = (dE/dt)/(I\Omega)$ where $I$ is the moment of inertia of the NS. The strong electromagnetic fields produce a wind of relativistic electrons and positrons which terminates in a shock wave, where the electrons and positrons are accelerated further. The particles advect and diffuse away from the shock to create a nebula of relativistic electrons and positrons which spiral around magnetic field lines and emit synchrotron radiation over a wide range of wavelengths, from radio to soft gamma rays, and through inverse Compton scattering, to the TeV band. The resulting nebula is called a pulsar wind nebula [80, 81].

Figure 4-6, Figure 4-7 and Figure 4-8 show three images representative of ~50 pulsar wind nebulae that have been detected with *Chandra*. Two morphological types can be distinguished: (1) torus-jet pulsar wind nebulae, which show a toroidal structure around a pulsar, and sometimes one or two jets along the torus axis, and (2) bow-shock-tail objects whose appearance is dominated by a cometary structure generated by the pulsar's rapid motion through the interstellar medium.

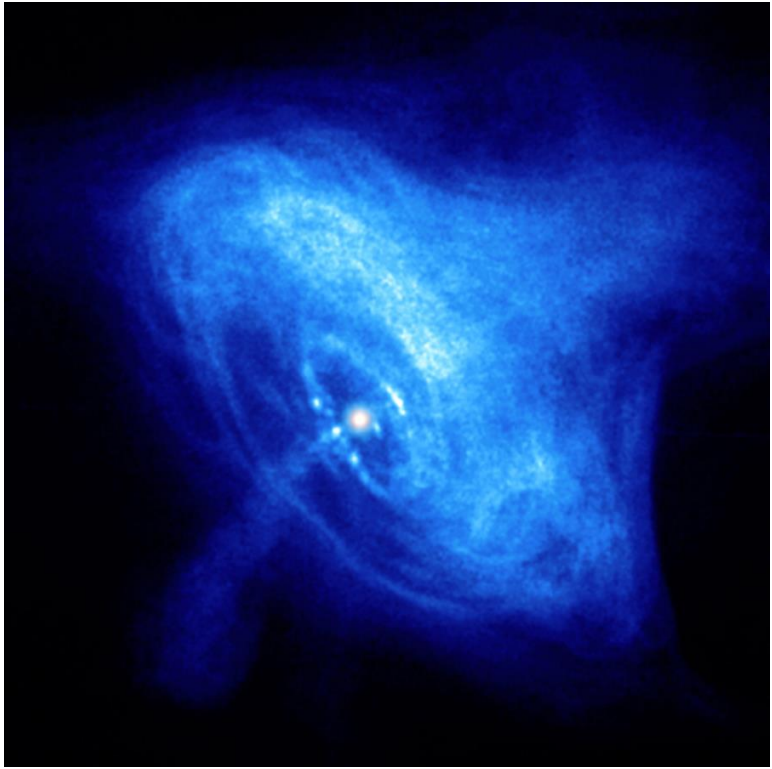

Figure 4-6. *Chandra* image of the Crab Nebula showing the pulsar (white dot near the center), the inner ring produced by the shocked relativistic flow, and jets of matter and anti-matter moving away from the north and south poles of the pulsar [82]. Scale: image is 1.6 arcmin across (about 1 pc). Credit: NASA/CXC/ASU/J. Hester et al.

The Crab Nebula (Figure 4-6), which is associated with the historical supernova of 1054 A.D., and the 34 millisecond pulsar B0531+21, is the archetypical example of a torus-jet pulsar wind nebula. The spin-down luminosity is ~$5\times10^{38}$ erg s$^{-1}$, about a factor 10 larger than the X-ray luminosity. The *Chandra* image of the Crab Nebula [82] shows an axisymmetric nebula with a tilted inner ring, which has been associated with the termination shock in the equatorial pulsar wind, a larger torus, and two jets emanating along the pulsar spin axis. Two-dimensional magneto-hydrodynamic simulations with prescribed radial energy flux density and magnetic field are able to reproduce axial and equatorial features similar to those observed [83, 84].

The Vela pulsar wind nebula is a hybrid object, showing both a torus, and evidence of a bow shock produced by rapid motion through the surrounding gas (Figure 4-7). The pulsar wind is generated by the relatively young ($t = 11$ kyr), nearby ($d = 300$ pc) pulsar B0833–45. The nebula created by the wind exhibits an unusual morphology with two arcs, possibly part of two rings above and below the equatorial plane, bright inner jets, and much fainter, strongly variable outer jets [81].

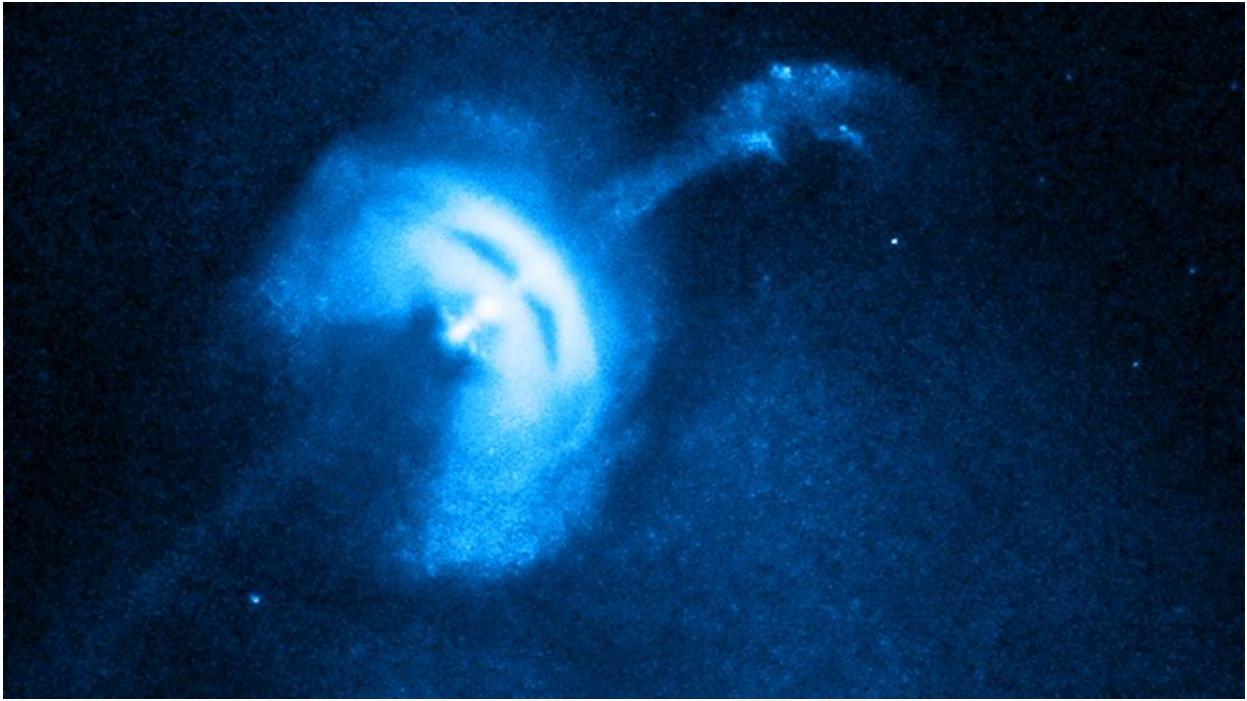

Figure 4-7. A *Chandra* image of the Vela pulsar, showing the helical jet, and tilted rings of X-ray emission from shock waves produced by high energy particles flowing away from the central NS. The image was constructed by adding data from eight different observations, so the precessing jet is blurred [85]. The image is 3.0 arcmin across (about 0.27 pc). Credit: NASA/CXC/Univ of Toronto/M.Durant et al.

The "Mouse" pulsar wind nebula shown in Fig. 4-7 is an example of a class of pulsar wind nebulas that exhibit a cometary structure strongly suggestive of supersonic motion at ~ 600 km/s through the interstellar medium. The Mouse is associated with pulsar J1747-2958, which has a spin down age ~ 25,000 yr. The brightest part of the pulsar wind nebula is associated with the shocked wind just outside the termination shock. The tail of the pulsar wind nebula represents the shocked wind confined by the swept-back ejecta. The inner arc marks a termination shock in the outflow from the pulsar [86, 87].

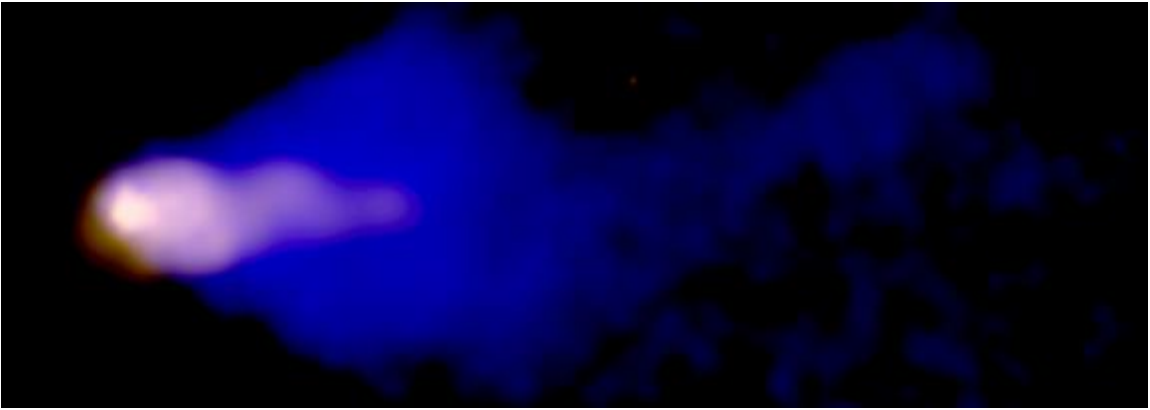

Figure 4-8. The Mouse, a.k.a. G359.23-0.82, gets its name from its appearance in radio images that show a compact snout, a bulbous body, and a remarkable long, narrow, tail that extends for about 17 pc. The image, a composite X-ray (gold) and radio (blue), shows a close-up of the head of the Mouse where a shock wave has formed as the young pulsar plows supersonically through interstellar space [86, 87]. Scale: Image is 2.5 arcmin wide (about 3.6 pc). Credit: X-ray: NASA/CXC/SAO/B. Gaensler et al; Radio: NSF/NRAO/VLA.

*4.4 Magnetars*

Observations of compact objects in some SNR have revealed the presence of slowly-spinning ($P \sim 0.3 - 10$ seconds), highly magnetized ($B \sim 10^{14}$ G) NS called magnetars The relatively large periods and their observed large rates of change require that the dissipation of magnetic energy (rather than magnetic dipole radiation) must be invoked to explain the steady X-ray luminosity ($\sim 10^{32}$-$10^{35}$ erg s$^{-1}$) and short, energetic flares ($10^{40}$-$10^{46}$ ergs) that characterize magnetars. According to the magnetar model, the large internal field ultimately deforms or cracks the crust of the NS, disturbing the external field, causing the X-ray bursts.

The source SGR 0418+5729 represents an anomaly among anomalies. It exhibits sporadic X-ray bursts and persistent X-ray pulsations characteristic of magnetars, yet the spin-down rate implies a magnetic field $\sim 6\times10^{12}$ G [88], 1-2 orders of magnitude less than for other magnetars. Monitoring the X-ray output of this source and modeling the magnetic and thermal evolution of the NS and its crust suggest that this object is old, with an estimated age of 550,000 yr. This advanced age has probably allowed the surface and internal magnetic field strength to decline over time. The crust also weakened, so magnetic outbursts could still occur. The example of SGR0418+5729 suggests that many of the pulsars in the Galaxy may exhibit magnetar behavior every 1000 yr or so, yielding a rate for the Galaxy of ~ one magnetar outburst per year.

*4.5 The Equation of State of Ultradense Matter*

NS are formed with temperatures of billions of degrees, and cool via a combination of neutrino and photon emission. Observing the cooling rates of young NS offers a unique method for investigating the behavior of matter at extreme densities.

The discovery of a centrally located NS in the center of the Cas A SNR has provided a rare opportunity to study the cooling of a young NS of known age and infer the conditions in its interior. An analysis of 10 yr of archival *Chandra* observations of the source shows that a model of a NS with a carbon atmosphere, a mass of 1.65 M$_\odot$, a surface temperature ~ 1.5 MK, and a low magnetic field produce a good fit to its X-ray spectrum [89]. The size of the emission region is consistent with the NS having a roughly uniform temperature, possibly explaining the absence of detectable X-ray pulsations from the source.

The complexity of the bright and varying supernova remnant background makes a definitive interpretation of archival Cas A *Chandra* observations difficult. A combined analysis of *Chandra* data on the Cas A NS, using all *Chandra* X-ray detectors and modes, shows a decline of 2.9% ± 0.5% statistical ± 1.0% systematic over 10 yr [90]. The observed decline can be explained if neutrons in the core have recently undergone a transition to a superfluid state, producing enhanced neutrino emission due to Cooper pair formation in the process [91, 92]. If confirmed, this provides the first direct evidence that superfluidity and superconductivity occur at supranuclear densities, and provides an important probe of the nuclear strong force. Over the next several years, observations of Cas A spaced by approximately one year will continue to monitor the surface temperature evolution of the NS and test the superfluid transition model, which predicts a continued decline in temperature.

*4.6 X-ray Binary Systems*

SNR and pulsar wind nebulae will fade away after several thousand years, but if a NS or BH is part of a binary star system, it may become a bright X-ray source once again as it accretes matter from its companion star. Indeed, X-ray binaries are the brightest X-ray sources in the galaxy, so they were the first extra-solar X-ray sources discovered, and they continue to be prime targets for X-ray missions.

Because most SNR are visible for only a few $10^4$ yr, studies of NS X-ray binaries within SNR probe the earliest stages in the life of accreting NS. However, such objects are exceedingly rare: until recently, none were known in the Galaxy. In 2013, *Chandra* and radio observations revealed the natal SNR of the accreting NS Circinus X-1. An upper limit of $t<4600$ yr was placed on its age, making it the youngest known X-ray binary. The young age is consistent with the observed rapid orbital evolution and the highly eccentric orbit of the system, providing the best opportunity yet for detailed study of the post-SN orbital evolution of X-ray binaries [93].

BHs in X-ray binaries are of prime interest, because they offer the potential of measuring with fairly high accuracy both the mass and spin of the BH, the two quantities that fully characterize the intrinsic nature of a black hole. The mass can be determined from dynamical modeling based on optical data on the binary system. Two independent methods have been used to measure the spin, both of which depend upon identifying the inner radius of the accretion disk with the radius of the innermost stable circular orbit, which is a function only of the mass and spin of the black hole. In the Fe Kα method, spin-dependent models of the profile of the relativistically-broadened iron line are compared with observations. The continuum-fitting model in essence uses the Stefan-Boltzmann law to measure the area of the X-ray emitting region, which depends on the mass and the spin.

Spectra obtained from *Chandra*, the *Rossi* X-ray Timing Explorer, and the *Swift*, and *Suzaku* observatories, were used in the continuum-fitting approach to show that the Cygnus X-1 BH is rotating at or near its maximum spin rate with a spin parameter $a/R_g > 0.983$ at the 3σ confidence level, where $a =J/Mc$ and $J$ is the angular momentum of a BH of mass $M$ and $a/R_g = 1$ corresponds to the maximum possible rate [94]. This result is consistent with a recent analysis that applied the Fe Kα method to a single *Suzaku* observation of Cygnus X-1, and found $a/R_g =0.97$ (-0.02, +0.014) (1σ) [95].

Further afield, at a distance of 800 kpc, the galaxy M33 was found to harbor an eclipsing BH X-ray binary, M33 X- 7 [96]. *Chandra* recorded the location with the accuracy needed for follow-up optical study and identification. This system was found to contain one of the most massive stellar BH known, with a mass of 15.7 ± 1.4 M$_\odot$. A model fit to the X-ray spectrum showed that M33 X-7 is rotating with $a/R_g = 0.77 ± 0.05$.

The study of accreting stellar mass BHs shows that the flow pattern around BHs is a mixture on accretion and outflow. A striking example is the BH binary system GRO J1655-40. A *Chandra* HETG observation revealed one of the richest X-ray spectra known (Figure 4-9), with more than 100 blueshifted X-ray absorption lines identified [97].

The absorption is primarily from H-and He-like ions with atomic number $Z$ ranging from 8 to 28, indicating that the ions are relatively near to the BH, while the blue shifts and column densities indicate the presence of a dense wind. Detailed modeling shows that the observed wind cannot be driven by radiation pressure or photoionization. Heating by magnetic turbulent viscosity in the accretion disk is a possible cause for the wind. The wind carries away orbital angular momentum, allowing a large fraction of the gas in the disk to spiral inward to the BH.

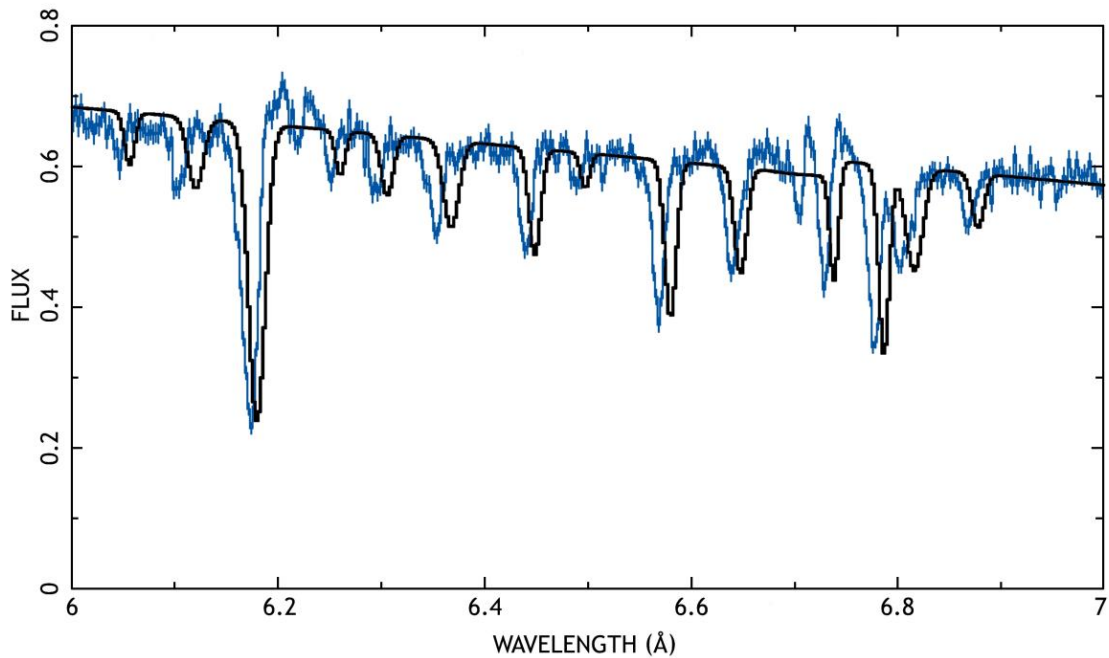

Figure 4-9. A portion of the *Chandra* HETG spectrum, showing evidence for absorption by a wind in the BH X-ray binary system GRO J1655-40. The most prominent lines in this wavelength band are due to Si XIV 2p-1s (6.17 Å), Mg XII 5p-1s (6.57 Å) and Fe XXIV 6p-2s (6.78 Å) transitions. The blue line shows the observed spectrum and the black line shows a model spectrum where the absorption lines are plotted at their rest wavelengths. Comparing the two spectra shows that the *Chandra* spectrum is Doppler-shifted towards shorter wavelengths giving evidence for a wind blowing towards the observer, with velocities ranging up to 1200 km s$^{-1}$. Credit: NASA/CXC/U.Michigan/J.Miller et al.

*4.7 Microquasars*

Some X-ray binaries produce jets of relativistic matter that travel for tens to hundreds of parsecs through the interstellar medium. In addition to being of great interest in their own right, these so-called microquasars provide a valuable stellar-scale analog of the jets associated with quasars, i.e., SMBH in the centers of galaxies. The mechanism for launching these jets is analogous to that for pulsar wind nebulae: rapid rotation combined with strong magnetic fields can collimate and power the outflow of high energy particles.

The most powerful known microquasar was recently discovered in *Chandra* archival observations of a large nebula (S26) in the nearby galaxy NGC 7793 [98]. The X-ray source displays a pair of collimated jets powered by a stellar mass BH. The source is similar to the Galactic microquasar SS 433, but its linear size of ~ 300 pc is twice as large. The power necessary to explain the expansion of the optical jet-driven bubble of S26 is estimated to be a few $10^{40}$ erg s$^{-1}$, $10^4$ times the X-ray luminosity. This suggests that there may be accretion modes that channel most of the power into jets rather than photons, even at very high mass accretion rates.

*4.8   Ultraluminous X-ray Sources (ULX)*

ULX are X-ray sources that are not associated with the nuclei of galaxies and have luminosities in the range of a few $\times 10^{39}$ erg s$^{-1}$ to $10^{42}$ erg s$^{-1}$. Assuming that they are radiating isotropically, the mass can be constrained by the requirement that the radiation pressure due to Compton scattering at the surface of the source is less than the gravitational force. This yields a limit on the luminosity (the Eddington limit): $L < L_{Edd} = 1.3 \times 10^{38}$ ($M/M_\odot$) erg s$^{-1}$.

The most luminous, well-studied stellar mass black holes have ($M/M_\odot$) ~ 10 and luminosities ~ $10^{38}$ erg$^{-1}$ so they are radiating at about 10% of the Eddington limit [99]. The inferred masses for ULX, if they behave similarly, are in the range 100-100,000 $M_\odot$.

A vigorously debated issue is whether this inference is correct. Do ULX represent a new class of BH with masses ~ $10^2$-$10^5$ $M_\odot$, intermediate between stellar mass BH and SMBH, or are they normal, albeit massive, stellar BH that are radiating above the Eddington limit through a combination of anisotropic radiation and super-Eddington accretion rates [100]?

If it is assumed that the bulk of the radiation from a ULX is coming from an accretion disk that is radiating approximately like a black body, and that most of the radiation comes from near the innermost stable orbit of the accretion disk which is a few times $R_g$, and the luminosity is a fraction of the Eddington limit, then $L \propto T^4 R_g^2 \propto T^4 M^2$. Since $L \propto M$ we find $T \propto M^{-1/4}$. The larger masses of Intermediate Mass Black Holes (IMBH) therefore imply that their accretion disks will be cooler than those of stellar mass BH.

Fits to X-ray spectra using *Chandra*, *XMM-Newton*, and Swift data for the most luminous ULX, M82 X-1 and ESO 243-49 HLX-1, provide evidence for BH with masses ~ 200-800 $M_\odot$ and 3000 – 100,000 $M_\odot$, respectively [101, 102, 103]. Observations with *Chandra*, *XMM-Newton* and Swift of spectral changes and the long-term variability of HLX-1 are not consistent with the source being in a super-Eddington accretion state, providing further support for the IMBH interpretation.

In contrast, studies of a large sample of *Chandra* and *XMM-Newton* data on ULX in nearby galaxies suggest, on the basis of their luminosity function and association with star-forming regions, that these object are probably the high-luminosity extension of the population of high-mass X-ray binaries [100, 103, 104].

Although the evidence is still largely circumstantial, and a consensus is still in the future, the following picture is emerging: (1) ULX are a diverse population; (2) massive (~30-100 $M_\odot$) stellar BH with moderate super-Eddington accretion seem to be the easiest solution to account for most sources up to luminosities ~ a few $10^{40}$ erg s$^{-1}$ (3) in the few exceptional cases where the luminosity reaches $L \sim 10^{42}$ erg s$^{-1}$, e.g. M82 X-1, and HLX-1, IMBH are preferred. As to the origin of these IMBH, we are still very much "at sea.

## 5 Supermassive Black Holes (Active Galactic Nuclei)

Understanding BH, how they are formed and grow, and how they affect their environment, continues to be a major quest for astrophysical research. One of the best ways to study BH is through the radiation emitted as they accrete. It is estimated that approximately a quarter of the total cosmic radiation emitted since the Big Bang comes from matter accreted by supermassive black holes (SMBH) with typical masses ranging from $10^6$ to several $\times 10^9$ $M_\odot$. A significant amount of this radiation is emitted as X-rays, which can penetrate the clouds of dust and gas shrouding many SMBH so that *Chandra* has proven to be an exceptionally valuable tool to find and study them. *Chandra's* arcsecond resolution and low background are critical for avoiding source confusion while enabling detection, location, identification and classification of the faintest SMBH in the nuclei of galaxies (active galactic nuclei, AGN). Deep exposures (Figure 5-1, [105]) have made it possible to investigate the evolution of the SMBH population and its accretion history. Wider-area, shallower surveys find more objects, particularly at higher luminosities, providing statistical samples to probe the population as a whole.

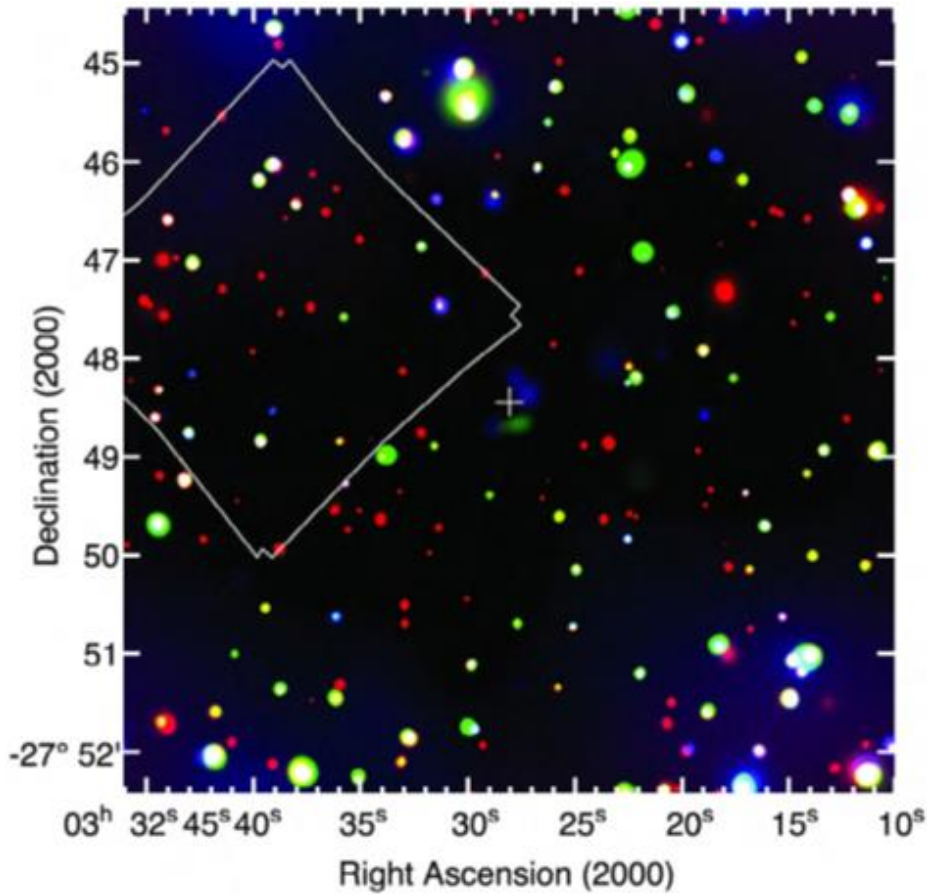

Figure 5-1. A *Chandra* image of the inner 8×8 arcmin of the 4 Ms *Chandra* Deep Field South [105], the deepest X-ray image currently available. This is a composite of smoothed images in the 0.5-2.0 keV (red), 2-4 keV (green) and 4-8 keV (blue) bands. The polygon shows the field of view of the Hubble Ultra Deep Field [106] and the plus sign shows the average aim point for the multiple *Chandra* observations comprising this image. Despite its depth, the image remains photon-limited rather than confusion or background limited. Credit: Adapted from [105].

*Chandra's* high-resolution has probed the innermost regions close to the SMBH enabling studies of the absorbing and emitting material surrounding the nuclear regions, and revealed complex, X-ray emitting jets. Deep imaging and grating spectroscopy have probed the detailed structure and radiation mechanisms of individual sources, investigated the interaction with their environment and the nature of their host galaxies, and revealed binary SMBH in the process of merging.

## 5.1 The Structure of the Nuclear Regions of Active Galaxies

It is well-established that accretion of surrounding material drives both the growth of the central SMBH and the powerful multi-wavelength emission which characterizes AGN. Due to their small size, generally ≤100 pc, the nuclear regions can only be directly imaged in the X-rays for a small number of nearby AGN. Models are developed via multi-wavelength and temporal observations of the emission that originates in the nuclear regions. The basic model is one of a geometrically-thin accretion disk surrounded by a more extended molecular torus with radiatively-driven, high-velocity outflows above and below the accretion disk/torus structure and relatively little obscuration within ~60° of the polar axis of the accretion disk. The X-rays, often originating in the hottest material, probe the regions closest to the SMBH event horizon. Constraints on the covering factors of the various components are placed by the relative numbers of unobscured, Compton-thin ($10^{22}$ cm$^{-2}$ <$N_H$<$1.5\times10^{24}$ cm$^{-2}$), and Compton-thick (CT, i.e. optically thick to Compton scattering, with a column density, $N_H$>$1.5\times10^{24}$ cm$^{-2}$) sources in obscuration unbiased samples.

### 5.1.1 The X-ray Source.

The primary X-ray power-law continuum in AGN is known to have a remarkably constant slope, $\Gamma$~1.9 (where flux is proportional to $E^{-\Gamma}$ photons cm$^{-2}$ s$^{-1}$) over a wide range of redshift and luminosity. It is thought to originate via Compton scattering of UV photons from the inner edge of the accretion disk very close to the central SMBH in a corona of ionized plasma within a few gravitational radii of the SMBH [107]. The origin of the corona remains unknown and observations of the primary spectrum, which could constrain its physical parameters, are difficult due to the presence of re-processed X-rays: reflected, absorbed and re-emitted X-rays from material further from the nucleus. *Chandra's* high sensitivity and longevity has facilitated deeper observations, grating spectroscopy and temporal monitoring to provide new constraints on the physical and dynamical conditions of the material in the nuclear regions [108].

*Chandra* sky surveys and studies of known AGN samples have demonstrated that the power law spectral index of unobscured AGN is independent of redshift (out to $z$>6) or luminosity and remains consistent with the well-established canonical value, $\Gamma$~1.9 with a small dispersion (~0.3) [109,110, 111]. For radio-loud AGN additional, non-thermal X-ray emission is commonly associated with radio structures: jets, lobes, and hot spots. Contributions from these extended regions can often be resolved from the nuclear X-ray emission with the high spatial resolution of *Chandra* [108, 112, 113] (Section 5.3). At low redshift ($z$<1) it has been confirmed that the nuclear jet-related X-ray component correlates with the core radio emission while accretion-related X-ray components are obscured to varying extent by material whose column density is related to a source's orientation to the observer's line-of-sight [114, 115, 116]. The lack of X-ray obscuration in *Chandra* observations of both Gigahertz-peaked and Compact Steep Spectrum radio sources [117] rules out earlier models for these small radio sources as being due to confinement by a surrounding medium in favor of young sources which have not yet expanded far (≤10 kpc). Their X-ray spectra are steeper and somewhat less luminous than the typical radio-loud AGN, consistent with weaker jet-related emission [118, 119]. *Chandra* observations have also confirmed long-standing expectations that high-ionization (dominated by high-ionization species such as NV$\lambda$1240, CIV$\lambda$1549) broad-absorption line quasars have weak X-ray emission due to strong X-ray absorption primarily by cold gas, while low-ionization (dominated by low-ionization species such as MgII$\lambda$2798, CII$\lambda$1335) broad-absorption line quasars are optically weak suggesting larger quantities of dust [120, 121].

5.1.2 *Outflows/Winds.*

*Chandra* grating spectroscopy of the absorption features in nearby AGN provides a unique probe of the plasma not visible in the optical and UV due to obscuration. Analysis reveals multi-phased, highly-ionized, outflowing material ("winds") with velocities ranging from $10^2$-$10^5$ km s$^{-1}$ within 0.01-0.1 pc of the SMBH in individual AGN (e.g. NGC3783 [122], MCG-6-30-15 [123], Mrk766 [124], NGC3516 [125]). Monitoring shows variations in the absorption features on timescales of days-months which constrain both the size of the absorbing gas cloud and its distance from the SMBH (e.g. NGC1365 [126], MCG-6-30-15 [127]). In contrast to these multi-phase outflows, recent *Chandra* HETG and multi-wavelength observations of the obscured quasar IRAS 13349+2438 [128] detect several X-ray/UV warm absorbers which, when using a realistic intrinsic X-ray-UV ionizing continuum as input [129], can be modeled as a smooth flow with a continuous distribution of ionization states [130]. Mass outflow rates, determined via an assumed filling factor, are generally estimated at ~0.1-1 $M_\odot$ yr$^{-1}$, ~1-10 times the accretion rate and with a kinetic energy small in comparison with the AGN's bolometric luminosity. This suggests the outflow does not lead to significant feedback, as discussed in section 6. However, estimates of the mass outflow rates at levels of ~10-100 $M_\odot$ yr$^{-1}$ in several sources known as mini-broad-absorption line quasars, are high enough to have a significant impact on the evolution of the host galaxy (e.g. PG1116+080 [131], APM 08279-5255 [132]). These rates remain highly uncertain given the level of variability seen in the absorption features and the single line-of-sight necessarily involved [133].

In a few nearby AGN, *Chandra's* high-spatial resolution can resolve and investigate larger-scale extended X-ray emission directly. For example, observations of the nearby AGN NGC 4151 show outflow rates of ~2 $M_\odot$ yr$^{-1}$ at ~130 pc and kinetic energy in the outflowing material of ~$1.7 \times 10^{41}$ erg s$^{-1}$, ~0.3% of the source's bolometric luminosity [134].

5.1.3 *Accretion Flow.*

Fe Kα is the strongest emission line in the X-ray spectra of AGN. It originates in reflection of the hard X-ray continuum off the inner edge of the accretion disk very close to the SMBH event horizon, and its line profile is used to estimate the relativistic broadening and thence the spin of the SMBH. *Chandra* grating observations have confirmed the relativistic broadening of the strong Fe Kα emission line in MCG-6-30-15. The top portion of Figure 5-2 shows a narrow Fe I emission line and blue-shifted absorption lines superimposed on the broad Fe Kα line. Detailed fits to the absorption and emission features in the full spectrum (e.g. Figure 5-2, lower) rule out absorption-related models for the continuum around Fe Kα and demonstrate that the X-ray emission is dominated by a relativistic accretion flow very near the BH (inner radius <1.9 $R_g$, spin $a/R_g > 0.95$ [123]), confirming previous deductions for BH spin based on the shape of broad Fe lines.

The recent launch of NASA's NuStar satellite extends CCD-like-resolution spectroscopy up to ~50 keV which significantly improves the constraints on the reflected X-ray continuum and emission line profile of Fe Kα when combined with lower-energy *Chandra*/*XMM-Newton* observations. This was recently demonstrated by coordinated *XMM-Newton*/NuStar observations which confirm a relativistically broadened line in NGC1365 and demonstrate that the SMBH is spinning at close to maximum value ($a$>0.84) and originating in material within ~2.5 $R_g$ of the SMBH [135].

As we have seen, the X-ray emission from an accreting SMBH comes from within a few gravitational radii, a region which typically subtend an angle of less than a micro-arcsecond. Notwithstanding this very small scale, microlensing produced by stars in a lensing galaxy along the line of site provides a natural telescope for studying the structure of the accretion disk. Gravitational microlensing in quasars, requires arcsecond telescope resolution and has been used in observations of the quasar RXJ1131-1231 [136] to fix the extent of the X-ray and optical emission length scales at ~10 and 70 $R_g$ respectively. The size of the optical emission region is larger than expected, and, when combined with the X-ray results, indicates that modifications to the standard thin disk model may be required, a result also found in microlensing studies of other quasars [137, 138].

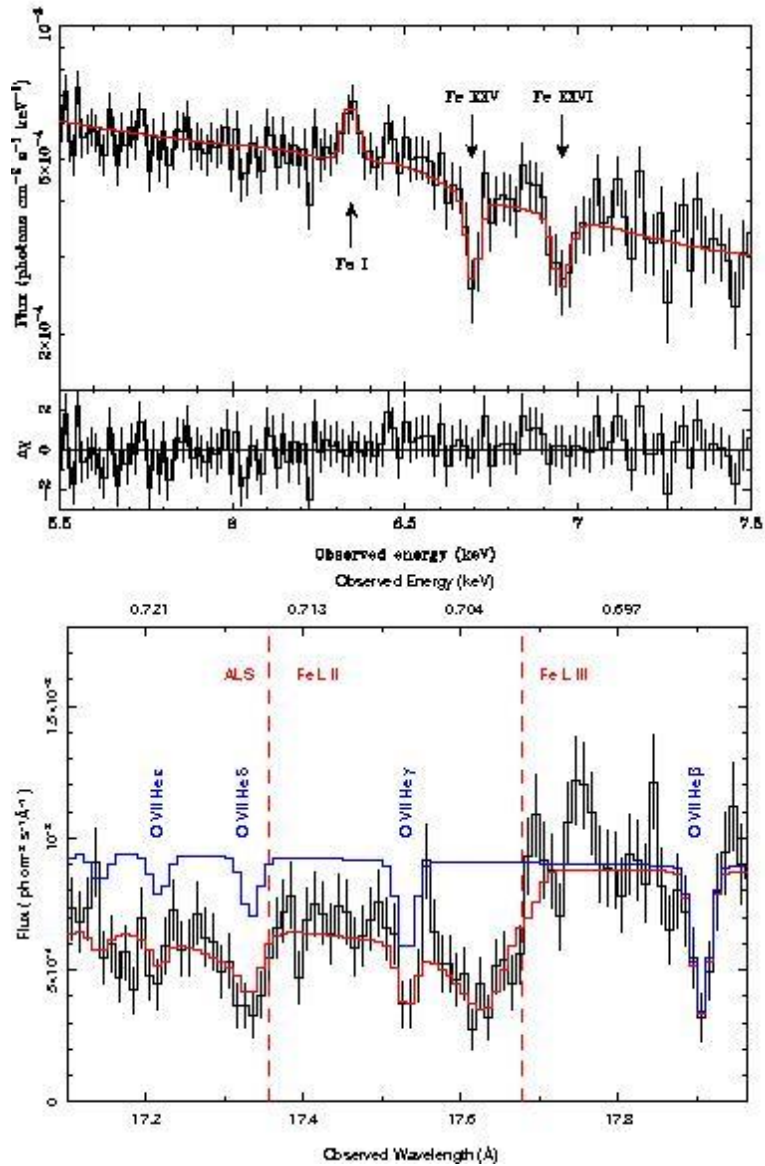

Figure 5-2. HETG spectrum of MCG-6-30-15 showing the Fe Kα line region (upper [123]) and the long-wavelength region (lower, courtesy J.C. Lee). Absorption due to highly-ionized oxygen (blue line) is combined with FeI in dust/molecules local to the source to generate the best fit spectrum (red line). Credit: upper [123] and lower J. C. Lee.

At the other end of the luminosity scale, *Chandra* observations of galaxies in the Virgo cluster [139] and the Sloan Digital Sky Survey (SDSS) [140] were able to resolve low-luminosity nuclear emission from surrounding sources and diffuse emission. These low-luminosity AGN radiate from $L/L_{Edd}$ <0.01 down to $L/L_{Edd}$ ~$10^{-8}$ where $L_{Edd}$ is the Eddington Luminosity (defined above in section 4.8). In addition, a *Chandra*/HETG study of the SMBH in M81 [141] shows that the models that have been applied to Galactic stellar mass black holes apply to this low-luminosity SMBH. The M81 result provides the first high resolution detection of X-ray line emission in a low-luminosity AGN, plausibly giving the first evidence for the hot plasma associated with advection-dominated accretion flow (i.e. a flow in which the infalling material crosses the SMBH event horizon without radiating much of its energy, also known as inefficient accretion) near a SMBH.

Sgr A*, the source at the center of our Milky Way Galaxy, has a mass $\sim 3\times 10^6$ $M_\odot$ [142] and is the nearest SMBH. *Chandra* has shown that its low quiescent luminosity $\sim 10^{-11}$ $L_{Edd}$ is due to inefficient accretion [143], as opposed to lack of fuel. *Chandra* observations of another, relatively nearby (distance $\sim$ 24 Mpc) elliptical galaxy NGC821 also show a low accreting luminosity, $\sim 10^{-8}$ $L_{Edd}$ [144], and lead to a similar conclusion for the quiescent SMBH in its nucleus. Similar results are found for a number of other nearby SMBH [145]. *Chandra* has also traced current [146] and past flaring activity for Sgr A*, the latter based on light echoes from nearby molecular clouds [147].

In the course of ground-based infrared monitoring of the motions of stars around SgrA*, astronomers discovered a dense, cold cloud of gas (G2) on a potential collision course with the SMBH. The cloud is estimated to have a mass about 3 times that of the Earth, with some uncertainty as to its actual makeup [148]. In spring 2013, G2 was observed to be stretching, with a small fraction (<10%) of its emission seen beyond the pericenter (which is estimated to be ~20 light-hours or ~4000 $R_g$ from Sgr A*) [149]. It was projected that pericenter passage would last at least a year as the stretching process continues. The stressing and disruption of G2 by its motion through the hot ambient gas may increase the X-ray luminosity to as much as 10 times the typical X-ray output from Sgr A*. Measuring the heating versus time may enable investigators using *Chandra* to trace the density of the ambient gas in the vicinity of the black hole. Various models for inefficient accretion and radiation have been invoked to explain the extremely low "nominal" luminosity for Sgr A*. These models predict different density profiles versus radius for the gas close to the SMBH, so the G2 data may determine which if any of these models are viable. With the compression and eventual tidal shredding of G2, a substantial amount of fuel should be liberated and potentially feed the SMBH for several years or more following the pericenter pass. Here, predictions are even more uncertain but enough gas may be available to raise the average X-ray output of Sgr A* by 3 or 4 orders of magnitude for as long as a decade. Just how this matter flows towards the SMBH and how much energy may be released and with what cadence will be the objective for ongoing monitoring campaigns, including *Chandra*, for a number of years to come.

*5.2 X-ray Surveys, Population Studies, and the Cosmic X-ray Background*

*5.2.1 X-ray Surveys*

The advent of NASA's Great Observatories has facilitated a number of major multi-wavelength surveys at a range of depths in order to sample both bright and faint populations of celestial sources (e.g. SWIRE [150], GOODS [151], Bootes [152], ChaMP [153], COSMOS [154], AEGIS [155]). Through the use of hard (up to 10 keV) X-ray observations, which probe deep into obscured regions, combined with mid-infrared selection, these surveys probe the AGN population, including the large fraction of obscured objects, more completely than traditional optical and near-infrared surveys [156]. Mid-infrared selection sees all the AGN but requires a secondary, usually X-ray, selection method to distinguish them from the larger infrared (IR) galaxy population [156, 157, 158]. Thus *Chandra* is a powerful AGN finder, able to see all but the most highly-obscured (CT) ones which are invisible over much of the electromagnetic spectrum.

*Chandra* is unique among current X-ray missions in being able to perform observations much deeper than those made previously. Because of *Chandra's* high spatial resolution and low background, the deepest *Chandra* surveys to date remain photon rather than background or confusion limited (Figure 5-1). The 4

Ms *Chandra* Deep Field South (CDFS), with limiting flux, $F$(0.5-2 keV) ~ $9\times10^{-18}$ erg cm$^{-2}$ s$^{-1}$, detects 776 X-ray sources [159]. About 97% of the 776 sources have reliable multi-wavelength counterparts showing that >75% are AGN in the range 0.2<$z$<5, providing the highest sky density (10,000 deg$^{-2}$) and most complete view of the AGN population to date. As expected based on previous studies, there is a trend of increasing absorption with decreasing X-ray flux.

Wider, shallower (limiting flux, F(0.5-2 keV) ~ $10^{-16}$ erg cm$^{-2}$ s$^{-1}$) surveys, such as COSMOS [160] along with archival data used to generate "serendipitous" surveys (Chandra Multi-wavelength Project; ChaMP [161]), yield larger samples including rarer, brighter sources. They generally confirm the trend towards a larger fraction of obscured and non-active galaxies at fainter flux levels. With increased numbers of sources and fainter optical flux limits, the AGN population found with Chandra reaches X-ray/optical luminosity ratios ($L_x/L_{opt}$) ~3× larger than in earlier surveys. A strong correlation between hard-band (2-10 keV) $L_x/L_{opt}$ and $L_x$ [162] for obscured AGN [160, 163] can be used to estimate redshifts for sources too faint for optical or near-infrared spectroscopy.

### 5.2.2 *Population Studies*

The combined *Chandra* surveys have provided a unique view of how SMBH grow over cosmic time. The data reveal that the most luminous AGN are more numerous at early times. If interpreted in terms of their mass, this result suggests that (counter-intuitively) the most massive BH ($\geq10^9$ M$_\odot$) grow at the earliest times, possibly driven by higher galaxy merger rates in the younger, smaller universe, and the less massive BH ($\leq10^7$ M$_\odot$) are still growing today [162]. This picture, where the less massive, lower luminosity AGN are seen preferentially at the current epoch is often characterized as cosmic down-sizing. An alternate interpretation suggests that the lower luminosities at the current epoch are due to lower accretion rates rather than being linked directly to the mass of the SMBH [164].

Recent wide-area and deep X-ray surveys have pushed AGN X-ray luminosity functions to higher redshifts ($z$>5) [165] and lower fluxes ($10^{-17}$ erg cm$^2$ s$^{-1}$) [105]. Studies combining the CDFS with the wider, shallower surveys AEGIS [166] and ChaMP [167] find a peak in the total luminosity density of AGN at $z$~1.2 and a luminosity function with constant shape but with strong luminosity evolution out to $z$~1 and negative density evolution at higher redshifts. Comparisons indicate that the SMBH mass accretion rate decreases with increasing redshift faster than the star formation rate at $z\geq1.5$. At high luminosity, the space density of AGN declines rapidly out to $z$>5, confirming the trend seen in optical surveys [165].

### 5.2.3 *The Obscured AGN Population and the Cosmic X-ray Background*

Current successful models for the cosmic X-ray background include roughly equal populations of unobscured and moderately obscured (log $N_H$ (intrinsic) ~22-23.2, Compton thin) AGN to account for the emission up to ~10 keV [168]. A population of CT AGN comparable to that of Compton thin AGN is required to explain the higher energy (~30 keV) portion of the background. *Chandra* X-ray surveys have now resolved ~75-95% of the background at energies 0.3-10 keV into a mix of moderately obscured and unobscured AGN in proportions consistent with the models [161, 169, 170] but the heavily-obscured population remains elusive.

The ratio of obscured (both Compton thin and CT) to all AGN is critical to estimating the total accretion power and thus the AGN contribution to the energy budget of the universe. This ratio remains a matter of debate as multiple studies in different wavebands draw a variety of conclusions, complicated by the fact that optical/IR studies measure dust obscuration while X-ray studies measure the cold gas column density.

Optical and near-IR surveys estimate that 65-75% of all AGN are obscured [171, 172, 173, 174]. The general consensus from *Chandra* and *XMM-Newton* results includes a strong luminosity dependence with 10-20% of AGN obscured at high luminosity ($L_X$~$10^{44-46}$ erg s$^{-1}$), while at lower-luminosity the numbers are ~80% [175, 176]. This luminosity dependence is combined with a redshift dependence, such that the obscured fraction increases with redshift [177, 178, 179], although perhaps only at high luminosities ($L_X$~$10^{44}$ erg s$^{-1}$) [180, 181]. Notably, one X-ray study concludes that there is no luminosity or redshift dependence [182], indicating that still more work is required to fully reconcile all of the observations and models.

Radio-selected samples indicate a high, 50-60% obscured fraction [183, 184] at high luminosity, which again increases towards lower luminosity [185]. Since low-frequency radio selection has no obscuration bias, these results might be reconciled with the X-ray results quoted above if the most highly-obscured sources are missed in the X-ray surveys [118, 186, 187]. This interpretation is supported by the recognition that the apparent $L_X$ of the obscured sources extends a factor of ~300 below that of unobscured sources with the same intrinsic luminosity. A luminosity dependence can be explained by a physical model in which the opening angle of an obscured disk/torus increases as the luminosity increases (the "receding torus model" [188, 189]) or by contamination of the AGN samples by sources which are not actively accreting (e.g. low-ionization emission-line radio galaxies) [190, 191].

The CT population remains mostly undetected individually. They are difficult to find at *Chandra* and *XMM-Newton* X-ray energies (≤10 keV) and even at higher X-ray energies covered by Swift, INTEGRAL and now NuSTAR. Direct light from NGC 1068, the "Rosetta-stone" CT source, is undetected to much higher energies (≤100 keV) [192]. The X-rays detected in the few keV band are scattered and/or reflected into our line of sight.

Estimates of the fraction of CT AGN, which are based on the small number of detections and/or on X-ray stacking (summing the counts found at the location for each of the galaxies to treat the collection as if it is a single target) techniques [110, 156, 170, 193, 194, 195, 196, 197, 198, 199, 200] yield a wide range from 0.05 to twice the rest (unobscured and Compton thin) of the AGN population. A study via a mix of detections and stacking analysis of the X-ray properties of massive galaxies in the 4 Ms CDFS [201] yields an estimate of the space density of CT AGN at $z$ ~1.4-2.6 comparable to values derived for unobscured AGN [166, 167]. At lower redshifts of $z$ ~0.5-1 stacking the CDFS data on 23 galaxies with excess infrared (IR) emission over similar star-forming galaxies in the central 6′ show a hard effective X-ray spectrum indicative of heavy obscuration [170]. Monte Carlo simulations indicate that >50% of these 23 IR galaxies are heavily obscured with 80% of these being CT. A recent *Chandra* survey of high-redshift (1<$z$<2) radio-loud 3CRR sources, selected based on extended low-frequency radio emission and so unbiased by obscuration, concludes that 21% of the sample are CT and the overall obscured fraction (CT and Compton-thin) is 50% [118].

In cases where sufficient counts are available, one can employ spectral fitting to detect the reflection-dominated X-ray continuum and strong 6.4 keV Fe Kα emission that are the X-ray signatures of CT absorption, where Compton scattering within the absorption column makes it difficult for X-rays to escape directly. X-ray spectroscopy has confirmed the identification of several individual CT-AGN at high redshift: $z$ = 2.6 and 2.9 [202] and the most distant heavily obscured AGN confirmed to date at $z$ = 4.76 [203]. Although the numbers of CT objects amenable to detailed X-ray spectral fitting are small, they strengthen the evidence that a significant population of highly obscured AGN exists at high $z$.

### 5.3 *Jets and Extended Radio Structure*

The primary mode by which the SMBH transmits energy over galactic and intergalactic distances is thought to be via relativistic jets that originate from the innermost regions of an accretion disk around the BH. Extended radio structures including jets, lobes and hotspots have been known for decades, but it was only with *Chandra* that it became apparent that jets are prominent at X-ray wavelengths as well. The first target observed by *Chandra*, PKS 0637-755 (a radio-loud quasar at $z$= 0.651) provided the first detection of a long (100 kpc) X-ray jet co-spatial with the radio jet [204, 205]. The emission mechanisms (see Section 1.) proposed for the extended X-ray/radio structures include synchrotron radiation and inverse Compton scattering of cosmic microwave background (CMB) photons or radio synchrotron electrons, depending on the size and physical conditions of each structure. The observations require high velocity plasma with short radiative lifetimes ~years, so the radiating electrons must be continually re-accelerated [206]. *Chandra's* high resolution in combination with current, high-resolution radio observations have resulted in major strides in our understanding of jet physics and provided a tool to study the physical conditions in hot spots and lobes [207].

Centaurus A, the nearest radio galaxy at 3.7 Mpc, is a low-power radio source with a powerful X-ray jet (~5 kpc long) and affords the best opportunity for detailed study. The X-ray jet includes many discrete knots (Figure 5-3) which appear to be the sites of standing shocks and in situ particle acceleration required to generate sustained X-ray synchrotron emission [207]. The changing spectrum of the knots transverse to the jet suggests varying shock strength and possible knot migration [208]. The power law spectrum of the X-ray emission at the outer edge of the south-west radio lobe suggests non-thermal synchrotron emission as a result of particle acceleration by the expanding radio lobe with speeds estimated at 2600 km s$^{-1}$ (Mach 8 relative to the ambient medium i.e. 8 times the sound speed) [209]. This has implications for understanding particle acceleration processes, and directly indicates the transfer of the energy from the jets to the surrounding gas.

*Chandra* observations of NGC 4151 have found soft diffuse X-ray emission ($L$(0.3-2 keV) ~$10^{40}$ erg s$^{-1}$) within the inner ~150 pc spatially correlated with the radio outflow and [OIII]$\lambda$5007 optical line emission [210]. Analysis of the emission lines and X-ray emission reveals collisional ionization in addition to photoionization by the central AGN. The thermal energy suggests that ≥0.1% of the jet power is deposited in the hot interstellar medium in the dense circumnuclear region.

At higher redshift and luminosity, radio-loud AGN often show extended emission line regions aligned with the radio structure [211, 212, 213]. *Chandra* observations have shown associated, aligned X-ray emission, most likely thermal emission from shocked hot gas (e.g. PKS 1138-262 [214], 3C305 [215, 216]), which dominates the energy budget in extended material. The optical emission lines probe the dynamics of the gas and demonstrate that the radio source drives a large-scale outflow with speeds ~100s km s$^{-1}$ [217], providing direct evidence for interaction between the jet and its environment.

X-ray emission is produced by inverse Compton scattering of the CMB off the relativistic electrons in large radio sources. Inverse Compton/CMB (IC/CMB) X-rays can be detected out to high redshift due to the $(1+z)^4$ enhancement of the CMB energy density which compensates for the $(1+z)^{-4}$ decrease in surface brightness [218]. *Chandra* has detected a few high-redshift ($z$>4) sources to date, the current highest is a 24 kpc X-ray/radio jet in GB 1428+4217 at redshift 4.72 [219]. All are consistent with inverse Compton scattering off the CMB for the X-rays with somewhat lower Lorentz factors (~3-5) than their lower-redshift counterparts (~5-10). Although the high-redshift sample is small, this difference suggests slower large-scale jets in the more inhomogeneous high-redshift environments. At lower redshifts, where the structures can be resolved, X-rays are associated with lobes and hot spots as well as jets [112, 220, 221]. Modeling of the spectral energy distributions shows X-ray emission from internal IC (synchrotron self-Compton) and direct synchrotron emission as well as IC/CMB in different sources.

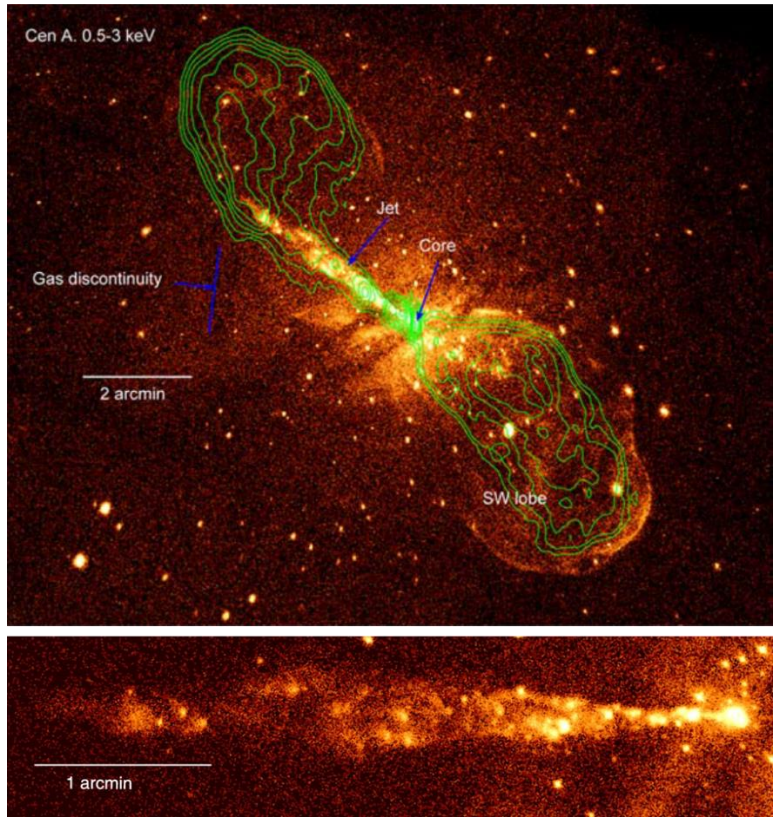

Figure 5-3. (Top) Radio contours superposed on a deep *Chandra* ACIS image of Centaurus A [207] showing the core and the northeast jet crossed by absorption stripes corresponding to NGC 5128's dust lanes, the southwest lobe, structures associated with the northeast lobe and the position of a merger-related gas discontinuity. (Bottom) 0.8-3 keV *Chandra* ACIS image of the northeast Centaurus A jet showing the complex, knotty structure [208]. Credit: top: [207] and bottom: [208].

Powerful radio galaxies offer a view of the interaction of AGN with their environment on a larger scale and out to relatively high redshifts. *Chandra* observations [222] of the $z\sim2.5$ quasar 4C 23.56 show X-ray emission, presumably due to IC scattering of CMB photons, extending for ~500 kpc along the radio structure. Offsets between the radio and X-ray structure provide evidence for multiple outbursts suggesting that they have a long-lasting influence on ongoing structure formation processes.

## 5.4 Growth of SMBH and Host Galaxies

In the past decade, astronomers have gathered a substantial body of evidence which indicates that many and perhaps all elliptical galaxies and spiral bulges host SMBH, and that the growth of these BH and their host galaxies is closely interconnected [223, 224, 225]. A major challenge is to better understand the nature of the co-evolution of SMBH and galaxies, and the astrophysical processes which drive it. Among the critical questions: when did this co-evolution begin, how long does it continue, and how does it depend on galaxy type? A fundamental constraint on all theories modeling the interplay of SMBH growth and galaxy evolution is the fraction and type of galaxies that host actively accreting SMBH for which X-ray emission is the most reliable signature. *Chandra's* extremely high sensitivity to point sources, combined with existing data bases for these sources generated from complementary, deep exposures in other wavelength bands has produced significant advances in the ongoing investigation of SMBH and galaxy co-evolution.

Extensive observational and theoretical research on the growth of SMBH has identified several plausible mechanisms: (1) major mergers involving collisions between two roughly equal-sized galaxies channel gas into the SMBH which eventually merge [226], (2) more gradual, so-called secular processes such as accretion of filaments or clumps of gas from galactic halos, bar formation, or minor mergers involving satellite galaxies, followed by instabilities internal to the galaxy drive accretion onto a centrally located SMBH [227], and (3) cosmological accretion of baryons from dark matter filaments [228, 229]. Evidence is accumulating for a picture in which all scenarios may play important roles at different stages in the co-evolution of galaxies and SMBH.

BH mergers are most common at high $z$, but we can only study relatively nearby examples in detail. *Chandra's* ability to detect and resolve rare, close pairs of accreting SMBH is critical to this research. X-ray observations allow for a clear identification of BH accretion, less hampered by obscuration and merger-induced star formation that can severely impact optical emission-line studies, especially for galaxies in pairs. Combined *Chandra* and optical data have been used recently to discover 4 pairs of SMBH with angular separations ranging from 0.6" to 5.8", with physical separations of ~ 0.15 kpc to 21 kpc [230, 231, 232, 233] (Figure 5.4). Surveys of the spatial distribution of AGN can also be used to investigate merger-induced SMBH growth. An analysis of the observed clustering properties of AGN in the *Chandra* survey of the Bootes field [234] found that the distribution of AGN is broadly consistent with simulations [235] that suggest that AGN are triggered by mergers of similarly-sized galaxies near the centers of galaxy groups. A higher incidence of moderate luminosity AGN ($10^{42}$ erg s$^{-1}$ <$L_x$ <$10^{44}$ erg s$^{-1}$) was found [236] in galaxy pairs for a sample of galaxies with 0.25 <$z$ <1.05, supporting a merger scenario. However, the same study found that about 80% of moderate luminosity AGN are not in pairs, leaving open the question of what physical processes are responsible for their fueling.

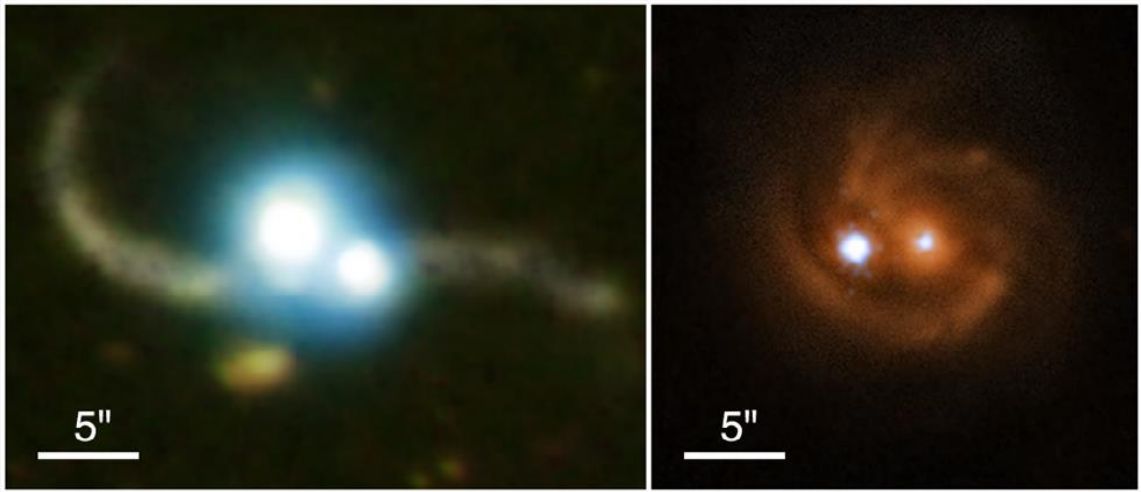

Figure 5-4. The left image shows a binary quasar in SDSS 1254+0846 [233], where the *Chandra* image (blue) shows the two quasars (projected separation 21 kpc) and the optical Magellan image (yellow) shows tidal tails around the host galaxy. The right image shows the binary AGN in Mrk 739 [231] where the *Chandra* image (blue-white) shows the 2 AGN (projected separation 3.4 kpc) and an SDSS optical image (red) shows the host galaxy. Credit: left: X-ray: NASA/CXC/SAO/P. Green et al.; Optical: Carnegie Obs./Magellan/W. Baade Telescope/J. S. Mulchaey et al.; right: adapted from [231].

*Chandra* and *XMM-Newton* observations combined with HST imaging of galaxies harboring moderate luminosity AGN in the CDFS [237] and COSMOS [238] show that about 80% of the host galaxies have undisturbed, disk-dominated light profiles indicating that internal secular processes and minor interactions dominate BH growth in the redshift range 0.3<$z$<3. This strengthens the suggestion that a significant fraction of BH growth is not due to major mergers. Additionally, sub-millimeter and far-infrared with (ESA's Herschel Space Observatory) studies of X-ray selected AGN in the redshift range $0.5<z<3$ in the *Chandra* Deep Field North, CDFS and the extended CDFS find that the infrared host galaxy luminosity (a measure of the star formation rate) in moderately luminous AGN mirrors the redshift dependence of the mean galaxy, suggesting that star formation is decoupled from activity in galactic nuclei [239, 240, 241].

X-ray stacking analysis in the CDFS suggests that galaxies and SMBH have grown with a constant ratio of SMBH to stellar mass, $M_{BH}/M_*\sim0.001$, since at least $z\sim2$ over a wide range of host galaxy mass [242] consistent with the observed $M_{BH}$–$M_{bulge}$ relation [223]. However, in higher luminosity AGN ($L_X \geq 10^{44}$ erg s$^{-1}$) a strong correlation between X-ray and IR luminosities suggests a direct relationship and points towards merger-driven SMBH growth [243, 244].

The discovery [245] of a quasar at redshift $z = 7.07$, subsequently detected by *Chandra* with an implied SMBH mass $\sim 2\times10^9$ M$_\odot$ places severe constraints on models to explain the rapid SMBH growth required (< 1 Gyr) [246]. However, the very low space density of luminous quasars indicates that they represent only a fraction of early SMBH growth in the range $z\sim$1- 6. Instead, a significant amount of growth is thought to occur in lower mass SMBH cloaked by dust and gas, and this obscuration could persist for 100 Myr to several Gyr during which time SMBH could accumulate much of their mass [247, 248]. *Chandra* observations of distant sub-millimeter galaxies, extremely powerful starburst galaxies at $z\sim$1.5-3, generally detect moderate-luminosity, obscured AGN [249]. Thus sub-millimeter galaxies appear to represent the primary epoch of spheroid formation in galaxies and the presence of obscured AGN indicates that the SMBH are growing concurrently in these sources.

The recent evidence for a SMBH in a nearby dwarf galaxy suggests that a galactic bulge is not required for a $10^6$ M$_\odot$ SMBH to form, in contrast to expectations based on the $M_{BH}$-$M_{bulge}$ relation. Specifically, *Chandra* archival data have been used to pinpoint the location of a hard X-ray source coincident with a compact radio source at the dynamical center of the nearby dwarf starburst galaxy Henize 2-10 [250]. The combined radio and X-ray data, together with the fundamental plane of black hole activity, which links observed correlations between radio and X-ray luminosities and SMBH mass [251] imply a SMBH mass of $\sim 2\times10^6$ M$_\odot$. Nearby dwarf galaxies with extreme starburst activity and concurrent SMBH growth could be analogs of galaxies in the early universe, when galaxies were colliding and merging more frequently than at present, and so could provide insight into the formation of SMBH at very high redshifts.

In summary, X-ray observations show that both major mergers and secular growth contribute to the growth of SMBH at the centers of galaxies. In order for sufficient growth to occur to fuel luminous AGN at high redshift, it is likely that both major mergers and cosmological accretion contribute at early times.

## 6 Galaxies, Groups, and Clusters

The X-ray emission from galaxies generally consists of three components: (1) a point source in the galactic nucleus due to a SMBH; (2) point sources due to X-ray binaries; and (3) diffuse emission from the hot interstellar gas, a hot corona, and faint unresolved point sources including individual stars. For very nearby galaxies at a distance less than a few Mpc, it is also possible to resolve the extended emission from SNR. In groups and clusters of galaxies there is an additional, often dominant, component due to the diffuse, intergalactic hot gas bound to the group or cluster by the gravity of dark matter. *Chandra's* unique sub-arcsecond angular resolution, combined with its capability for spatially resolved spectroscopy, has led to major advances in the study of these components, with profound implications for understanding the evolution of galaxies, groups and clusters. The SMBH component is discussed in Section 5. In this section we focus on point sources outside the nucleus, hot galactic coronae, and intergalactic hot gas associated with groups and clusters.

### 6.1 Normal and Starburst Galaxies

#### 6.1.1 Point Sources in Normal Galaxies.

Starting within our own Milky Way Galaxy, a long (900 ks) *Chandra* observation of a field about 1° south of Sgr A* at the Galactic center, with a low density absorbing column of hydrogen along the line of sight, has resolved ~80% of the previously unexplained X-ray emission from the Galactic Ridge along the plane of the Galaxy. The detected point sources are primarily accreting white dwarfs and coronally active stars [252]. This discovery argues against the existence of a significant amount of diffuse, unconfined, 100 MK gas in that region of our Galaxy.

Individual X-ray sources with luminosities comparable to those of X-ray binaries in the Galaxy can now be detected at the distance of the Virgo Cluster (~20 Mpc) and beyond. While studies of X-ray binaries in the Galaxy often suffer from significant distance uncertainties and interstellar absorption due to cold gas in the Galaxy, these difficulties are minimized for X-ray sources in nearby galaxies. Absorption and scattering effects are less problematic along lines of sight away from the Galactic disk, and effects of distance uncertainties are reduced since all X-ray sources in an external galaxy are essentially equidistant from the observer. This makes it possible to construct luminosity functions over two-three decades of luminosity for X-ray binary populations with luminosities in excess of ~ $10^{36}$ erg s$^{-1}$.

The luminosity of a star varies non-linearly with mass, $L \propto M^x$ with x ~ 3-5 for stellar masses in the range $0.1 < M/M_\odot < 100$, so the evolutionary lifetime of a star is $\propto (M/L) \propto M^{(1-x)}$, decreasing rapidly with increasing mass. Typically, X-ray binaries fall into two classes: high mass X-ray binaries (HMXB), in which a NS or BH is accreting gas from a relatively young and massive ($\geq 2.5 M_\odot$) star, and low mass X-ray binaries (LMXB) in which the donor is a low mass star. HMXB are therefore found in galaxies (mostly spirals), where stars are still forming, whereas the LMXB trace the older, less massive stellar populations of the host galaxies [253, 254].

One of the most important quantities in determining the appearance and evolution of a galaxy is the rate at which stars are forming in the galaxy. By constructing model galaxies that incorporate the spectra and luminosities of stars of different masses and comparing these models with actual optical and infrared measurements, astronomers determine the star formation rate (SFR) of a galaxy [255]. However, star-forming galaxies usually contain a large amount of dust, which causes absorption and scattering for optical and near-IR radiation and introduces uncertainty in the determination of the SFR. Since massive stars have short lifetimes, ~ a few × $10^7$ y or less, their collective luminosity is a good indicator of the rate at which they are currently being formed. For spiral galaxies with vigorous star formation, HMXB should dominate the stellar X-ray luminosity. Thus, HMXB can be a good proxy for the SFR, both locally and out to redshift $z\sim$1-2 [256, 257, 258].

For more distant galaxies, *Chandra* cannot detect individual HMXB or the galaxies themselves. However, *Chandra's* angular resolution enables stacking X-rays for galaxies in the CDFS, showing that the $L_X$ – SFR correlation remains invariant over the redshift range 0<$z$<4 [259]. It has been suggested that for $z$>6, the lower-metallicity environment could lead to the formation of more stellar-mass BH and thereby more HMXB with accreting BH, leading to higher overall X-ray output. If the X-ray emissivity for these high star-forming galaxies were to be boosted by an order of magnitude, the X-rays could be responsible for the re-ionization of the intergalactic medium at $z$ ~6 [260].

A *Chandra* study of blue compact dwarf galaxies provides suggestive evidence in support of this proposed role of X-ray binaries in re-ionizing the intergalactic medium. An order of magnitude enhancement of X-ray emissivity over that predicted by the usual $L_X$ – SFR relation is found [261]. It has been suggested by a number of authors that blue compact dwarf galaxies are analogs to unevolved galaxies in the early universe. If so, then enhanced production of X-ray binaries in blue compact dwarf galaxies supports the idea that X-ray binaries play a significant role in re-ionization at high $z$. Much deeper (factor 3) exposures with *Chandra* are needed to test this idea.

In gas-poor elliptical and lenticular (central bulge, flattened disk and no spiral arms) galaxies, star formation has effectively ceased. The massive stars have run their evolutionary course so the point-like X-ray sources are expected to be predominantly LMXB, as now confirmed by *Chandra* [253]. Moreover, *Chandra* observations of these early-type galaxies show that a significant fraction (30-70%) of LMXB is found in dense, gravitationally bound collections of stars called globular clusters [253]. As self-gravitating systems, globular clusters require a supply of energy to avoid collapse. Their evolution is characterized by three stages: (1) core contraction where gravity shrinks the central (or core) region, (2) binary burning where energy is drawn from stellar collisions/interactions and binary orbits to stabilize the core, and (3) core collapse where energy from such stellar interactions is insufficient causing the core to shrink dramatically, with the central density increasing to the point where new binaries may form.

The LMXB in globular clusters are thought to have formed by dynamical interactions between ordinary binary star systems and individual NS and BH. Thus, the number of X-ray sources in a cluster can be used to infer the frequency of such encounters, which in turn is related to the dynamical history and current state of the cluster. Surprisingly, *Chandra* data show an excess in the number of X-ray sources expected for a few clusters thought to have already undergone core-collapse [262, 263]. The interpretation of this result, combined with theoretical and optical work, is that these clusters have not reached core-collapse but are in the quasi-steady binary burning state. A further implication is that most globular clusters are even less evolved than previously thought and are still in the initial stage of core contraction. This view represents a significant shift in our understanding of the dynamical evolution of globular clusters.

6.1.2  *Hot Gas in Elliptical and Star-Burst Galaxies.*

It had been speculated for years that elliptical and lenticular galaxies might contain substantial amounts of hot, interstellar gas that remained undetected because of its high temperature (~2-10 MK). This gas was discovered by the Einstein X-ray Observatory in the late 1970's, but the relative contributions of hot gas and unresolved point sources were the subject of debate for at least two decades. With *Chandra*, it is possible to resolve individual point sources, and to disentangle the relative contributions of X-ray binaries, white dwarfs, active stellar coronas and the diffuse hot gas. With this done, the mass, temperature, and elemental abundances of hot gas present in a galaxy can be investigated [254, 264, 265].

For example, *Chandra* data for the galaxy NGC1316 demonstrate the need to account fully for the LMXB before tracing the hot gaseous halo density and temperature versus radius and comparing with optical light profiles. With the LMXB properly accounted for, spectral fits to the hot gas show that its elemental abundances are similar to solar, in agreement with predictions of stellar evolution models, resolving an issue raised by earlier reports of extremely low abundances for heavier elements in the hot gas in this galaxy [266].

Prior to *Chandra*, X-ray surveys of early-type galaxies showed a surprisingly large scatter in the correlation between X-ray luminosity and optical (blue band) light. With *Chandra's* ability to sort out point sources and to survey over a sufficient range of X-ray luminosity, the data show the presence of a break-point around an X-ray luminosity of order $10^{40}$ erg s$^{-1}$. Below this level the X-ray luminosity of an early-type galaxy is dominated by LMXB point sources and above it a hot, diffuse gas component dominates [253].

For samples restricted to galaxies with truly diffuse X-ray luminosity above $10^{40}$, the perplexingly large scatter in X-ray luminosity versus optical light (and/or stellar mass) substantially disappears. Furthermore, an even tighter correlation is now found between diffuse X-ray luminosity and total galaxy mass, including dark matter. One interpretation is that galaxies deficient in hot gas also have relatively small dark matter halos that are unable to retain hot gas which then flows out of the galaxy. In contrast, galaxies with massive dark matter halos retain their hot gas, which then dominates their X-ray emission [267].

*Chandra* observations of the extended gaseous halo in two galaxies have provided new insights into the nature of two previously reported outliers to the well-established correlation between mass of a galaxy's bulge region and mass of its central SMBH (cf Section 5.4). These galaxies, NGC4342 and NGC4391, have substantially more massive (by factors ~15 and ~50 respectively) SMBH than predicted by the bulge-SMBH correlation. It was hypothesized that in these galaxies, most of the bulge stars had been stripped from the host galaxy via interactions with other nearby galaxies, thereby lowering the host's brightness and the associated bulge mass.

*Chandra* has now disproved the tidal stripping explanation by detecting extended X-ray emission associated with hot gaseous coronae. These hot coronae require substantial dark matter halos to bind the gas. Dark matter halos would be more vulnerable to tidal stripping than stars in the bulge, hence their presence rules out tidal stripping of the stars. It is likely that at least in these galaxies the central BH grew faster while star formation was suppressed, perhaps due to outbursts from the central AGN which now appears as a quiescent SMBH [268].

In starburst galaxies, the heating of the gas can be so intense that the gas cannot be confined to the galaxy. These galaxies undergo intense bursts of star formation, often as a result of collisions between galaxies. The most massive stars race through their evolution and explode as supernovae. If the supernova rate is high enough, the combined effects of the many supernova shock waves create and drive a galactic-scale superwind that blows gas out of the galaxy. A prime example of this phenomenon is seen in the relatively nearby (distance ~3.5 Mpc) galaxy known as M82. Galactic superwinds such as the one in M82 are rare today, but they were much more common billions of years ago when collisions of galaxies were more frequent.

*Chandra* observations of M82's superwind provide the first direct measurement of the energy efficiency of supernova and stellar wind feedback, and their supply of metals to enrich the interstellar medium [269]. One consequence is the generation of new models, which more accurately represent the energetics of starburst-driven superwinds and their roles in galaxy evolution as well as the enrichment of intergalactic space with metals synthesized in stars (in this context, astronomers characterize all elements heavier than helium as metals).

*Chandra* observations of the more distant and more energetic Antennae starburst galaxy show that it is filled with a hot interstellar medium, which is enriched in metals (e.g., Ne, Mg, Si, Fe). The abundances in certain places are much larger than solar, consistent with production in Type II supernovae typical of a young stellar population rich in massive stars [270]. As illustrated in Figure 6-1, two galaxy-sized loops of hot gas are also seen in the Antennae, embedded in a more tenuous diffuse hot halo. While the specific geometry of these features is yet to be explained, their presence suggests outflows, which may disperse the metals into inter-galactic space. Their long cooling times suggest that the loops may persist to form the hot X-ray halo of an emerging elliptical galaxy [270].

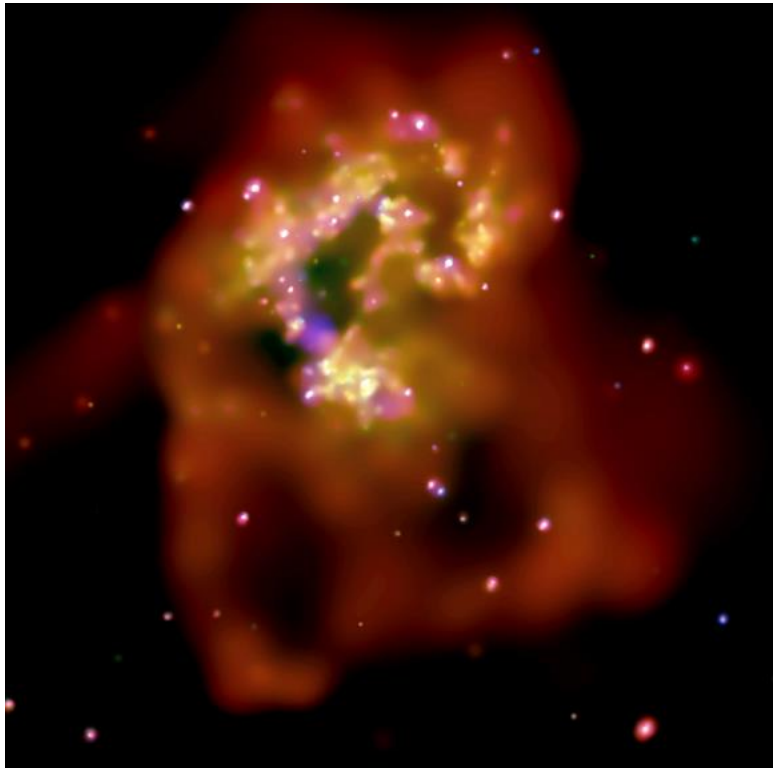

Figure 6-1. *Chandra* view (4.8 arcmin or ~27 kpc across) of the Antennae showing spectacular loops of hot gas [270]. Red corresponds to the lowest energy (0.3-0.65 keV) X-rays, green intermediate (0.65-1.5 keV), and blue the highest (1.5-6.0 keV). Credit: NASA/CXC/SAO/G. Fabbiano et al.

## 6.2    AGN Feedback in Galaxies and Clusters

Galaxies are not randomly distributed in space but are usually found in collections of several to dozens of galaxies called groups, or of hundreds to as many as a thousand called clusters, which are the largest gravitationally bound structures in the universe. Groups and clusters contain large amounts of hot, X-ray emitting gas and dark matter. X-ray observations have established that the dominant baryonic component of clusters is hot gas with a temperature ~10-100 MK, and a mass as high as $10^{14}$-$10^{15}$ $M_\odot$. The mass of the hot gas in a cluster is roughly 6 times that of the mass in stars. The dark matter component has a density ~5 times that of the baryonic matter so it is primarily responsible for gravitationally binding groups and clusters. Since the hot gas radiates its energy very slowly, it retains a "fossil-like" record over hundreds of millions of years, tracing "cosmic violence" which has involved enormous injections of energy powered by outbursts from central SMBH and high velocity collisions between galaxies and sub-clusters.

Starting in the early 1970's with cluster observations from UHURU – the first satellite dedicated to extra-solar X-ray studies – astronomers began to model the properties of the hot, X-ray emitting gas. It was argued that the energy radiated as X-rays would lead to cooling of the central gas allowing pressure from the hotter surrounding regions to drive a mass infall or flow [271, 272, 273]. With the advent of the Einstein X-ray Observatory, observations of many clusters showed centrally peaked (core) X-ray emission, often centered on the brightest member galaxy [274]. The X-ray emissivity $j \propto n^2 P(T)$ erg s$^{-1}$ cm$^{-3}$, where $n$ is the electron density, $T$ is the temperature, and $P(T)$ is the cooling function which takes into account the total radiative loss at all wavelengths. For gases having approximately the cosmic abundances of elements, line emission following collisional excitation of iron dominates $P(T)$ in the temperature range 10-20 MK, with thermal bremsstrahlung dominating above ~20 MK. The thermal energy density $u \propto nT$ erg cm$^{-3}$ so the radiative cooling time $t_{rad} \propto (u/j) \propto T/nP(T)$, where $T/P(T)$ varies slowly with $T$ in the range of interest. With the Einstein Observatory, it was possible to determine the variation of $n$ and $T$ with radius, and to show $t_{rad}$ decreases with radius and becomes $<<10^{10}$ yr in the central regions of a number of clusters [275, 276, 277]. Such cooling times are less than the age of the clusters, so the gas should cool, leading to a flow towards the center. The gas temperature at the centers of some systems was indeed measured to be lower than in the less dense regions, seeming to confirm this picture. However, the ultimate fate of the cooling gas remained a puzzle as little evidence could be found for the disposition of the mass associated with flow rates as high as hundreds of solar masses per year either in the form of newly formed stars or as substantial amounts of very cold, molecular gas [276, 278, 279, 280].

X-ray observations of the Perseus Cluster taken in the 1990's with the *ROSAT* observatory showed evidence for the interaction of the central radio source (an AGN) with the cluster gas [281]. These data were compatible with suggestions that radiative cooling of the gas in the cluster center could be offset by energy input from an accreting supermassive black hole (i.e. AGN), but proof of this idea was lacking. Thus, the cooling flow problem remained a puzzle for more than 2 decades until the launch of the *Chandra* and The European Space Agency's X-ray Multi-Mirror Mission (*XMM-Newton*) observatories in 1999. *XMM-Newton* spectra provided strong evidence [282, 283] that cooling flows do not achieve runaway status. At the same time, radio observations and *Chandra* images revealed jets, cavities, bubbles, ripples, shells and filaments, indicating repeated episodes of AGN activity in many galaxies, groups and clusters. The near equality of the average AGN jet power and the radiative cooling in the intracluster gas suggests that heating and cooling are coupled in a feedback loop, which suppresses star formation and stunts the growth of luminous galaxies at the centers of clusters. The detailed *Chandra* images and the balancing of jet power against radiative cooling plus work done in clearing the cavities make for a compelling case.

In simplified terms, the AGN feedback cycle can be summarized as follows:

1. accreting gas falls towards a SMBH and is heated, converting gravitational potential energy to radiation including X-rays;
2. jets are launched from the SMBH (BH spin may play an important role here), re-heating radiating gas to prevent runaway cooling and pushing aside infalling gas;
3. gas supply is diminished, jets turn off, and the SMBH returns to inactive state;
4. accretion resumes and the cycle starts over.

Elucidating the details of this AGN feedback in clusters and elliptical galaxies is one of *Chandra's* most significant accomplishments to date.

### 6.2.1 *Feedback in Early-Type Galaxies and Groups.*

Over the past decade, a number of lines of evidence have strengthened the view that the evolution of a galaxy, and even a cluster of galaxies, can be profoundly influenced by an accreting SMBH at its center, which may or may not manifest itself as a "classical" AGN characterized by optical emission lines and a point source of X-ray emission. The existence of a relationship between the mass of SMBH and their host galaxy's bulges [223, 224, 225] strongly suggests that most SMBH growth is self-regulated.

*Chandra* observations of nine nearby luminous elliptical galaxies find a tight correlation between the simple spherical or Bondi accretion rate and the power emerging in the relativistic jets. The data indicate that the central black hole is fueled by the cooling interstellar gas, and that there is sufficient feedback energy in the jets to stem cooling and star formation [284].

Observations with *Chandra* of NGC5813, the dominant central galaxy in a nearby galaxy group, provide details on the interaction between a SMBH and the surrounding gas. Three pairs of co-linear cavities with bright rims of X-ray emission have been detected at 1 kpc, 8 kpc and 20 kpc from the central $2.8 \times 10^8$ M$_\odot$ SMBH. Measured temperature and density jumps associated with the rims around the inner and intermediate cavities unambiguously identify these features as shocks with Mach numbers ~1.5 and indicate that 3 distinct outbursts at intervals ~10 Myr produced these cavities. Although previous studies have found clusters where there is enough total shock energy to offset radiative cooling (e.g. M87 [285] and Hydra A [286]), the temperature jumps measured by *Chandra* for NGC5813 explicitly show that the fraction of shock energy that goes into heating the gas (5-10%) is sufficient to balance radiative cooling locally at the shock fronts, while the interval between outbursts is short enough for such shocks to offset cooling over much longer timescales [287].

### 6.2.2 *Feedback in Clusters of Galaxies.*

The *Chandra* image of the Perseus cluster, on the left side of Figure 6-2, shows two large (~10's kpc across), bubble-shaped cavities extending north and south of the SMBH at the center of the giant galaxy NGC1275 [288, 289]. The radio data show that these cavities are filled with energetic particles and magnetic fields, which appear to be pushing aside most of the X-ray emitting hot gas as the cavities rise buoyantly through the gas and expand. Special processing of the *Chandra* image unveils a series of quasi-spherical, ripple-like structures, which if interpreted as sound waves, could provide energy sufficient to prevent a runaway cooling flow. The spacing of these ripples and the sound speed in the rarefied cluster gas is used to estimate a typical recurrence time scale of 10 Myr between outbursts of the central AGN [290].

At the relatively nearby distance of ~20 Mpc, the giant elliptical galaxy M87 in the Virgo cluster of galaxies graphically illustrates the interactions between the activity of the central SMBH and X-ray emitting gas (see right side of Figure 6-2). A wealth of complex structure is apparent, including jets, bright knots, cavities, shocks, and filaments. The filamentary gas shows evidence of the influence of magnetic fields. Images in the harder (3.5-7.5 keV) *Chandra* X-ray band reveal nested shock fronts, presumably from a series of outbursts [291].

On a much larger scale, a *Chandra* image of the cluster MS0735.6+7421 ($z$=0.216), shown in the central panel of Figure 6-2, reveals two opposing X-ray cavities, each coincident with radio emission and more than 200 kpc in diameter. Based on their linear dimensions, the estimated age of the features is ~$10^8$ yr, and the energy required to produce them is $U$~ $6 \times 10^{61}$ ergs [292]. If η is the conversion efficiency of rest mass to energy for accreted matter, then the accreted mass $M_{acc}=U/\eta c^2 \sim 3\times10^7 \eta^{-1}$ $M_\odot$. The maximum accretion efficiency is for a maximally rotating BH, in which case $\eta=0.4$, so at least $7.5\times10^7$ $M_\odot$ of material must have been accreted by the BH over the last $10^8$ y. This startling result indicates that, at least in some clusters, the SMBH are still growing at a rapid rate during the current epoch despite the fact that the central galaxy in this cluster shows no apparent signature of harboring a luminous, "active" BH in either the optical or X-ray band. The energetics of the galaxy - at least as presently observed - are dominated by the outflow of highly energized particles in the radio jets and the cavities they clear in the cluster gas.

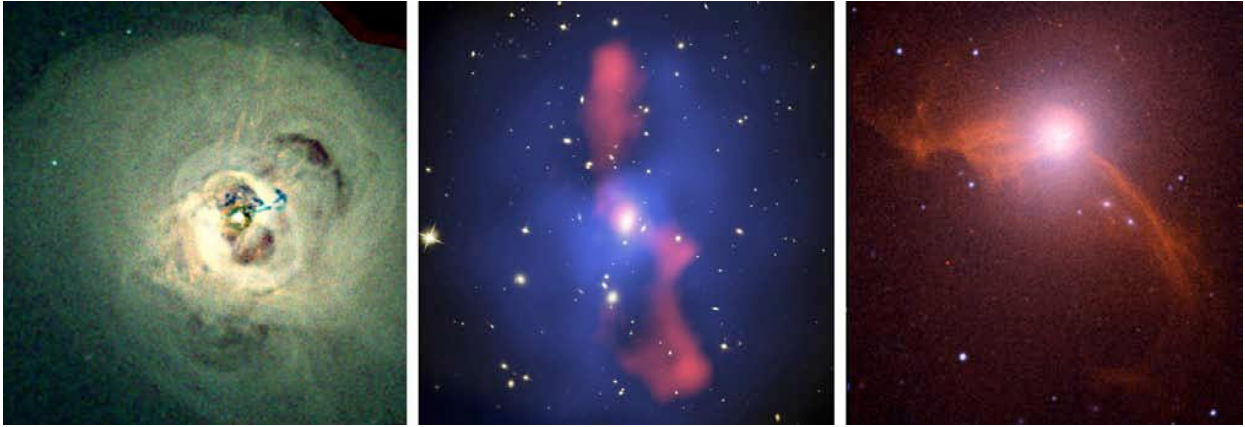

Figure 6-2. . *Chandra* images of three clusters of galaxies which exhibit cavities and jets indicating the influence of SMBH on their surroundings. **Left:** *Chandra* observation (4.7 arcmin or ~105 kpc across) of the Perseus Cluster. X-ray energies represented by red (0.3-1.2 keV), green (1.2-2.0 keV), and blue (2.0-7.0 keV). A 10 arcsec smoothed image, scaled to 80%, has been subtracted from the original to highlight fainter features here [289]. Credit: NASA/CXC/IoA/A. Fabian et al. **Center:** Multi-wavelength observations (3 arcmin or ~630 kpc across) of the cluster MS 0735.6+7421 [292]. *Chandra* X-ray (blue), Hubble optical (yellow/white), and VLA radio (red). Credit: X-ray: NASA/CXC/U Waterloo/B. McNamara et al.; Optical: NASA/ESA/STScI/B. McNamara et al.; Radio: NRAO/VLA/B. McNamara et al. **Right:** *Chandra* X-ray and DSS optical observations (8 arcmin or ~40 kpc across) of M87 in Virgo Cluster [291]. *Chandra* X-ray (red) and DSS optical (white/blue). Credit: X-ray: NASA/CXC/SAO/W. Forman et al.; Optical: DSS.

*Chandra* observations of the Hydra A cluster provide a quantitative description of AGN feedback. The mechanical energy expended by the radio jet to inflate the observed cavities is estimated as the product of the volume of the X-ray cavities times the surrounding gas pressure. The total enthalpy is the sum of this mechanical energy plus the internal energy required to support the cavities, and is 2.5-4 times greater than the work done to inflate the cavities. Comparison of the enthalpy with the energy required to explain the observed synchrotron radiation from the radio jets and lobes in Hydra A (and a number of other sources) shows that relatively modest radio jets can be indicators for very large amounts of mechanical energy liberated by an AGN outburst from the central SMBH [293].

A statistical study using archival *Chandra* images of X-ray cavities in 76 moderately distant clusters of galaxies ($0.3<z<0.7$) provides insight into the behavior of AGN feedback at earlier epochs. No evidence is found for evolution in the properties of the cavities to $z$ of at least 0.6. The cavities of powerful (~$10^{45}$ erg s$^{-1}$) outbursts are not larger (nor smaller) at higher redshift than in the nearby universe, and the energetics of the outbursts are the same. This suggests that feedback in these clusters started as early as 7-8 Gyr after the Big Bang and has operated at roughly the same maximum level of power since then. Since significant SMBH growth would be accompanied by strong AGN feedback and evolution in cavity properties, the lack of such evolution implies that the SMBH masses must be large, ~$10^8$ M$_\odot$ by $z$~0.6 [294]. This same study also provides evidence for evolution of the radiative properties of the central AGN, many of which have a bright X-ray point-like core of non-thermal emission that is rarely seen in clusters at lower redshifts ($z$<0.3). This could indicate a transition between quasar mode (rapid SMBH growth with efficient radiation) and radio mode (sub-Eddington accretion, low radiation efficiency, and powerful radio jets) feedback.

The cluster SPT-CLJ2344-4243 shows that in some cases AGN feedback is insufficient to limit star-formation. Nicknamed the "Phoenix Cluster" this rather distant cluster ($z$=0.6) is among the most massive ($M_{200}$~$2.5\times10^{15}$ M$_\odot$, where $M_{200}$ is the mass enclosed within $R_{200}$ and $R_{200}$ is the radius at which the density is 200 times the average mass density of the universe) and X-ray luminous ($L_X$~$8.2\times10^{45}$ erg s$^{-1}$) known. The central galaxy contains a powerful AGN, and Hubble Space Telescope data are used to demonstrate the presence of a massive population of young stars with an amazingly high star formation rate of ~800 M$_\odot$ y$^{-1}$. In this case, the very high cooling rate and the resultant large reservoir of cooling gas appears to overwhelm the energy supplied by the central radio source and feedback loses out to star formation [295, 296].

6.2.3    *Enrichment of the Intracluster Medium.*

The ultimate source of the metal enrichment of the intracluster gas is generally thought to be both Type Ia and core-collapse supernovae. Prevalent models predict Type Ia products such as Fe will be more centrally concentrated than silicon-group elements such as Si and S, which are produced by both supernova types. A deep *Chandra* observation of the core of the Virgo Cluster extending to 40 kpc finds that the abundance profiles of Si and S are more centrally peaked than Fe, posing a challenge to the standard picture of enrichment dominated by Type Ia supernovae [297]. Future deep *Chandra* observations of relatively nearby clusters such as Fornax will provide additional abundance profiles to test and refine models for metal enrichment.

AGN feedback also plays a role in the removal of metal-enriched gas from galaxies and the redistribution of these metals in the intracluster gas. A *Chandra* observation of the galaxy cluster Hydra A, illustrated in Figure 6-3, shows that the metal content of the intracluster gas is enhanced by up to a factor of 1.6 along the radio jets and lobes. These enhancements are detected over the region extending from ~20-120 kpc from the central AGN. An estimated 10-20% of the iron produced in the central galaxy is transported out of the galaxy, establishing AGN jets as an important mechanism for dispersing metals into the intracluster gas [298]. Similar *Chandra* data on the spatial distribution of metal-rich gas in 9 more galaxy clusters provide further evidence that metals have been carried beyond the extent of the inner cavities in all of these clusters, suggesting that this is a common and long-lasting effect sustained over multiple outburst cycles [299].

Ram-pressure stripping is another process that can enrich the intracluster medium with metals. In massive clusters with a dense intracluster medium, the ram pressure induced by the motion of a galaxy through the core can strip the enriched gas from the galaxy; for example, the ram-pressure-stripped tail mapped by *Chandra* in the Virgo cluster galaxy M86 [300].

6.3    *Growth and Evolution of Groups and Clusters of Galaxies*

The Lambda-Cold Dark Matter (ΛCDM) cosmological model predicts that massive galaxy clusters were built up hierarchically as smaller groups and clusters collided and merged on a timescale governed by the details of the cosmological model. The kinetic energy of colliding sub-clusters can be enormous, ~$10^{65}$ erg. Following a merger, much of this energy is dissipated by shocks and turbulence that further heat the cluster gas. Some of the energy may also go into accelerating ultra-relativistic particles and amplifying magnetic fields in the intracluster medium [301].

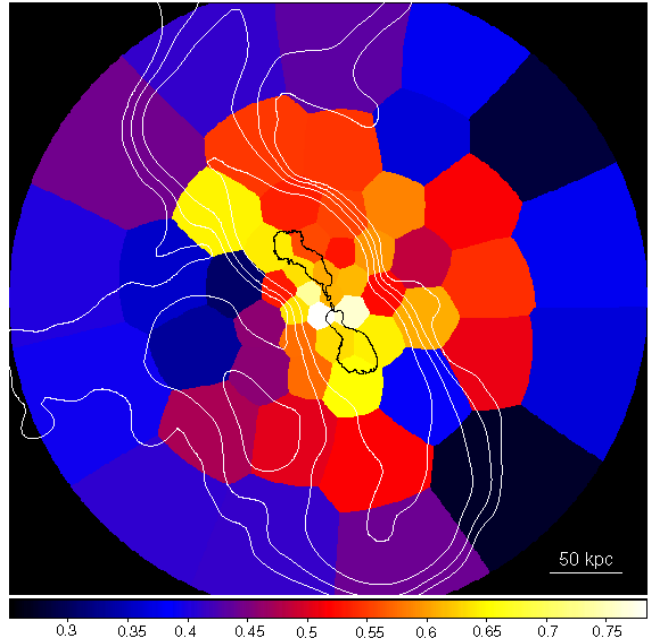

Figure 6-3. *Chandra* metallicity map of the central 5×5 arcmin (or ~320×320 kpc) of the Hydra A cluster [298]. Yellow and white regions have the highest metallicity at ~0.65-0.75 solar and blue the lowest at ~0.35. X-ray map produced via Voronoi tessellation to fit *Chandra* spectra with comparable statistical precision in each bin. Radio contours at two different frequencies are shown in black (1400 MHz) and white (330 MHz). Credit: [298].

*Chandra's* ability to map the cluster gas and its temperature in fine detail over much of a cluster's expanse has led to a better understanding of "cluster weather". Prior to *Chandra*, the few sharp edges and discontinuities seen in X-ray images of clusters were interpreted as shock fronts driven by mergers and collisions. However, *Chandra* now demonstrates that most of these features do not correspond to actual shocks. In the case of the cluster A2142, temperature measurements across two surface brightness edges show that the higher surface brightness regions are both denser and cooler than the lower surface brightness regions [302]. The absence of a substantial gas pressure change across these edges indicates that they are contact discontinuities between hot, diffuse gas and colder, denser gas rather than shocks. Such features are called "cold fronts". In many cases, cold fronts are explained by treating the colder gas cloud as a remnant cooling core from a merging subcluster. However, for A2142, deeper *Chandra* observations suggest that sloshing of cooler gas from the cluster core (see more below regarding sloshing) brings it into contact with hotter, less dense cluster gas creating the contact discontinuity or cold front [301].

The detections of cold fronts provide opportunities to probe physical processes in clusters not accessible by other means. For example, pressure continuity across a cold front requires the gas pressure on the one side to balance with the gas pressure plus ram pressure on the other. *Chandra* observations measure the gas density and temperature on both sides of the front, with the difference then attributed to ram pressure, thereby determining the velocity of the moving gas in the plane of the sky.

The efficiency of thermal conductivity and the orientation of the magnetic field across the cold fronts can also be investigated. Over 50 yr ago, Spitzer [303] calculated the thermal conductivity for an unmagnetized plasma based on the mean free time between Coulomb collisions for the plasma electrons. This "classical" Spitzer conductivity scales as $T^{5/2}$ and could play an important role in gas cooling and equilibration timescales in clusters. However, in A2142, the temperature jump across the cold front - covering a relatively small distance of only a few kpc - requires suppression of thermal conductivity by as much as a factor of 100 compared to the classical Spitzer value. For the cluster A3667, the X-ray data show that the gas density discontinuity occurs over a region small compared to the electron mean free path, requiring suppression of Coulomb diffusion by at least a factor of 3. This suppression of diffusion and collisional thermal conductivity is likely due to magnetic effects with the field predominantly aligned with the cold front rather than crossing [301].

Mergers with off-center impacts can cause the gas in the cluster to slosh or oscillate about the cluster potential minimum. With each oscillation the gas core moves against its own trailing gas, heating the gas and producing an edge in the X-ray surface brightness that expands out from the cluster core. The resultant structures evolve and in some cases remain detectable for at least $10^8$ yr. The fronts may form a spiral structure when the sloshing direction is near the plane of the sky and the merger has a nonzero angular momentum which prevents the displaced gas from simply falling back radially to the cluster core [304]. The sloshing gas carries a substantial amount of kinetic energy, so it can provide another mechanism in addition to AGN feedback for reheating the gas in the core of a cluster. Sloshing may also heat the gas by transporting hotter gas from outside the core back to the cooler central region [301].

Structures that are likely the result of sloshing have been observed in a number of clusters. The *Chandra* image of Abell 2052, shown in Figure 6-4, provides an especially good example of this phenomenon. The abundance of metals is enhanced in the vicinity of a large spiral structure in A2052, evidence that sloshing also can play a role in redistributing metal-enriched gas over large distances in the cluster [305].

Although some of the sharply defined features associated with cluster mergers are cold fronts, the more energetic mergers do produce shock waves, which can also be studied with *Chandra*. The colliding objects (clusters, sub-clusters, or groups) are moving in the same general gravitational potential, so the velocity of a colliding or infalling secondary object is not that different from the sound speed in the primary cluster gas. Consequently, the Mach numbers $M_s$ of the shocks are relatively small, with $M_s \leq 3$. The "Bullet Cluster" (also discussed in section 7.3) exhibits both a sharp density edge and temperature jump with a large pressure change. This feature can be unambiguously identified as a shock front with $M_s = 3.0 \pm 0.4$. The merger is primarily in the plane of the sky, and the collision occurred recently enough that the shock has not yet moved to the lower density outskirts of the cluster. From the measured Mach number and the speed of sound in the pre-shocked gas, the relative velocities of the merging components are determined. Given the distance to the cluster and the angular separation of the merging components, it is estimated that the collision occurred about 150 Myr ago [301].

Given the importance of clusters for cosmological studies, a better understanding of the outer limits of clusters beyond the virial radius (within which virial equilibrium holds with the average kinetic energy equal to – ½ the average potential energy, and often approximated by $R_{200}$), is needed to accurately determine their mass profiles. In this region, the gas is not in hydrostatic equilibrium, and simulations indicate that the gas distribution could be clumpy [306]. Data on the Perseus cluster from *Chandra* and the Japanese-led *Suzaku* mission together measure the temperature and metallicity of the intracluster gas with high precision out to the virial radius [307, 308]. These data show that at the virial radius, the temperature has dropped to approximately one-third of the peak temperature, and that the metallicity is about one-third solar. Beyond about $0.5R_{200}$, *Suzaku* data show that the gas mass fraction deduced from a smoothly distributed gas rises above the average cosmic mean value (cf. Sections 7.1 and 7.4).

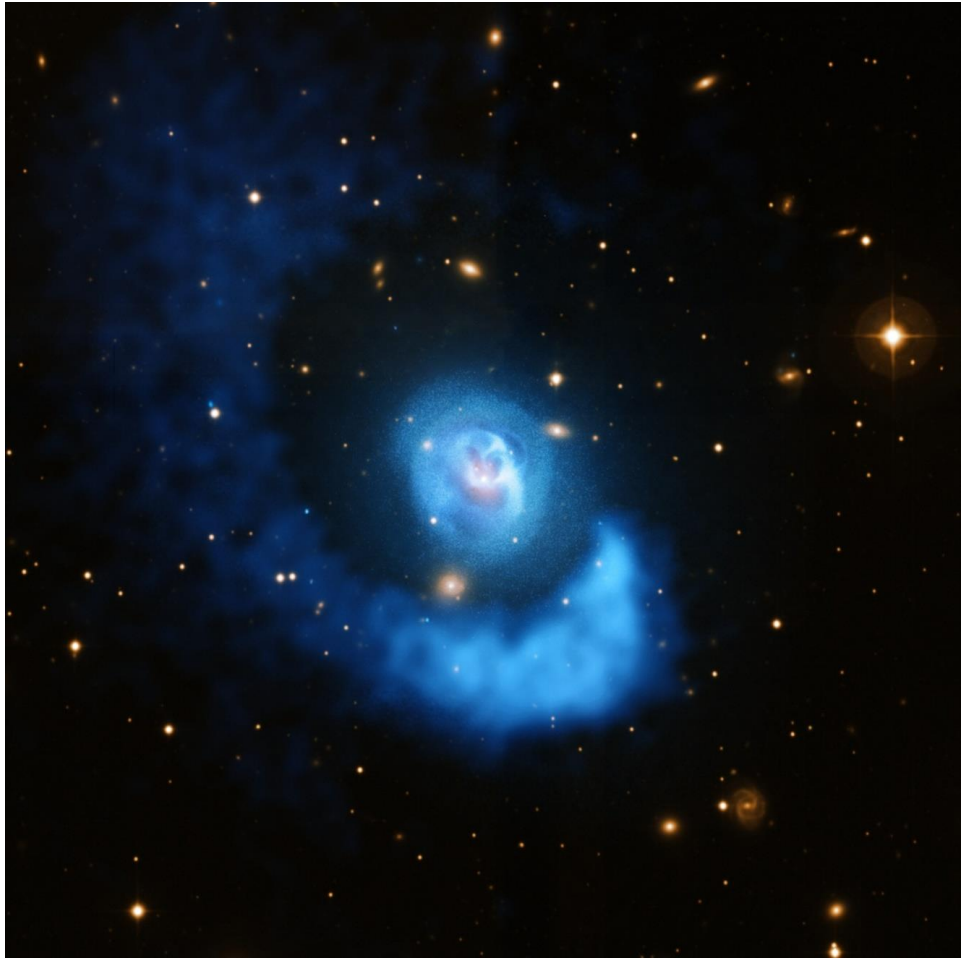

Figure 6-4. *Chandra* and VLT optical image of the central region (9.3 arcmin or ~400 kpc across) of cluster Abell 2052, showing complicated structures including bubbles and shocks in central region driven by AGN feedback, plus spiral-like outer structure ~300 kpc across resulting from sloshing of gas driven by off-center collision [305]. *Chandra* X-rays (blue) and VLT optical (orange). Credit: X-ray-NASA/CXC/BU/E. Blanton et al.; Optical-ESA/VLT.

To better understand these data, deeper *Chandra* observations of the Perseus and Abell 133 (A133) clusters were undertaken to directly image any clumps in the outer regions. For A133, the *Chandra* resolution is used to eliminate obvious foreground stars and background AGN enabling *Chandra* to reach a fainter surface brightness than *Suzaku*. A number of extended or clumped sources are detected, at least some of which are likely associated with galaxies and/or sub-clusters in the outskirts of A133. Some larger-scale filamentary structures are then detectable and a possible edge to the cluster emission is seen for the first time. A key interpretation from this rich data set is that fitting the observed X-ray emission in the outer regions for this cluster does not require a baryon/dark matter ratio higher than the cosmic mean [309].

# 7 Large-Scale Structure, Dark Matter, Dark Energy & Cosmology

*Chandra* observations are important for addressing a broad range of open questions in cosmology, ranging from the problem of missing baryons in galaxies to exploring the nature of dark matter and dark energy.

## 7.1 Missing Baryons and Intergalactic Medium

Cosmological hydrodynamic simulations [310] predict the gradual formation of a local ($z <1$) filamentary web of gas in a low density ($n_b = 10^{-6}-10^{-5}$ cm$^{-3}$), warm-hot ($T = 0.1$-$10$ MK) intergalactic medium (WHIM) connecting galaxies, galaxy groups and clusters of galaxies and permeating the large–scale structures to which these systems belong. The WHIM is predicted to account for a sizable fraction (~50%) of all the baryons in the local ($z <1$) universe [310]. It is thus considered the best candidate to host the baryons detected in the intergalactic gas at high redshift [311] and missing from the low redshift census. Due to the high-predicted temperature of the bulk of the WHIM, only X-ray observations can provide strong constraints on temperatures, ionization states and, statistics permitting, metallicities.

Deep *Chandra* grating spectrometer observations, combined with earlier *XMM-Newton* and *Chandra* observations, give a 4σ detection of the OVII Kα absorption line in the spectrum of a blazar at redshift $z = 0.165$ behind the Sculptor Wall, a large-scale superstructure of galaxies at $z$ ~0.03 [312]. Because the redshift of the Sculptor Wall is known, absorption signatures can be regarded as significant even if they are less prominent than those found in a random search. For a metallicity $Z = 10$% of the solar value, the implied baryon over-density $\delta$ is ~30 ($\delta \equiv \rho/<\rho> - 1$, where $\rho$ is the density of the WHIM at a given location, and $<\rho>$ is the mean density of the universe) [312], consistent with cosmological simulations. This work can be extended by observing other bright blazars behind large-scale structures to search for more absorbers along the lines of sight.

Observations of nearby galaxies at radio, IR and optical wavelengths indicate a deficit of baryons relative to the cosmological average [313]. Hot gas in the extended haloes of galaxies could account for some of the as yet unseen baryons. An extensive hot (~6 MK) gaseous halo is detected by *Chandra* around the giant spiral galaxy NGC 1961 out to a radius $r$ of ~50 kpc. Extrapolation of β-models (surface brightness $S(r) = S_0[1+(r/r_0)]^{(0.5-3\beta)}$) to $r$ ~500 kpc – corresponding to ~$R_{200}$ – leads to an estimate for the halo gas mass in the range 1-6×10$^{11}$ M$_\odot$ [314]. When the stellar and cold gas masses are included, the baryonic fraction $f_b = \Omega_b/\Omega_m$ within $R_{200}$ is ~0.02-0.05, which falls significantly below the cosmological mean of 0.17 [315].

This low value may be indicative of a very early phase of heating before the baryons collapse into a dark matter halo. Such heating may have prevented the baryons from falling deeply into the potential well of the galaxy. *XMM-Newton* observations of the massive spiral galaxy UGC 12591 [316] detect X-rays to ~110 kpc. A β-model fit extrapolated to ~500 kpc determines a similar value for $f_b$ ~0.03-0.04. In contrast, *Chandra* and *Suzaku* studies of the isolated elliptical galaxy NGC 720 detect X-rays out to ~100 kpc [317], and in this instance, model fits extrapolated to radii of ~300−400 kpc yield a larger value of $f_b$~0.10−0.15, closer to the cosmological mean. These results suggest that it is possible for a galaxy with a mass similar to the Milky Way to have a baryonic fraction equal to the cosmic mean, at least if it is isolated.

## 7.2 Large-Scale Structure

As stated in section 6.3, the ΛCDM scenario predicts that massive galaxy clusters were built up hierarchically as smaller groups and clusters collided and merged on a timescale governed by the details of the cosmological model. Computer simulations show that the most massive galaxy clusters should grow in regions where large-scale filaments of intergalactic gas, galaxies, and dark matter intersect and material falls inward along the filaments. A spectacular example of this process is the system MACS J0717.6+3745 (see Figure 7-1) where four clusters are colliding. The collision history can be traced by means of optical measurements of the speed and directions of motion of the clusters and *Chandra* observations of the gas temperatures and offsets between the galaxies and the hot gas. The data indicate that at least two of the clusters are associated with a large filamentary structure [318]. The X-ray observations show an extended (~1Mpc) hot region at the cluster-filament interface with a jump in gas density and temperature along the interface, which could be evidence for accretion of gas into the cluster along the filament. Deeper observations would allow this hypothesis to be tested and may yield the first credible characterization of the gas content of large-scale filaments.

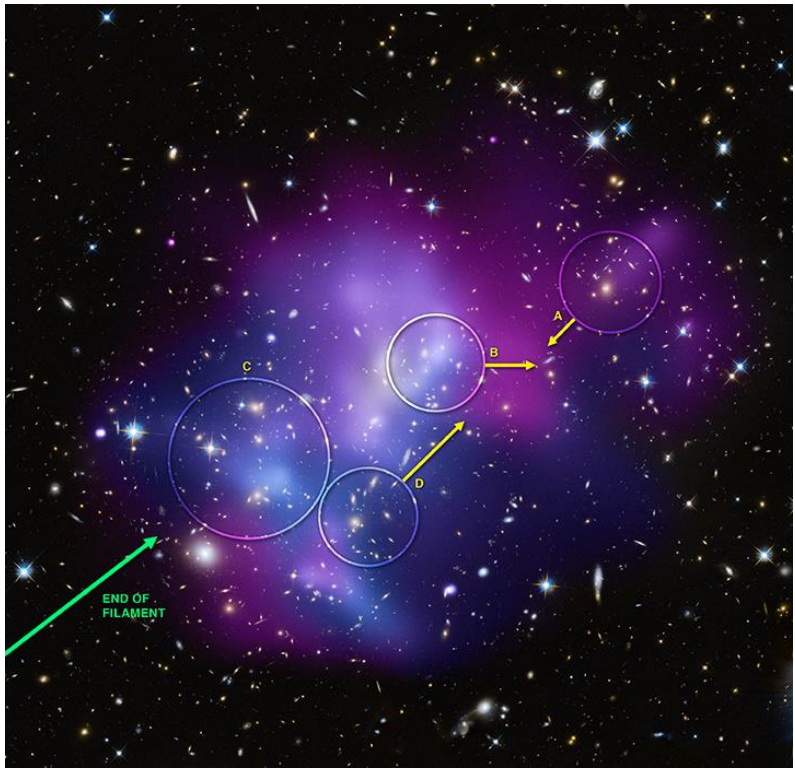

Figure 7-1. Composite *Chandra* and optical HST image of MACS J0717+3745, a merger of four separate galaxy clusters. The clusters are circled with velocities for A, B, and D indicated by the direction and size of the arrows. No velocity is shown for C because there is no offset seen between the X-ray peak and the brightest galaxy in the cluster. This offset is used to derive the projected direction of motion of each cluster. X-ray emission is color-coded with red showing the coolest, and blue showing the hottest, gas. The position of the cluster-filament interface is indicated [318]. The image is 4.5 arcmin across, corresponding to 1.7 Mpc. Credit: X-ray: NASA/CXC/IfA/C. Ma et al.; Optical: NASA/STScI/IfA/C. Ma et al.

*7.3   Dark Matter*

As noted in Section 6.2, X-ray observations establish that the dominant baryonic component of clusters of galaxies is hot gas with a temperature ~10-100 MK, and that the hot baryons are confined by a dark matter component with a density ~5 times that of the baryonic matter. In addition to accurately measuring the total matter content of clusters [319], *Chandra* images, in concert with optical data, are instrumental in confirming the existence of dark matter and constraining its nature.

A deep *Chandra* observation of 1E 0657-56 (the "Bullet Cluster") – a merging system with a clear-cut geometry in which projection effects are small – was combined with deep optical data from HST and ground-based telescopes to investigate the distribution of baryonic and dark matter. The observed separation between the sub-clusters of dark matter, whose positions were derived using gravitational lensing, and the sub-clusters of hot gas – the dominant baryonic-matter component in the cluster – provides one of the most compelling pieces of evidence yet for the existence of dark matter (see Figure 7-22). This finding removes the main motivation for alternative gravity hypotheses designed to avoid the need for dark matter [320].

The stars in the galaxies are collisionless, so the lack of detection of an offset between the peaks in total mass of the sub-clusters and the peaks in the light from the stars in the galaxies, together with an estimate of merger velocities from *Chandra* data, places a significant upper limit on the cross-section for self-interaction of the dark matter particles. The dark matter is consistent with being collisionless [321, 322], with the self-interaction cross section per unit mass $\sigma/m < 1.25$ cm$^2$ g$^{-1}$. The derived limit rules out most of the range for the self-interaction cross section invoked to explain some of the difficulties with the cold dark matter model, such as the lack of sharply-peaked density profiles in the centers of galaxies.

Several other systems with properties similar to the Bullet Cluster have been discovered. Observations of MACS J0025.4-1222, produced by a merger between two galaxy clusters with similar mass, at *z*=0.586, also show a clear separation between the total mass distribution and the baryonic matter, providing further confirmation of the collisionless nature of dark matter [323].

*Chandra* data, together with strong and weak lensing optical data, show that the Abell 2744 cluster is undergoing a complex merger [324]. A2744 shows several instances of separation between dark and baryonic matter, which may lead, with a full numerical simulation, to the determination of similar or improved constraints on the collisional cross-section of dark matter compared to those obtained with the Bullet Cluster.

Studies of the extragalactic background light observed at multiple wavelengths further probe the nature of dark matter. For example, dark matter annihilation could produce electrons and positrons, and CMB photons could up scatter off the electrons and protons to produce a broad spectrum extending through the X-ray band up to gamma-rays. Observations ranging from soft X-rays measured with *Chandra*, to gamma-rays measured with the Fermi Gamma-Ray Space Telescope place significant constraints on the contribution of dark matter annihilation to the extragalactic background light [325]. Interestingly, the results do not support an explanation based only on dark matter annihilation for the positron excess measured by the Payload for Antimatter Exploration and Light-Nuclei Astrophysics satellite [326].

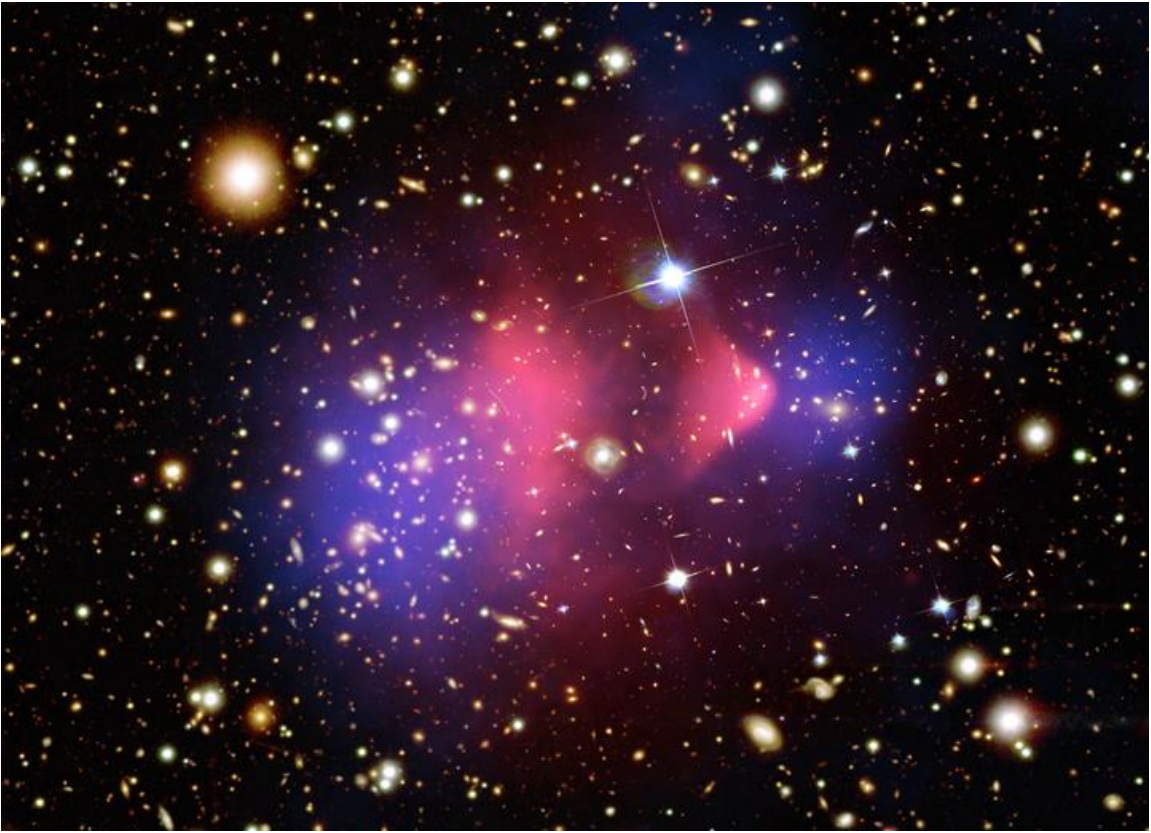

Figure 7-2. Composite image of 1E 0657-56 (the "Bullet Cluster") shows hot gas (pink) detected by *Chandra* and the inferred location of most of the cluster mass (blue) from gravitational lensing [320]. The image is 7.5 arcmin across, corresponding to ~2 Mpc. Credit: X-ray: NASA/CXC/CfA/M. Markevitch et al.; Optical: NASA/STScI; Magellan/U.Arizona/D. Clowe et al.; Lensing Map: NASA/STScI; ESO WFI; Magellan/U. Arizona/D. Clowe et al.

Archival data on M31 are used to set limits on line emission from sterile neutrinos, a particle that is included in some extensions to the standard model of particle physics and is a possible candidate for dark matter. *Chandra's* high spatial resolution allows the removal of a large number of X-ray point sources from a 12-28 arcmin (2.8-6.4 kpc) annular region of the galaxy [327]. This procedure reduces the uncertainties in the estimate of the dark matter mass in the field of view and improves the signal-to-noise ratio for prospective sterile neutrino decay signatures relative to hot gas and unresolved stellar emission. The result is the most stringent limit yet on the mass, $m_s$, of the sterile neutrino. In the context of the Dodelson-Widrow model [328], the observations imply $m_s$ <2.2 keV. This mass limit still falls within the range that can best explain the dark matter-dominated core of the Fornax dwarf spheroidal galaxy [329], so careful searches of these objects for evidence of radiative decay of sterile neutrinos should be pursued in the future.

## 7.4   Dark Energy and Constraining Cosmological Parameters

*Chandra* continues to play a critical role in the endeavor to understand the makeup of the universe, by exploiting independent probes of cosmological models. One powerful method ties the number of massive clusters versus redshift to predictions of growth-of-structure models. The growth of structure, particularly of clusters as the largest bound objects in the universe, is critically dependent on cosmological parameters.

*Chandra's* spatially-resolved imaging and spectral data provide high-quality estimates of the cluster mass. A study of the mass function of clusters as a function of redshift, using a sample of 86 clusters divided into low redshift and more distant subsets (see Figure 7-3), confirms the existence of dark energy and constrains the dark energy equation of state parameter, $w = p/\rho$, where $p$ is the dark energy pressure, and $\rho$ is its density. This constraint is further improved by fitting the cluster data jointly with data from Type Ia supernovae, from the CMB using the Wilkinson Microwave Anisotropy Probe (WMAP), and from baryonic acoustic oscillations (BAO; oscillations in the density of baryonic matter, caused by acoustic waves in the early universe and showing up today as large-scale fluctuations in the galaxy density). The result is $w = -0.991 \pm 0.045$ (statistical) $\pm 0.039$ (systematic) [330]. This result is consistent with a value of $w = -1$, which corresponds to dark energy being described by a cosmological constant term in General Relativity (GR). This analysis reduces the statistical uncertainties by a factor of 1.5 and the systematic uncertainties by almost a factor of 2 when compared with previous constraints obtained without clusters. The *Chandra* data, when combined with WMAP data, also provide the most accurate measure (uncertainty <2%) yet on $\sigma_8$, the linear amplitude of density perturbations at the length scale of 8 $h^{-1}$ Mpc at $z=0$, where $h$ is the Hubble constant in units of 100 km s$^{-1}$ Mpc$^{-1}$.

A previous study that used the evolution with $z$ of the cluster mass function to constrain the parameters of the ΛCDM model [330] has been independently confirmed and extended. Analysis of 238 galaxy clusters from the *ROSAT* All-Sky Survey [331], with *Chandra* follow-up observations of 94 clusters, measures the mass function of clusters for redshifts $z<0.5$. *Chandra's* excellent spatial resolution, spectral measurements, and good sensitivity are important for making accurate mass estimates, especially at redshifts $z>0.2$. Combining the X-ray data with CMB, Type Ia supernovae, gas mass fraction $f_{gas}$ (see below), and BAO data gives $\Omega_m=0.27\pm0.02$, $\sigma_8 = 0.79\pm0.03$ and $w = -0.96\pm0.06$ for spatially flat models (where $\Omega_m+\Omega_\Lambda=1$) with a constant $w$ (see Figure 7-4).

The best available constraints are also derived on time dependent $w$ models with $w = w_0 + w_a(1-a)$, where $a = 1/(1+z)$, $w_0=-0.88\pm0.21$, and $w_0+w_a =-1.05\pm0.23$ [331].

An important constraint from X-ray observations of clusters is the limit that can be placed on the mass of the neutrino because of the prominent role these particles play in the evolution of the early universe. If neutrinos have non-zero mass they can suppress the formation of cosmic structure on small and intermediate scales. Therefore, a comparison of the CMB, reflecting large-scale structure at early times, with the growth of structure on small and intermediate scales in the local universe, provides a way to constrain the mass of the neutrino. Joint analysis of the four data sets mentioned above [330] gives a conservative upper limit on the mass of light neutrinos of 0.33 eV. Identical constraints were derived by an independent team [331]. Although these constraints are formally weaker than those derived from analysis of Sloan Digital Sky Survey data [332], they represent a completely independent technique.

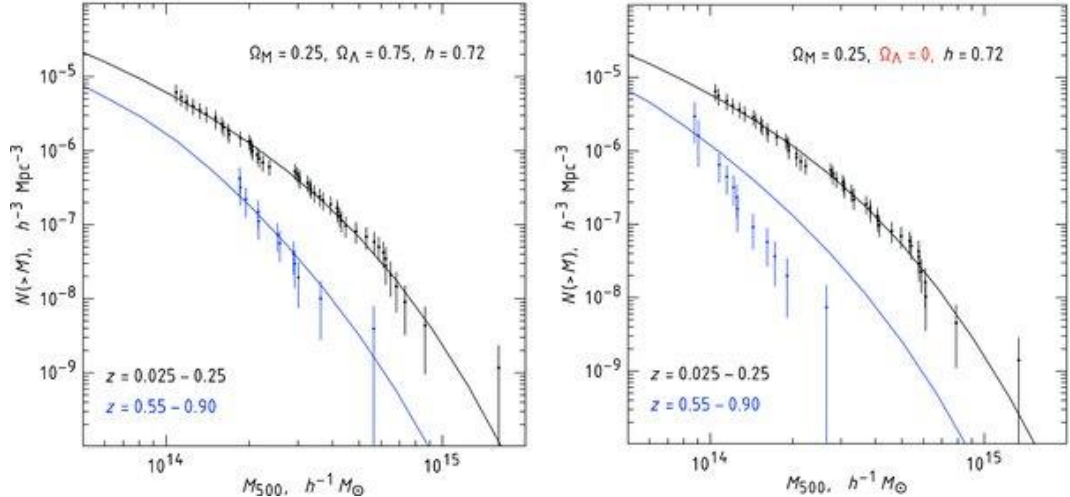

Figure 7-3. Left: Measured cluster mass functions in two redshift ranges showing excellent fits to models with $\Omega_\Lambda = 0.75$. Right: Data and models computed with $\Omega_\Lambda = 0$. The measured mass function changes because it is derived for a different distance-redshift relation. For a model normalized to the low $z$ function, the predicted number density for $z>0.55$ clusters shows strong disagreement with the data [330]. Credit: [330].

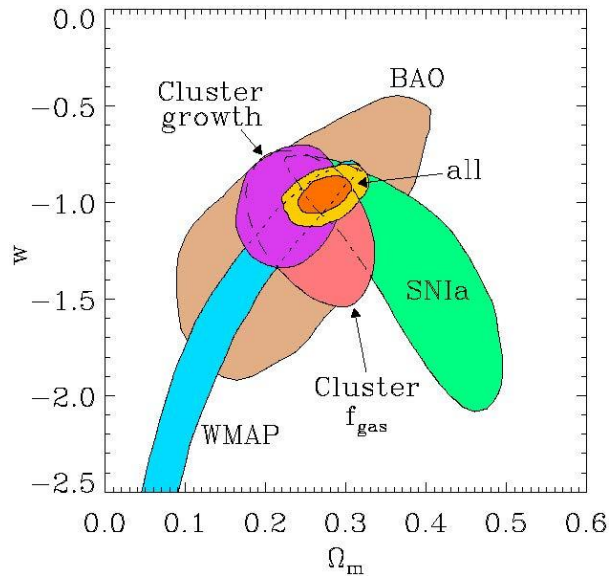

Figure 7-4. Constraints on $w$ and $\Omega_m$ for the cluster growth results [331]. Joint 68.3% confidence regions are shown for this work and for $f_{gas}$ data [319], 5-yr Wilkinson Microwave Anisotropy Probe (WMAP) data [315], SN Ia data [333] and BAO data [334]. Both 68.3% and 95.4% confidence regions, when combining the five data sets, are shown in gold. Credit: adapted from [331].

Another approach to study dark energy, called the $f_{gas}$ method, relies on the measurement of the X-ray gas mass fraction in clusters of galaxies. It proceeds from the assumption, supported by hydrodynamic simulations for the largest dynamically relaxed clusters, that $f_{gas}$ = [mass in the X-ray gas]/ [total mass, including dark matter] is nearly constant with redshift. The measured $f_{gas}$ depends on the assumed distance to a cluster, which, in turn, depends on the cosmological model. This method was used on a sample of 42 hydrodynamically relaxed clusters observed by *Chandra* to detect the effects of dark energy on the expansion of the universe at ~99.99% confidence level [319]. The data also show that the evolution of the dark energy density over redshifts <1 is consistent with that expected for models in which the dark energy is a cosmological constant.

The Sunyaev-Zeldovich (SZ) effect [335] is a distortion in the CMB spectrum caused by inverse-Compton scattering of CMB photons off electrons in the hot intracluster medium. The total SZ effect is independent of cluster redshift and is tightly correlated with the cluster mass as determined by X-ray observations. The accuracy with which the observed mass proxies (gas temperature $T_X$ and Comptonization parameter $Y = M_X T_X$, where $M_X$ is the gas mass) can be linked to the true cluster mass places a fundamental limit on the precision of cosmological constraints from a given survey.

The importance of *Chandra* for calibrating SZ surveys is illustrated by a recent, detailed study of a sample of 15 clusters observed by the South Pole Telescope (SPT) in the redshift range $0.29 < z < 1.08$ [336, 337]. *Chandra* data for 13 of these clusters confirm that the scaling relations of the SZ effect-selected clusters are consistent with the properties of other X-ray-selected samples of massive clusters. Moreover, by constraining the SZ mass-observable relation, these *Chandra* measurements improve the cosmological constraints for the initial sub-sample of SPT massive clusters. This early work establishes a proof-of-concept and lays out the methodology for larger studies of clusters using *Chandra* and SPT. One important study is a mass-limited sample of 80 galaxy clusters selected from a 2000 deg$^2$ survey by the SPT. The redshifts of the clusters range between 0.4 and 1.2, encompassing the epoch ($z$~0.5) where the effects of dark energy begin to be important. The *Chandra* observations were completed in February 2013, and the analysis for an initial set of papers was nearing completion as this review article was being written.

In 2015, the Russian Spectrum-Roentgen-Gamma satellite is scheduled to carry aloft the German-led extended ROentgen Survey with an Imaging Telescope Array (eROSITA) instrument. The driver for eROSITA is an all-sky survey in the 0.5–10 keV energy band to detect 50-100 thousand clusters of galaxies out to a redshift of ~1.3 to study large-scale structure and test cosmological models, especially those describing dark energy. The dark energy tests depend on the determination of the cluster mass function over cosmic time. The angular resolution and sensitivity of *Chandra* (along with weak lensing measurements in the optical band) will be critically important for calibrating the scaling of X-ray proxies for cluster mass based on gas density and temperature profiles to the more distant clusters detected by eROSITA. With the X-ray proxies, shorter *Chandra* exposures for hundreds of luminous, high redshift clusters will determine cluster masses and provide a powerful extension to measures of the growth of structure with cosmic time, enabling astronomers to distinguish among different cosmological models.

### 7.4.1 *Testing General Relativity (GR)*

The growth-of-structure method depends on the gravitational effect on density perturbations, so it opens up the prospect of using cosmological observations to explore the possibility that cosmic acceleration arises not from dark energy but from a modification of GR. Observations of the CMB involve the linear regime of growth of structure, so they provide only weak constraints on departures from GR. In contrast, in the nonlinear regime that applies to the formation of galaxy clusters, the effects can be significant and measurable.

Modified gravitational force models that seek to explain cosmic acceleration without dark energy typically predict cluster mass distributions that differ substantially from those predicted by cold dark matter models. An example is the modified gravity model known as *f(R)*, where *R* is the curvature [338]. This model introduces an extra force designed to mimic the cosmological constant and reproduce the background expansion in the linear regime. Combining *Chandra* cluster results with geometric constraints from the CMB, Type Ia supernovae, the Hubble parameter, and BAO shows that for the *f(R)* model no deviations from GR were found on scales ranging from 40 Mpc up to the size of the observable universe [339].

The growth-of-structure method also promises to be a good probe of other modified gravity scenarios, such as models motivated by higher-dimensional theories and string theory, as the systematic uncertainties in the cluster mass function are reduced by the ongoing multiwavelength cluster surveys. Improved data and simulations on the abundance of massive clusters could push this limit on *f(R)* theories even lower, rivaling solar system tests of gravity, but in a very different low-curvature regime [339].

Galaxy cluster data from *ROSAT* and *Chandra* have been combined with CMB data from WMAP and galaxy clustering data from the WiggleZ Dark Energy Survey, the 6-degree Field Galaxy Survey and the Sloan Digital Sky Survey III to test the growth of structure predicted by GR and the cosmic expansion history predicted by ΛCDM [340]. The growth rate of density perturbations on large scales, $g(a)$, can be modeled as a power law so that $g(a) = \Omega_m (a)^\gamma$. For their most general model allowing γ to vary, they find that $\gamma = 0.604 \pm 0.078$, showing excellent agreement with GR ($\gamma = 0.55$).

### 7.4.2 *Tests of the ΛCDM Cosmological Model*

The standard ΛCDM cosmological model posits a bottom-up sequence of structure formation largely dominated by the gravity of cold dark matter. A hierarchical series of mergers of galaxies, groups and small clusters culminates in the formation of clusters of galaxies, the most massive gravitationally bound systems in the universe. Large fluctuations, and correspondingly large cluster masses, should be extremely rare in the early universe. The space density of massive, high-redshift clusters is a particularly sensitive probe of large perturbations and the Gaussian nature of perturbations in the primordial matter density field, providing a strong test of ΛCDM as well as alternative models. To carry out such tests, *Chandra* observations are being used to determine masses for clusters detected by a new generation of dedicated millimeter-wave telescopes, including the SPT, the Atacama Cosmology Telescope, and the Planck space observatory.

Combined *Chandra*-SPT observations were used to discover the most massive known cluster at redshift $z>1$: SPT–CLJ2106-5844, with a redshift $z = 1.13$ [341]. The mass estimate from the SZ effect and X-ray data is $M_{200} = 1.27 \times 10^{15}$ M$_\odot$ (where $M_{200}$ is the mass enclosed within a radius $R_{200}$). Under the assumption of ΛCDM cosmology and only Gaussian perturbations, there is only a 7% chance of finding a cluster this massive at such a high redshift in the SPT survey region. Furthermore, only one such cluster is expected in the entire sky, thus hinting at possible tension with the ΛCDM model. Indeed, some authors are already reporting a possible conflict with the ΛCDM model in that the probability of finding even a few clusters with masses $>10^{15}$ M$_\odot$ at $z>1$ is small [342, 343] but others disagree [344, 345].

Use of the SZ effect led to the discovery of another extreme cluster called ACT-CL J0102-4915, or "El Gordo" (see Figure 7-5), using the Atacama Cosmology Telescope. This may be a higher redshift ($z = 0.87$) analog of the Bullet Cluster. Current measurements suggest that it is the most massive ($M_{200} = 2.16 \pm 0.32 \times 10^{15}$ M$_\odot$), hottest ($T_X = 14.5 \pm 0.1$ keV) and most X-ray luminous ($L_X = 2.19 \pm 0.11 \times 10^{45}$ erg s$^{-1}$) cluster known at $z > 0.6$ and that it is undergoing a major merger [346]. This cluster is consistent with ΛCDM cosmology as long as its mass is in the lower portion of its statistically allowed range.

Another constraint on ΛCDM comes from the detection of the highest redshift galaxy cluster yet discovered with *Chandra*, CXO J1415.2+3610 at $z \sim 1.5$ [347]. The properties of this cluster agree with expectations for ΛCDM using the WMAP cosmology [348], but place it among the sample of massive, distant clusters that may be used to test the standard ΛCDM cosmology in the future.

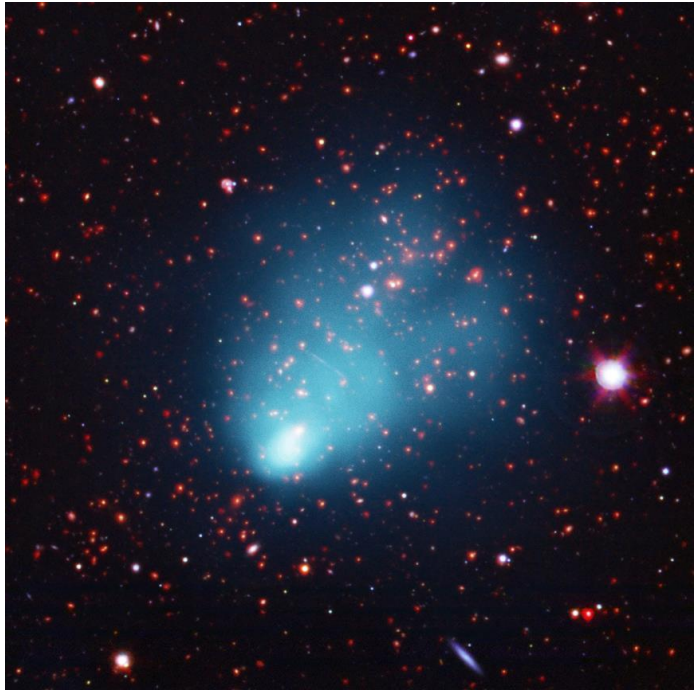

Figure 7-5. A composite image of ACT-CL J0102-4915, a.k.a. El Gordo, likely the most massive, hottest and most X-ray luminous cluster known at *z* >0.6 [346]. *Chandra* data (blue) shows the hot gas in this merging system and optical/IR images (red) – Southern Astrophysical Research telescope (SOAR), Very Large Telescope and Spitzer – show the galaxies. The image is 5.3 arcmin across, corresponding to 2.5 Mpc. Credit: X-ray: NASA/CXC/Rutgers/J.Hughes et al,; Optical: ESO/VLT/Pontificia Universidad. Catolica de Chile/L.Infante & SOAR (MSU/NOAO/UNC/CNPq-Brazil)/Rutgers/F.Menanteau; IR: NASA/JPL/Rutgers/F.Menanteau.

Clusters of galaxies are also an optimal place to test two fundamental predictions of N-body simulations for the $\Lambda$CDM model, namely that clusters are triaxial (i.e. having three axes) and the logarithmic slope $\beta$ of the dark matter density profile ($\rho_{DM} \propto r^{-\beta}$) in the inner part of a cluster approaches a shallow power law with $\beta=1$ [349]. Whereas lensing probes the projected mass contained in cylinders (2-dimensional), the X-ray data are sensitive to the spherically-averaged (3-dimensional) enclosed masses, so the combination provides information on the line of sight geometry that could not be obtained from optical data alone.

*Chandra* and optical lensing and stellar velocity dispersion data show that the dark matter distribution in Abell 383 from ~2 kpc to ~1.5 Mpc is triaxial, but with $\beta<1$ (95% confidence) in the inner region of the cluster [350]. This result suggests a need for a revision of either the dark matter model, or the relevant baryonic physics for shaping of the cluster core. Cosmological hydrodynamical simulations incorporating baryonic physics (cooling and feedback) indicate that these effects steepen the dark matter density profile, increasing the expected value for $\beta$ and worsening the discrepancy with observations. A similar analysis for radii greater than 25 kpc confirmed the triaxial structure of the cluster, but found $\beta = 1.02\pm0.06$, consistent with $\Lambda$CDM [351]. Possible explanations for the difference between the two studies are that they are measuring a different part of the density profile, or that projection effects in the stellar velocity dispersion data in the former study [350] might bias the mass estimates at small radii. Future work on density profiles for a sample of clusters will be needed to resolve this issue and further test the $\Lambda$CDM models.

## 8 *Chandra's* Continuing Science Impact

*Chandra* has had a major impact on many areas of astronomical research. Proposals for *Chandra* observing time are open to scientists world-wide as is access to the *Chandra* data archive which contains all of the data within one year after receipt on the ground and processing. As of January 2014, more than 3600 individual Principal Investigators and Co-Investigators have participated in successful *Chandra* proposals, and ~5700 *Chandra* papers have appeared in refereed journals. This article has summarized only a small fraction of the *Chandra* results, while endeavoring to highlight some of the most surprising and significant discoveries.

When several of us first engaged in building X-ray astronomy payloads for balloon and rocket flights and then some of the early satellites, it was inconceivable to talk about a 25-yr or longer mission. Even when *Chandra* was proposed, sizing the expendables for as long as 10 yr seemed extremely optimistic. Now with 14 yr in the books and a very healthy observatory along with the enthusiasm driven by the science results described in this review, it is not so much of a "pipe-dream" to talk about extending *Chandra* through the 2020's. With its sub-arcsecond angular resolution and unmatched sensitivity for faint sources, *Chandra* is unique among all currently operating X-ray observatories as well as any planned for development and possible launch for at least another decade and likely longer. As emphasized throughout this article, angular resolution on this scale is essential for much of the cutting-edge science being done with *Chandra*. It is possible that a true successor to *Chandra* with comparable (or better) angular resolution, something like 50-100 times the throughput, and a suite of more modern, state-of-the-art instruments could be approved for development in the 2020's with a launch around 2030, at which time it might be appropriate to consider "retirement" for *Chandra*.

## 9 Acknowledgments

We feel privileged to be members of the team responsible for the development and operation of *Chandra* and wish to acknowledge the continued support for this mission by NASA over the past 35+ yr. We gratefully acknowledge the contributions of the thousands throughout the world who have worked so hard to make *Chandra* so successful.

# 10  List of Acronyms in the Text

| | | |
|---|---|---|
| ΛCDM | - | Lambda Cold Dark Matter |
| ACIS | - | Advanced Camera for Imaging Spectroscopy |
| AGN | - | Active Galactic Nucleus (or Nuclei as appropriate) |
| ApJ | - | Astrophysical Journal |
| BH | - | Black Hole (or Holes as appropriate) |
| BU | - | Boston University |
| CCD | - | Charge Coupled Device |
| CDFS | - | *Chandra* Deep Field South |
| ChaMP | - | *Chandra* Multi-wavelength Project |
| CMB | - | Cosmic Microwave Background |
| CXC | - | *Chandra* X-ray Center |
| CT | - | Compton-Thick |
| DSS | - | Digitized Sky Survey |
| ESA | - | European Space Agency |
| ESO WFI | - | European Southern Observatory Wide Field Imager |
| GR | - | General Relativity |
| GSFC | - | Goddard Space Flight Center |
| GTO | - | Guaranteed Time Observer |
| HETG | - | High Energy Transmission Grating |
| HMXB | - | High Mass X-ray Binary (or Binaries as appropriate) |
| HRC | - | High Resolution Camera |
| HRMA | - | High Resolution Mirror Assembly |
| HST | - | Hubble Space Telescope |
| IfA | - | Institute for Astronomy |
| IMBH | - | Intermediate Mass Black Hole (or Holes as appropriate) |
| INT | - | Isaac Newton Telescope |
| IoA | - | Institute of Astronomy |
| IPHAS | - | A Photometric Hα Survey of the Northern Galactic Plane |
| IR | - | InfraRed |
| JPL | - | Jet Propulsion Laboratory |
| LETG | - | Low Energy Transmission Grating |
| LMXB | - | Low Mass X-ray Binary (or Binaries as appropriate) |
| MHD | - | Magneto Hydrodynamics |
| MIT | - | Massachusetts Institute of Technology |
| MSFC | - | Marshall Space Flight Center |
| MSU | - | Michigan State University |
| CNPq Brazil | - | Conselho Nacional de Desenvolvimento Científico e Tecnológico |
| NASA | - | National Aeronautics and Space Administration |
| NOAO | - | National Optical Astronomy Observatories |
| NS | - | Neutron Star (or Stars as appropriate) |
| NuStar | - | Nuclear Spectroscopic Telescope Array |
| PI | - | Principal Investigator |
| PSU | - | Pennsylvania State University |
| ROSAT | - | Rontgensatellit |
| SAO | - | Smithsonian Astrophysical Observatory |
| SDSS | - | Sloan Digital Sky Survey |
| SOAR | - | Southern Astrophysical Research telescope |
| SFR | - | Star Formation Rate |
| SMBH | - | SuperMassive Black Hole (or Holes as appropriate) |
| SNR | - | SuperNova Remnant (or Remnants as appropriate) |
| STScI | - | Space Telescope Science Institute |
| U.Arizona | - | University of Arizona |
| ULX | - | Ultra-Luminous X-ray source |
| UNC | - | University of North Carolina |
| UV | - | UltraViolet |
| VLA | - | Very Large Array |
| VLT | - | Very Large Telescope |
| XMM | - | X-ray Multi-Mirror Mission |

[1] Friedman H, Lichtman S and Byram E 1951 *PhRv* **83** 1025-1030
[2] Giacconi R, Gursky H, Paolini F and Rossi B 1962 *PhRevL* **9** 439-443
[3] Cohen D H, Leutenegger M A, Wollman E E, Zsargó J, Hillier D J, Townsend R H D and Owocki S P 2010 *MNRAS* **405** 2391-2405
[4] Cohen D H, Gagné M, Leutenegger M A, MacArthur J P, Wollman E E, Sundqvist J O, Fullerton A W and Owocki S P 2011 *MNRAS* **415** 3354-3364
[5] Gagné, M, Oksala M E, Cohen D H, Tonnesen S K, ud-Doula A, Owocki S P, Townsend R H D and MacFarlane J J 2005 *ApJ* **628** 986-1005
[6] Mitschang A W et al. 2011 *ApJ* **734** article id 14
[7] Wolk S J, Bourke T L, Smith R K, Spitzbart B and Alves J 2002 *ApJ* **580** L161-L165
[8] Townsley L K, Feigelson E D, Montmerle T, Broos P S, Chu Y -H and Garmire G P 2003 *ApJ* **593** 874-905
[9] Figer D F 2005 *Nature* **434** 192-194
[10] Crowther P A, Schnurr O, Hirschi R, Yusof N, Parker R J, Goodwin S P and Kassim H A 2010 *MNRAS* **408** 731-751
[11] Townsley L K, Broos P S, Feigelson E D, Garmire G P and Getman K V 2006 *AJ* **131** 2164-2184
[12] Townsley L K et al. 2011 *ApJS* **194** 1
[13] Townsley L K et al. 2011 *ApJS* **194** 15
[14] Roccatagliata V, Bouwman J, Henning T, Gennaro M, Feigelson E, Kim J S, Sicilia-Aguilar A, Lawson W A 2011 *ApJ* **733** 113
[15] Guarcello M G, Micela G, Peres G, Prisinzano L, Sciortino S 2010 *A&A* **521** A61
[16] Moriguchi Y, Onishi T, Mizuno A and Fukui Y 2002 in 8th Asian-Pacific Regional Meeting, Volume **II**, ed S Ikeuchi et al. 173–174
[17] Getman, K V, Feigelson E D, Grosso N, McCaughrean M J, Micela G and Broos P, Garmire G. and Townsley L 2005 *ApJS* **160** 353
[18] Sicilia-Aguilar A, Kóspál Á , Setiawan J, Ábrahám P, Dullemond C, Eiroa C, Goto M, Henning T and Juhász A 2012 *A&A* **544** A93
[19] Guarcello M G, Drake J J, Wright N J, García-Álvarez D, Drew J E, Aldcroft T, Fruscione A, Kashyap V L 2013 *RMxAC* **42** 5-7
[20] Kastner J H et al. 2012 *AJ* **144** 58
[21] Kastner J H, Soker N, Vrtilek S D and Dgani R 2000 *ApJ* **545** L57-L59
[22] Yu Y S, Nordon R, Kastner J H, Houck J, Behar E, and Soker N 2009 *ApJ* **690** 440-452
[23] Nordon R, Behar E, Soker N, Kastner J H and Yu Y S 2009 *ApJ* **695** 834-843
[24] Schröter S, Czesla S, Wolter U, Müller H M, Huber K F and Schmitt J H M M 2011 *A&A* **532** A3
[25] Scharf C A 2010 *ApJ* **722** 1547-1555
[26] Dennerl K 2008 *P&SS* **56** 1414-1423
[27] Bhardwaj A, Randall Gladstone G, Elsner R F, Østgaard N, Hunter Waite J, Cravens T E, Chang S -W, Majeed T and Metzger A E 2007 *JASTP* **69** 179-187
[28] Wargelin B J, Markevitch M, Juda M, Kharchenko V, Edgar R and Dalgarno A 2004 ApJ 607 596-610
[29] Dennerl K 2006 *SSRv* **126** 403-433
[30] Branduardi-Raymont G, Elsner R F, Galand M, Grodent D, Cravens T E, Ford P, Gladstone G R and Waite J H 2008 *JGRA* **113** 2202
[31] Elsner R F, Ramsey B D, Waite J H, Rehak P, Johnson R E, Cooper J F and Swartz D A 2005 *Icar* **178** 417-428
[32] Bhardwaj A et al. 2007 *P&SS* **55** 1135-1189

[33] Ness J -U, Schmitt J H M M, Wolk S J, Dennerl K and Burwitz V 2004 *A&A* **418** 337-345
[34] Bhardwaj A, Elsner R F, Waite J H, Jr, Gladstone G R, Cravens T E and Ford P G 2005 *ApJ* **624** L121-L124
[35] Bhardwaj A, Elsner R F, Waite J H Jr., Gladstone G R, Cravens T E and Ford P G 2005 *ApJ* **624** L121-L124
[36] Bodewits D, Christian D J, Torney M, Dryer M, Lisse C M, Dennerl K, Zurbuchen T H, Wolk S J, Tielens A G G M and Hoekstra R 2007 *A&A* **469** 1183-1195
[37] Ewing I, Christian D J, Bodewits D, Dennerl K, Lisse C M and Wolk S J 2013 *ApJ* **763** 66
[38] Gladstone G R et al. 2002 *Nature* **415** 1000–1003
[39] See Koutroumpa D, Lallement R, Kharchenko V and Dalgarno A 2009 *SSRv* **143** 217-230 and references therein
[40] Woosley S, and Janka H 2005 *Nature Physics* **1** 147–154
[41] Janka H 2012 *Ann. Rev. Nucl. Part. Sci*. **62** 407-451
[42] Nomoto K 1982 *ApJ* **253** 798–810
[43] Iben I and Tutukov A 1984 *ApJS* **54** 335-372
[44] Schmidt B 2012 *RMP* **84** 1151-1163
[45] Reiss A et al. 1998 *Astron. J.* **116** 1009-1038
[46] Perlmutter S et al. 1999 *ApJ* **517** 565-586
[47] Vink J 2012 *A&A Rev* **20** 1-78
[48] Lopez L A, Ramirez-Ruiz E, Badenes C, Huppenkothen D, et al, 2009 *ApJ* **706** L106–L109
[49] Lopez L A, Ramirez-Ruiz E, Huppenkothen D, Badenes C, et al. 2011 *ApJ* **732** 114–132
[50] Krause O, Tanaka M, Usuda T, Hattori T, et al. 2008 *Nature* **456** 617–619
[51] Rest A et al. 2008 *ApJ* **680** 1137–1148
[52] Kosenko D, Helder EA and Vink J 2010 *A&A* **519** A11-22
[53] Hachisu I, Kato M and Nomoto K 1999 *ApJ* **522** 487–503
[54] Chiotellis A et al. 2011 *Proceeding of the IAU Symposium* **281** 331-334; see also Chiotellis A, Schure K M and Vink J 2012 *A&A* **537** id A139
[55] Reynolds S P, Borkowski K J, Hwang U, Hughes J P et al. 2007 *ApJ* **668** L135–L138
[56] Sullivan M et al. 2010, *MNRAS* **406**, 782-802
[57] Borkowski K et al. 2013 arXiv:1305.7399
[58] Morris T and Podsiadlowski P 2009 *MNRAS* **399** 515–538
[59] Chita S M, Langer N, van Marle A J, García-Segura G and Heger A 2008 *A&A* **488** L37-L41
[60] Zhekov S, Racusin J, and Burrows D, 2010 *MNRAS* **407** 1157-1169
[61] Dewey D, Dwarkadas V, Haberl F, Sturm R, and Canizares C 2012 *ApJ* **752**, 103-135
[62] Helder E, Broos P, Dewey D, Dwek E, McCray R, and Park S, 2013 *ApJ* **764,**11-18
[63] DeLaney T, Rudnick L, Stage M D, Smith J D, et al. 2010 *ApJ* **725** 2038–2058,
[64] Hwang U et al. (2004) *ApJ* **615** L117–L120,
[65] Schure K, Vink, J, Garcia-Segura G and Achterberg A 2008 *ApJ* **686** 399-407
[66] Lee J-J, Park S, Hughes, J and Slane, P 2013 arXiv:1304.3973
[67] Lee J-J et al. 2010 *ApJ* **711** 861-869
[68] Park S et al. 2007 *ApJ* **670** L121–L124
[69] Burrows A 2013 *RMP* **85** 245-261
[70] Heger A et al. 2003 *ApJ* **591** 288–300
[71] Lopez L, Ramires E, Castro D and Pearson S, 2013 arXiv:1301.0618
[72] Blandford R and Eichler D 1987 *Phys. Rep*. **154** 1-75
[73] Giordano F et al. 2012 *ApJ* **744** L2-L7
[74] Atoyan A and Dermer C 2012 *ApJ* **749** L26-L30
[75 ] Eriksen K et al. 2012 *ApJ* **728** L28-L32

[76] Bykov A et al. 2011 *ApJ* **735** L40-L50
[77] Warren J et al. 2005 *ApJ* **634** 376–389
[78] Pacini F 1967 *Nature* **216** 567-568
[79] Spitkovsky A 2006 ApJ **684** L51-L54
[80] Gaensler B M and Slane P O 2006 *ARAA* **44** 17–47,
[81] Kargaltsev O and Pavlov G G 2008 *AIP Conf. Series* **983** 171–185
[82] Weisskopf M, Hester J, Tennant, A et al. 2000 *ApJL* **536**, L81–L84 & Hester J J, Mori K, Burrows D, Gallagher J S, Graham J R, Halverson M, Kader A, Michel F C and Scowen P 2002 *ApJL* **577**, L49
[83] Komissarov S and Lyubarsky Y 2003 *MNRAS* **344** L93-L96
[84] Del Zanna L, Volpi D, Amato E and Bucciantini N 2006 *A&A* **453** 621-633
[85] Durant, M, Kargaltsev O, Pavlov G G, Kropotina J and Levenfish K 2013 *ApJ* **763** 72
[86] Gaensler B et al. 2004 *ApJ* **616** 383-402
[87] Yusef-Zadeh F and Gaensler B 2005 *AdvSpRes* **35** 1129-1136
[88] Rea, N et al. 2013 *ApJ* **770** 65-78
[89] Ho W and Heinke C 2009 *Nature* **462** 71-72
[90] Elshamouty K et al 2013 *ApJ* **777**, article id. 22
[91] Shternin P, Yakovlev D, Heinke C, Ho W and Patnaude D 2011 *MNRAS* **412** L108-L112
[92] Page D, Parakash M, Lattimer J and Steiner A 2011 *PhRevL* **106** 1101-1114
[93] Heinz S et al. 2013 *ApJ* **779** article id 171
[94] Gou L et al. 2013, *arXiv 1308.4760v1*
[95] Fabian A et al. 2012 *MNRAS* **724** 217-223
[96] Liu J et al. 2008 *ApJL* **679** L37-L40
[97] Miller J et al 2008 *ApJ* **680** 1359-1377
[98] Pakull M, Soria R and Motch C 2010 *Nature* **466** 209-212
[99] McClintock J, Narayan R and Steiner J 2013 arXiv:1303.1583
[100] Feng H and Soria R 2011 *NewAR* **55** 166-183
[101] Feng H and Kaaret P 2009 *ApJ* **696** 1712-1726
[102] Servillat M et al. 2011 *ApJ* **743** 6-18
[103] Poutanen J, Fabrik S, Valeev A, Sholukhova O and Greiner J 2013 *MNRAS* **432** 506-518
[104] Swartz D, Tennant A and Soria R 2009 *ApJ* **703** 159-168
[105] Xue Y Q et al. 2011 *ApJS* **195** 10
[106] Beckwith S V W et al. 2006 *AJ* **132** 1729-1755
[107] Haardt F and Maraschi L 1991 *ApJ* **380** L51-L54
[108] Turner T J and Miller L 2009 *A&ARv* **17** 47-104
[109] Just D W, Brandt W N, Shemmer O, Steffen A T, Schneider D P, Chartas G and Garmire G P 2007 *ApJ* **665** 1004-1022
[110] Tozzi P et al. 2006 *A&A* **451** 457-474
[111] Mathur S, Wilkes B J and Ghosh H 2002 *ApJ* **570** L5-L8
[112] Wilkes B J et al. 2012 *ApJ* **745** 84
[113] Harris D E and Krawczynski H 2006 *ARA&A* **44** 463-506
[114] Hardcastle M J, Evans D A and Croston J H 2009 *MNRAS* **396** 1929-1952
[115] Evans D A, Worrall D M, Hardcastle M J, Kraft R P and Birkinshaw M 2006 *ApJ* **642** 96-112
[116] Belsole E, Worrall D M and Hardcastle M J 2006 *MNRAS* **366** 339-352
[117] Siemiginowska A, LaMassa S, Aldcroft T L, Bechtold J and Elvis M 2008 *ApJ* **684** 811-821
[118] Wilkes et al. 2013 *ApJ* **773** 15
[119] Tengstrand O, Guainazzi M, Siemiginowska A, Fonseca Bonilla N, Labiano A, Worrall D M, Grandi P and Piconcelli E 2009 *A&A* **501** 89-102

[120] Green P J, Aldcroft T L, Mathur S, Wilkes B J and Elvis M 2001 *ApJ* **558** 109-118
[121] Brotherton M S, Laurent-Muehleisen S A, Becker R H, Gregg M D, Telis G, White R L and Shang Z 2005 *AJ* **130** 2006-2011
[122] Netzer H et al. 2003 *ApJ* **599** 933-948
[123] Young A J, Lee J C, Fabian A C, Reynolds C S, Gibson R R and Canizares C R 2005 *ApJ* **631** 733-740
[124] Miller L, Turner T J, Reeves J N, George I M, Kraemer S B, and Wingert B 2007 *A&A* **463** 131-143
[125] Turner T J, Kraemer S B, George I M, Reeves J N and Bottorff M C 2005 *ApJ* **618** 155-166
[126] Risaliti G, Elvis M, Fabbiano G, Baldi A, Zezas A and Salvati M 2007 *ApJ* **659** L111-L114
[127] Holczer T, Behar E and Arav N 2010 *ApJ* **708** 981-994
[128] Lee J C, Kriss G A, Chakravorty S, Rahoui F, Young A J, Brandt W N, Hines D C, Ogle P M and Reynolds C S 2013 *MNRAS* **430** 2650-2679
[129] Chakravorty S, Misra R, Elvis M, Kembhavi A K and Ferland G 2012 *MNRAS* **422** 637-651
[130] Holczer T, Behar E and Kaspi S 2007 *ApJ* **663** 799-807
[131] Chartas G, Brandt W N, Gallagher S C and Proga D 2007 *AJ* **133** 1849-1860
[132] Chartas G, Saez C, Brandt W N, Giustini M and Garmire G P 2009 *ApJ* **706** 644-656
[133] Giustini M and Proga D 2012 *ApJ* **758** 70
[134] Wang J, Fabbiano G, Elvis M, Risaliti G, Karovska M, Zezas A, Mundell C G, Dumas G and Schinnerer E 2011 *ApJ* **742** 23-40
[135] Risaliti G et al. 2013 *Nature* **494** 449-451
[136] Chartas G, Kochanek C S, Dai X, Moore D, Mosquera A M and Blackburne J A 2012 *ApJ* **757** 137
[137] Pooley D, Blackburne J A, Rappaport S and Schechter P L 2007 *ApJ* **661** 19-29
[138] Chartas G, Kochanek C S, Dai X, Poindexter S and Garmire G 2009 *ApJ* **693** 174-185
[139] Gallo E, Treu T, Jacob J, Woo J-H, Marshall P J and Antonucci R 2008 *ApJ* **680** 154-168
[140] Constantin A, Green P, Aldcroft T, Kim D-W, Haggard D, Barkhouse W and Anderson S F 2009 *ApJ* **705** 1336-1355
[141] Markoff S et al. 2008 *ApJ* **681** 905-924
[142] Melia F, Bromley B C, Liu S and Walker C K 2001 *ApJ* **554** L37-L40
[143] Quataert E 2002 *ApJ* **575** 855-859
[144] Pellegrini S, Siemiginowska A, Fabbiano G, Elvis M, Greenhill L, Soria R, Baldi A and Kim D W 2007 *ApJ* **667** 749-759
[145] Pellegrini S 2005 *ApJ* **624** 155-161
[146] Markoff S, Falcke H, Yuan F and Biermann P L 2001 *A&A* **379** L13-L16
[147] Cramphorn C K and Sunyaev R A 2002 *A&A* **389** 252-270
[148] Gillessen S et al. 2012 *Nature* **481** 51-54
[149] Gillessen et al. 2013 *ApJ* **774** article id 44
[150] Lonsdale C J et al. 2003 *PASP* **115** 897-927
[151] Giavalisco M et al. 2004 *ApJ* **600** L93-L98
[152] Hickox R C et al. 2007 *ApJ* **671** 1365-1387
[153] Kim M et al. 2007 *ApJS* **169** 401-429
[154] Scoville N et al. 2007 *ApJS* **172** 1-8
[155] Davis M et al. 2007 *ApJ* **660** L1-L6
[156] Polletta M d C et al. 2006 *ApJ* **642** 673-693
[157] Donley J L et al. 2012 *ApJ* **748** 142
[158] Barmby P et al. 2006 *ApJ* **642** 126-139
[159] Xue Y Q et al. 2012 *ApJ* **758** 129
[160] Civano F et al. 2012 *ApJS* **201** 30

[161] Kim M, Wilkes B J, Kim D-W, Green P J, Barkhouse W A, Lee M G, Silverman J D and Tananbaum H D 2007 *ApJ* **659** 29-51
[162] Barger A J and Cowie L L 2005 *ApJ* **635** 115-122
[163] Fiore F et al. 2003 *A&A* **409** 79-90
[164] Babić; A, Miller L, Jarvis M J, Turner T J, Alexander D M and Croom S M 2007 *A&A* **474** 755-762
[165] Civano F et al. 2011 *ApJ* **741** 91
[166] Aird J et al. 2010 *MNRAS* **401** 2531-2551
[167] Silverman J D et al. 2008 *ApJ* **679** 118-139
[168] Gilli R, Comastri A and Hasinger G 2007 *A&A* **463** 79-96
[169] Hickox R C and Markevitch M 2007 *ApJ* **671** 1523-1530
[170] Luo B et al. 2011 *ApJ* **740** 37
[171] Maiolino R and Rieke G H 1995 *ApJ* **454** 95
[172] Huchra J and Burg R 1992 *ApJ* **393** 90-97
[173] Lawrence A and Elvis M 1982 *ApJ* **256** 410-426
[174] Simpson C 2005 *MNRAS* **360** 565-572
[175] Sazonov S et al. 2012 *ApJ* **757** 181
[176] Burlon D, Ajello M, Greiner J, Comastri A, Merloni A and Gehrels N 2011 *ApJ* **728** 58
[177] Hasinger G 2008 *A&A* **490** 905-922
[178] Treister E and Urry C M 2006 *ApJ* **652** L79-L82
[179] La Franca F et al. 2005 *ApJ* **635** 864-879
[180] Iwasawa K et al. 2012 *A&A* **546** A84
[181] Gilli R, Comastri A, Vignali C, Ranalli P and Iwasawa K 2010 *AIPC* **1248** 359-364
[182] Dwelly T and Page M J 2006 *MNRAS* **372** 1755-1775
[183] Willott C J, Rawlings S, Blundell K M and Lacy M 2000 *MNRAS* **316** 449-458
[184] Barthel P D 1989 *ApJ* **336** 606-611
[185] Grimes J A, Rawlings S and Willott C J 2005 *MNRAS* **359** 1345-1355
[186] Polletta M, Weedman D, Hönig S, Lonsdale C J, Smith H E and Houck J 2008 *ApJ* **675** 960-984
[187] Treister E, Krolik J H and Dullemond C 2008 *ApJ* **679** 140-148
[188] Falcke H, Gopal-Krishna and Biermann P L 1995 *A&A* **298** 395
[189] Lawrence A 1991 *MNRAS* **252** 586-592
[190] Ogle P, Whysong D and Antonucci R 2006 *ApJ* **647** 161-171
[191] Hardcastle M J, Evans D A and Croston J H 2006 *MNRAS* **370** 1893-1904
[192] Matt G et al. 1997 *A&A* **325** L13-L16
[193] Fiore F et al. 2012 *A&A* **537** A16
[194] Fiore F et al. 2009 *ApJ* **693** 447-462
[195] Daddi E et al. 2007 *ApJ* **670** 156-172
[196] Bassani L et al. 2006 *ApJ* **636** L65-L68
[197] Panessa F, Bassani L, Cappi M, Dadina M, Barcons X, Carrera F J, Ho L C and Iwasawa K 2006 *A&A* **455** 173-185
[198] Cappi M et al. 2006 *A&A* **446** 459-470
[199] Risaliti G, Maiolino R and Salvati M 1999 *ApJ* **522** 157-164
[200] Cappelluti N et al. 2012 *MNRAS* **427** 651-663
[201] Alexander D M et al. 2011 *ApJ* **738** 44
[202] Feruglio C, Daddi E, Fiore F, Alexander D M, Piconcelli E and Malacaria C 2011 *ApJ* **729** L4
[203] Gilli R et al. 2011 *ApJ* **730** L28
[204] Chartas G et al. 2000 *ApJ* **542** 655-666
[205] Schwartz D A et al. 2000 *ApJ* **540** L69
[206] Celotti A, Ghisellini G and Chiaberge M 2001 *MNRAS* **321** L1-L5

[207] Worrall D M 2009 *A&ARv* **17** 1-46
[208] Worrall D M et al. 2008 *ApJ* **673** L135-L138
[209] Croston J H et al. 2009 *MNRAS* **395** 1999-2012
[210] Wang J, Fabbiano G, Elvis M, Risaliti G, Mundell C G, Karovska M, and Zezas A 2011 *ApJ* **736** 62-70
[211] McCarthy P J, van Breugel W, Spinrad H and Djorgovski S 1987 *ApJ* **321** L29-L33
[212] Best P N, Röttgering H J A and Longair M S 2000 *MNRAS* **311** 23-36
[213] Longair M S, Best P N and Röttgering H J A 1995 *MNRAS* **275** L47-L51
[214] Carilli C L, Harris D E, Pentericci L, Röttgering H J A, Miley G K, Kurk J D and van Breugel W 2002 *ApJ* **567** 781-789
[215] Massaro F et al. 2010 *AIPC* **1248** 473-474
[216] Hardcastle M J, Massaro F, Harris D E, Baum S A, Bianchi S, Chiaberge M, Morganti R, O'Dea C P and Siemiginowska A 2012 *MNRAS* **424** 1774 1789
[217] O'Dea C P et al. 2002 *AJ* **123** 2333-2351
[218] Schwartz D A 2002 *ApJ* **569** L23-L26
[219] Cheung C C, Stawarz Ł, Siemiginowska A, Gobeille D, Wardle J F C, Harris D E and Schwartz D A 2012 *ApJ* **756** L20
[220] Massaro F et al. 2012 *ApJS* **203** 31
[221] Massaro F, Harris D E, Tremblay G R, Liuzzo E, Bonafede A and Paggi A 2013 *ApJS* **206** 7
[222] Blundell K M and Fabian A C 2011 *MNRAS* **412** 705-710
[223] Gültekin K et al. 2009 *ApJ* **698** 198-221
[224] Ferrarese L and Merritt D 2000 *ApJ* **539** L9-L12
[225] Gebhardt K et al. 2000 *ApJ* **539** L13-L16
[226] Hopkins P F, Cox T J, Kereš D and Hernquist L 2008 *ApJS* **175** 390-422
[227] Kormendy J and Kennicutt R C Jr 2004 *ARA&A* **42** 603-683
[228] Dekel A, Sari R and Ceverino D 2009 *ApJ* **703** 785-801
[229] Bournaud F, Dekel A, Teyssier R, Cacciato M, Daddi E, Juneau S and Shankar F 2011 *ApJ* **741** L33
[230] Fabbiano G, Wang J, Elvis M and Risaliti G 2011 *Nature* **477** 431-434
[231] Koss M, Mushotzky R, Treister E, Veilleux S, Vasudevan R, Miller N, Sander D B, Schawinski K and Trippe M 2011 *ApJ* **735** L42-L48
[232] Comerford J M, Pooley D, Gerke B F and Madejski G M 2011 *ApJ* **737** L19
[233] Green P J, Myers A D, Barkhouse W A, Mulchaey J S, Bennert V N, Cox T J and Aldcroft T L 2010 *ApJ* **710** 1578-1588
[234] Starikova S et al. 2011 *ApJ* **741** 15
[235] Hopkins P F, Hernquist L, Cox T J and Kereš D 2008 *ApJS* **175** 356-389
[236] Silverman J D et al. 2011 *ApJ* **743** 2
[237] Schawinski K, Treister E, Urry C M, Cardamone C N, Simmons B and Yi S K 2011 *ApJ* **727** L31
[238] Cisternas M et al. 2011 *ApJ* **726** 57
[239] Xue Y Q, Brandt W N, Luo B, Rafferty D A, Alexander D M, Bauer F E, Lehmer B D, Schneider D P and Silverman J D 2010 *ApJ* **720** 368-391
[240] Lutz D et al. 2010 *ApJ* **712** 1287-1301
[241] Mullaney J R et al. 2012 *MNRAS* **419** 95-115
[242] Mullaney J R, Daddi E, Béthermin M, Elbaz D, Juneau S, Pannella M, Sargent M T, Alexander D M and Hickox R C 2012 *ApJ* **753** L30
[243] Rosario D J et al. 2012 *A&A* **545** A45
[244] Netzer H 2009 *MNRAS* **399** 1907-1920
[245] Mortlock D J et al. 2011 *Nature* **474** 616-619

[246] Volonteri M and Rees M J 2006 *ApJ* **650** 669-678
[247] Granato G L, Silva L, Lapi A, Shankar F, De Zotti G and Danese L 2006 *MNRAS* **368** L72-L76
[248] Hopkins P F, Hernquist L, Cox T J, Di Matteo T, Robertson B, Springel V 2006 *ApJS* **163** 1-49
[249] Alexander D M, Bauer F E, Chapman S C, Smail I, Blain A W, Brandt W N and Ivison R J 2005 *ApJ* **632** 736-750
[250] Reines A E, Sivakoff G R, Johnson K E and Brogan C L 2011 *Nature* **470** 66-68
[251] Merloni A, Heinz S and di Matteo T 2003 *MNRAS* **345** 1057-1076
[252] Revnivtsev M, Sazonov S, Churazov E, Forman W, Vikhlinin A and Sunyaev R 2009 *Nature* **458** 1142-1144
[253] Fabbiano G 2006 *ARA&A* **44** 323-366 and references therein
[254] Boroson B, Kim D-W and Fabbiano G 2011 *ApJ* **729** 12
[255] Kennicutt R C Jr. 1998 *ARA&A* **36** 189-232 and references therein
[256] Mineo S, Gilfanov M and Sunyaev R 2012 *MNRAS* **419** 2095-2115
[257] Lehmer B D, Alexander D M, Bauer F E, Brandt W N, Goulding A D, Jenkins L P, Ptak A and Roberts T P 2010 *ApJ* **724** 559-571
[258] Mineo S, Gilfanov M and Sunyaev R 2011 *AN* **332** 349
[259] Cowie L L, Barger A J and Hasinger G 2012 *ApJ* **748** 50
[260] Mirabel I F, Dijkstra M, Laurent P, Loeb A and Pritchard J R 2011 *A&A* **528** A149
[261] Kaaret P, Schmitt J and Gorski M 2011 *ApJ* **741** 10
[262] Fregeau J M 2008 *ApJ* **673** L25-L28
[263] Pooley D 2010 *PNAS* **107** 7164-7167
[264] Pellegrini S, Ciotti L and Ostriker J P 2012 *ApJ* **744** 21
[265] Posacki S, Pellegrini S and Ciotti L 2013 arXiv:1304.4085
[266] Kim D-W and Fabbiano G 2003 *ApJ* **586** 826-849
[267] Kim D-W and Fabbiano G 2013 arXiv:1304.6377
[268] Bogdán Á et al. 2012 *ApJ* **753** 140 and references therein
[269] Strickland D K and Heckman T M 2009 *ApJ* **697** 2030-2056
[270] Fabbiano G, Baldi A, King A R, Ponman T J, Raymond J, Read A, Rots A, Schweizer F and Zezas A 2004 *ApJ* **605** L21-L24
[271] Cowie L L and Binney J 1977 *ApJ* **215** 723-732
[272] Fabian A C and Nulsen P E J 1977 *MNRAS* **180** 479-484
[273] Mathews W G and Bregman J N 1978 *ApJ* **224** 308-319
[274] Jones C and Forman W 1984 *ApJ* **276** 38-55
[275] Stewart G C, Fabian A C, Jones C and Forman W 1984 *ApJ* **285** 1-6
[276] Fabian A C 1994 *ARA&A* **32** 277-318
[277] White D A, Jones C and Forman W 1997 *MNRAS* **292** 419
[278] Johnstone R M, Fabian A C and Nulsen P E J 1987 *MNRAS* **224** 75-91
[279] O'Dea C P, Baum S A, Maloney P R, Tacconi L J and Sparks W B 1994 *ApJ* **422** 467-479
[280] Voit G M and Donahue M 1995 *ApJ* **452** 164
[281] Boehringer H, Voges W, Fabian A C, Edge A C and Neumann D M 1993 *MNRAS* **264** L25-L28
[282] Tamura T et al. 2001 *A&A* **365** L87-L92
[283] Peterson J R, Paerels F B S, Kaastra J S, Arnaud M, Reiprich T H, Fabian A C, Mushotzky R F, Jernigan J G and Sakelliou I 2001 *A&A* **365** L104-L109
[284] Allen S W, Dunn R J H, Fabian A C, Taylor G B and Reynolds C S 2006 *MNRAS* **372** 21-30
[285] Forman W, Nulsen P, Heinz S, Owen F, Eilek J, Vikhlinin A, Markevitch M, Kraft R, Churazov E and Jones C 2005 *ApJ* **635** 894-906
[286] Nulsen P E J, McNamara B R, Wise M W and David L P 2005 *ApJ* **628** 629-636
[287] Randall S W et al. 2011 *ApJ* **726** 86

[288] Fabian A C, Sanders J S, Ettori S, Taylor G B, Allen S W, Crawford C S, Iwasawa K, Johnstone R M and Ogle P M 2000 *MNRAS* **318** L65-L68
[289] Fabian A C, Sanders J S, Taylor G B, Allen S W, Crawford C S, Johnstone R M and Iwasawa K 2006 *MNRAS* **366** 417-428
[290] Fabian A C, Sanders J S, Allen S W, Crawford C S, Iwasawa K, Johnstone R M, Schmidt R W and Taylor G B 2003 *MNRAS* **344** L43-L47
[291] Forman W et al. 2007 *ApJ* **665** 1057-1066
[292] McNamara B R, Nulsen P E J, Wise M W, Rafferty D A, Carilli C, Sarazin C L and Blanton E L 2005 *Nature* **433** 45-47
[293] McNamara B R and Nulsen P E J 2007 *ARA&A* **45** 117-175 and references therein
[294] Hlavacek-Larrondo J, Fabian A C, Edge A C, Ebeling H, Sanders J S, Hogan M T and Taylor G B 2012 *MNRAS* **421** 1360-1384
[295] McDonald M et al. 2012 *Nature* **488** 349-352
[296] McDonald M, Benson B, Veilleux S, Bautz M W and Reichardt C L 2013 *ApJ* **765** L37
[297] Million E T, Werner N, Simionescu A and Allen S W 2011 *MNRAS* **418** 2744-2753
[298] Kirkpatrick C C, Gitti M, Cavagnolo K W, McNamara B R, David L P, Nulsen P E J and Wise M W 2009 *ApJ* **707** L69-L72
[299] Kirkpatrick C C, McNamara B R and Cavagnolo K W 2011 *ApJ* **731** L23
[300] Randall S, Nulsen P, Forman W R, Jones C, Machacek M, Murray S S and Maughan B 2008 *ApJ* **688** 208-223
[301] Markevitch M and Vikhlinin A 2007 *PhR* **443** 1-53 and references therein
[302] Markevitch M et al. 2000 *ApJ* **541** 542-549
[303] Spitzer Jr L 1962 *Physics of Fully Ionized Plasmas* (2nd ed) Interscience New York
[304] Ascasibar Y and Markevitch M 2006 *ApJ* **650** 102-127
[305] Blanton E L, Randall S W, Clarke T E, Sarazin C L, McNamara B R, Douglass E M and McDonald M 2011 *ApJ* **737** 99
[306] Roncarelli M, Ettori S, Dolag K, Moscardini L, Borgani S and Murante G 2006 *MNRAS* **373** 1339-1350
[307] Sanders J S and Fabian A C 2007 *MNRAS* **381** 1381-1399
[308] Simionescu A et al. 2011 Sci **331** 1576
[309] Vikhlinin A et al. 2013, in preparation
[310] Cen R and Ostriker J P 1999 *ApJ* **514** 1-6
[311] Fukugita M, Hogan C J and Peebles P J E 1998 *ApJ* **503** 518
[312] Fang T, Buote D A, Humphrey P J, Canizares C R, Zappacosta L, Maiolino R, Tagliaferri G and Gastaldello F 2010 *ApJ* **714** 1715-1724
[313] Hoekstra H, Hsieh B C, Yee H K C, Lin H and Gladders M D 2005 *ApJ* **635** 73-85
[314] Anderson M E and Bregman J N 2011 *ApJ* **737** 22
[315] Dunkley J et al. 2009 *ApJS* **180** 306-329
[316] Dai X, Anderson M E, Bregman J N and Miller J M 2012 *ApJ* **755** 107
[317] Humphrey P J, Buote D A, Canizares C R, Fabian A C and Miller J M 2011 *ApJ* **729** 53
[318] Ma C-J, Ebeling H and Barrett E 2009 *ApJ* **693** L56-L60
[319] Allen S W, Rapetti D A, Schmidt R W, Ebeling H, Morris R G and Fabian A C 2008 *MNRAS* **383** 879-896
[320] Clowe D, Bradač M, Gonzalez A H, Markevitch M, Randall S W, Jones C and Zaritsky D 2006 *ApJ* **648** L109-L113
[321] Markevitch M, Gonzalez A H, Clowe D, Vikhlinin A, Forman W, Jones C, Murray S and Tucker W 2004 *ApJ* **606** 819-824
[322] Randall S W, Markevitch M, Clowe D, Gonzalez A H and Bradač M 2008 *ApJ* **679** 1173-1180

[323] Bradač M, Allen S W, Treu T, Ebeling H, Massey R, Morris R G, von der Linden A and Applegate D 2008 *ApJ* **687** 959-967
[324] Merten J et al. 2011 *MNRAS* **417** 333-347
[325] Zavala J, Vogelsberger M, Slatyer T R, Loeb A and Springel V 2011 *PhRvD* **83** 123513
[326] Adriani O et al. 2009 *Nature* **458** 607-609
[327] Watson C R, Li Z and Polley N K 2012 *JCAP* **3** 18
[328] Dodelson S and Widrow L M 1994 *PhRvL* **72** 17-20
[329] Strigari L E, Bullock J S, Kaplinghat M, Kravtsov A V, Gnedin O Y, Abazajian K and Klypin A A 2006 *ApJ* **652** 306-312
[330] Vikhlinin A et al. 2009 *ApJ* **692** 1060-1074
[331] Mantz A, Allen S W, Rapetti D and Ebeling H 2010 *MNRAS* **406** 1759-1772
[332] Seljak U, Slosar A and McDonald P 2006 *JCAP* **10** 14
[333] Kowalski M et al. 2008 *ApJ* **686** 749-778
[334] Percival W J et al. 2010 *MNRAS* **401** 2148-2168
[335] Sunyaev R A and Zeldovich Y B 1972 *CoASP* **4** 173
[336] Andersson K, et al. 2011 *ApJ* **738** 48
[337] Benson B A, et al. 2013 *ApJ* **763** 147
[338] Hu W and Sawicki I 2007 *PhRvD* **76** 064004
[339] Schmidt F, Vikhlinin A and Hu W 2009 *PhRvD* **80** 083505
[340] Rapetti D, Blake C, Allen S W, Mantz A, Parkinson D and Beutler F 2013 *MNRAS* **432** 973-985
[341] Foley R J et al. 2011 *ApJ* **731** 86
[342] Jee M J et al. 2011 *ApJ* **737** 59
[343] Hoyle B, Jimenez R, Verde L and Hotchkiss S 2012 *JCAP* **2** 9
[344] Mortonson M J, Hu W and Huterer D 2011 *PhRvD* **83** 023015
[345] Harrison I and Coles P 2011 *MNRAS* **418** L20-L24
[346] Menanteau F et al. 2012 *ApJ* **748** 7
[347] Tozzi P, Santos J S, Nonino M, Rosati P, Borgani S, Sartoris B, Altieri B and Sanchez-Portal M 2013 *A&A* **551** A45
[348] Komatsu E et al. 2011 *ApJS* **192** 18
[349] Diemand J, Zemp M, Moore B, Stadel J and Carollo C M 2005 *MNRAS* **364** 665-673
[350] Newman A B, Treu T, Ellis R S and Sand D J 2011 *ApJ* **728** L39
[351] Morandi A, Limousin M, Rephaeli Y, Umetsu K, Barkana R, Broadhurst T and Dahle H 2011 *MNRAS* **416** 2567-2573